\documentclass[pdftex,12pt]{iopart}
\usepackage{iopams}  
\usepackage{bm} 
\usepackage[pdftex]{color,graphicx}
\usepackage{graphics}

        \newcommand{\laeq}{\raisebox{-.7ex}{$\stackrel{\textstyle<}{\sim}$}\ }
	\newcommand{\gaeq}{\raisebox{-.7ex}{$\stackrel{\textstyle>}{\sim}$}\ }
	\newcommand{\exgra}{{\em e.g.}}
	\newcommand{\idest}{{\em i.e.}}
	
	\newcommand{\iint}{\int\!\!\!\!\int}
	
\begin{document}

\title[Nonlinear dynamics of phase space zonal structures]{Nonlinear dynamics of phase space zonal structures and energetic particle physics in fusion plasmas}

\author{F Zonca$^{1,2}$, L Chen$^{2,3}$, S Briguglio$^1$, G Fogaccia$^1$,
G Vlad$^1$, X Wang$^4$}

\address{$^1$ENEA C. R. Frascati, CP 65-00044 Frascati, Italy}
\address{$^2$Institute for Fusion Theory and Simulation and Department of Physics, Zhejiang University, Hangzhou, 310027 P.R. China}
\address{$^3$Department of Physics and Astronomy, University of California, Irvine CA 92697-4575, U.S.A.}
\address{$^4$Max-Planck-Institut f\"ur Plasmaphysik, Boltzmannstr. 2, Garching D-85748, Germany}
\ead{fulvio.zonca@enea.it}

\begin{abstract}
A  general theoretical framework for investigating nonlinear dynamics of phase space zonal structures is presented in this work. It is then, more specifically, applied to 
the limit where the nonlinear evolution time scale is smaller or comparable to the wave-particle trapping period. In this limit, 
both theoretical and numerical simulation studies show that non-adiabatic frequency chirping and phase locking could lead to secular resonant particle transport on meso- or macro-scales. 
The interplay between mode structures and resonant particles 
then provides the crucial ingredient to properly understand and analyze the nonlinear dynamics of Alfv\'en wave instabilities excited by non-perturbative
energetic particles in burning fusion plasmas. 
Analogies with autoresonance in nonlinear dynamics and with superradiance in free electron lasers are also briefly discussed.
\end{abstract}

\submitto{\NJP}
\maketitle

\section{Introduction}
\label{sec:intro} 

The important role of shear Alfv\'en waves (SAW) instabilities and their interaction with energetic particles/charged fusion products (EP) in burning fusion plasmas is widely acknowledged (cf., \exgra, \cite{chen07a,fasoli07,heidbrink08}), and, historically, it was realized since the pioneering works by Belikov \etal~\cite{belikov68,belikov69}, Rosenbluth and Rutherford~\cite{rosenbluth75} and Mikhailovskii~\cite{mikhailovskii75a,mikhailovskii75b}.

Alfv\'enic oscillations can be excited by EPs as well as thermal plasma components~\cite{zonca99,nazikian06}, and, thus, are characterized by a broad spectrum of wavelengths, frequencies and growth rates, which are generally non-trivially interlinked due to the fact that short wavelength kinetic SAW can be excited by resonant mode conversion~\cite{hasegawa76a}. Therefore, these 
drift Alfv\'en waves (DAW) are expected to play important roles in complex behaviors of burning fusion plasmas~\cite{zonca08b,zonca08d,qiu12,zonca14e}; and may have the features of broad band turbulence with ${\cal O}(1)$ ratio between linear instability growth rate and mode frequency, $|\gamma_L/\omega_0| \sim 1$, or nearly periodic (quasi-coherent) nonlinear fluctuations with $|\gamma_L/\omega_0| \ll 1$~\cite{chen07a}. Meanwhile, such two components of the SAW/DAW spectrum have different effects on EP transports. In this work, the focus is on the nearly periodic (quasi-coherent) nonlinear fluctuations with $|\gamma_L/\omega_0| \ll 1$.

Alfv\'enic fluctuations in burning plasmas have typically low amplitudes, $\epsilon_\delta \equiv |\delta \bm B_\perp|/B_0 \laeq 5 \times 10^{-4}$~\cite{heidbrink08}, where the $\perp$ subscript stands for perpendicular to the equilibrium magnetic field direction $\bm b \equiv \bm B_0/B_0$. Therefore, resonant EPs, more than non resonant ones, are expected to play crucial roles in transport processes~\cite{white83,chen84,chen99,white10a,white10b}.
Here, in particular, we consider transport processes in two-dimensional (2D) magnetized plasma equilibria 
with two periodic angle-like coordinates, $(\theta,\zeta)$, one of which ($\zeta$) defines the equilibrium symmetry.
Thus, 
phase space structures that are of direct interest are those obtained by averaging out dependences on angle-like variables. In fact, in the general procedure used for deriving kinetic
equations in weakly-nonideal systems \cite{vanhove55,prigogine62,balescu63}, these structures are those dominating the ``principal series'' of secular terms obtained by formal perturbation 
expansion in powers of fluctuating fields; which can be summed up and yields the solution of the corresponding Dyson equation \cite{altshul65,altshul66}. 
These phase space structures may be generally referred to as ``phase space zonal structures''~\cite{zonca14e,zonca13,chen14} (PSZS), by analogy with 
the meso-scale configuration space patterns spontaneously generated by drift-wave turbulence (DWT); \idest, zonal flows and currents/fields or, more generally, zonal structures (ZS)~\cite{hasegawa79,lin98,diamond05,itoh06}. Here and in the following, meso-scales are intermediate length scales between the SAW/DAW/DWT micro-scales, typically ordered as the 
perpendicular fluctuation wavelength, $\lambda_\perp$, and the system-size macro-scales, characteristic of equilibrium plasma profiles and macroscopic fluctuations. In magnetized fusion plasmas, at the leading order,  ZS are independent of $(\theta,\zeta)$; due to finite magnetic shear, which suppresses convective cells
\cite{taylor71,okuda73}. Hence, ZS do not significantly contribute to cross-field transport \cite{shukla84}. 

Both ZS and PSZS in fusion plasmas are transverse fluctuations, since they have $\bm k \cdot \bm b = 0$ everywhere, with $\bm k$ the wave vector. Meanwhile, they describe
the corrugations of equilibrium radial profiles~\cite{zonca05,zonca06b,chen07b} as consequence of fluctuation induced transport processes, due to emission and reabsorption of 
(toroidal equilibrium) symmetry breaking perturbations \cite{zonca14e}. In general, the time scale of ZS and PSZS evolution is the same as that of the underlying fluctuations,
with important consequences on the meso-scale dynamics of SAW/DAW and DWT, which may become nonlocal and be characterized, \exgra, by avalanches \cite{wang07b,difpradalier10,jolliet12,gurcan13}.
For understanding and explaining different behaviors of PSZS, it is important to compare the wave-particle trapping time, $\tau_B$, with the characteristic time of nonlinear evolution of PSZS, $\tau_{NL}$. When $\tau_B \ll \tau_{NL}$, there exists an adiabatic (action) invariant; and phase space density is preserved inside the structure separatrix. When this occurs, phase space holes and clumps~\cite{bernstein57,berk70,dupree70,dupree72,dupree82,berman83,tetreault83} are limiting cases of PSZS, and the nonlinear evolution is referred to as ``adiabatic''. Meanwhile, when $\tau_B \sim \tau_{NL}$, no adiabatic (action) invariant exists in the phase space, resonant particle motion can be secular, and the dynamics is ``non-adiabatic''. Thus, different nonlinear behavior and EP transport are expected, in general, depending on the relative ordering of $\tau_B$ and $\tau_{NL}$. This also suggests that qualitative and quantitative differences may be expected depending on whether fluctuations are externally controlled, such as in a ``driven'' system, or are resulting as consequence of plasma instabilities. 
In fact, while, in the former case, phase space dynamics is typically adiabatic; the evolution, in the instability case, often becomes non-adiabatic such that, at saturation, $\tau_{NL} \gaeq \gamma_L^{-1}$ and $\tau_B \gaeq \gamma_L^{-1}$ \cite{chen14}.

In this work, we analyze the case of nonlinear EP distribution functions that dynamically evolve on characteristic ``transport'' time scales, which are generally of the same order of the nonlinear time scale of the underlying instabilities~\cite{chen14,zonca05,zonca06b}; \idest, typically in the non-adiabatic regime. Furthermore, we assume that resonant wave-particle interaction between EPs and Alfv\'enic fluctuations in burning plasmas is non-perturbative; that is, EPs play significant roles in determining mode frequency and mode structure. Non-perturbative EP response to SAW/DAW instabilities is the crucial element  characterizing PSZS nonlinear dynamics discussed in this work. By addressing the general non-adiabatic case, $\tau_{NL}\sim \gamma_L^{-1} \sim \tau_B$, the present approach can handle both the nonlinear saturation of EP driven Alfv\'enic instabilities, $\tau_{NL} \sim \gamma_L^{-1}  \laeq \tau_B$, as well as longer time scale nonlinear dynamics, $\tau_{NL} \gg \tau_B$, such as wave-particle trapping, quasi-linear diffusion, and collisional effects. The
general theoretical framework is illustrated in Sec.~\ref{sec:theory}, which demonstrates that ``radial decoupling'', due to resonant EPs exploring finite radial structures of fluctuations in nonuniform plasmas, becomes increasingly more important than and eventually dominates ``resonance detuning'', due to changing wave-particle phases. Such transition occurs when the size of nonlinear particle orbits is comparable with the radial fluctuation wavelength and corresponds to the onset of nonlocal behaviors; \idest, of meso-scale dynamics. Mode frequency chirping and ``phase locking''~\cite{white83,chen84,zonca13,chen14,zonca05,wang12,vlad13,chen14b,zonca14d,briguglio14} are shown to be the nonlinear dynamics by which secular resonant particle transport can occur on meso- and macro-scales. In order to illustrate theoretical framework and underlying concepts, Sec.~\ref{sec:theory} focuses on the obtained results and physics implications, while detailed analytical derivations are given in \ref{app:res2D}. Furthermore, analogies with ``autoresonance''~\cite{meerson90,fajans01} in nonlinear systems are pointed out in order to illustrate similarities and differences in the behaviors of ``driven'' systems and unstable fusion plasmas in the presence of non-perturbative EP free energy sources.

This work aims at presenting fundamental aspects of PSZS nonlinear dynamics to a broad readership; and, thus, presents these issues from different perspectives. For this reason, after discussing
the cornerstones of theoretical framework in Sec.~\ref{sec:theory}, we adopt numerical simulations to give examples of how these physics manifest themselves in practice. Thus,
meso-scale dynamics of radially localized Energetic Particle Modes (EPMs \cite{chen94}) are illustrated in Sec.~\ref{sec:EPMloc}, which demonstrates non-perturbative EP behaviors, phase locking and non-adiabatic frequency chirping by numerical simulation results with the extended Hybrid MHD Gyrokinetic Code (XHMGC)~\cite{briguglio95,wang11}. Section~\ref{sec:goveq}, meanwhile, provides specific derivations and discussions of PSZS nonlinear dynamics within the framework of nonlinear gyrokinetic theory \cite{frieman82,brizard07}. In Sec.~\ref{sec:dyson}, the PSZS evolution equations are specialized
to the simple case of precessional resonance, for which the nonlinear dynamics is that of 1D non-autonomous nonuniform system. This case allows to derive the general form of the 
corresponding Dyson equation; and to illustrate the onset of nonlocal behaviors as novel feature in contrast to the case of a 1D uniform plasma.
As application of the general theoretical framework of Sec.~\ref{sec:dyson}, Sec.~\ref{sec:epmaval} analyzes convective amplification of EPM wave packets as avalanches; soliton-like structures in active media, accompanied by secular EP transports on macro-scales. EPM avalanches can be taken as paradigmatic example of ``radial decoupling'' and non-perturbative EP behaviors. In fact, because of ``phase locking'', wave-particle resonance is limited only by the finite radial mode structure. At the same time, non-perturbative EP response causes the mode structure to adapt to the modified EP source. Ultimately, the spatiotemporal structure of EPM avalanches is, thus, the self-consistent result of ``radial decoupling'' and non-perturbative EP response. In Sec.~\ref{sec:epmaval}, as  further evidence of general implications of these physics, analogies with ``superradiance''~\cite{dicke54} operation regime~\cite{bonifacio90} in free electron lasers (FEL) are also discussed.
Section~\ref{sec:summary} finally gives summarizing remarks and conclusions, while \ref{app:GFLDR} and~\ref{app:Laplace} are devoted to illustrating detailed expressions and technical derivations, and are proposed to interested readers only.

\section{Theoretical framework}
\label{sec:theory}

We consider an axisymmetric toroidal magnetized plasma, and use $(r,\theta,\zeta)$ toroidal flux coordinates with straight magnetic fields lines. Here, $r$ is 
a radial-like magnetic flux coordinate, $\zeta$ is the toroidal symmetry coordinate (cf. Sec. \ref{sec:intro}), $\bm B_0$ is represented as in Eq.~(\ref{eq:B0}), and the magnetic field line pitch $q \equiv \bm b \cdot \bm \nabla \zeta/ \bm b \cdot \bm \nabla \theta = q(r)$ is a flux function, given in Eq.~(\ref{eq:straight}). Details about the adopted equilibrium representation and flux coordinates are given in  \ref{app:res2D}. Introducing $\beta = 8\pi P_0/B_0^2$ as the ratio of kinetic to magnetic pressures, we consider sufficiently low-$\beta$ plasma equilibria 
with $|\bm \nabla_\perp|\gg |\bm \nabla_\parallel|$, such that compressional Alfv\'en waves are suppressed and 
magnetic field compression can be solved explicitly via perpendicular pressure balance
\begin{equation}
\bm \nabla_\perp  \left( B_0 \delta B_\parallel + 4\pi \delta P_\perp \right ) \simeq 0 \;\; . \label{eq:perpprebal}
\end{equation} 
Thus, $\delta B_\parallel$ can be eliminated from governing equations of SAW/DAW and DWT, which are then described in terms of two fluctuating scalar fields; \idest, the scalar potential $\delta \phi$ and the parallel (to $\bm b$) vector potential $\delta A_\parallel$. 
As usual, subscript $0$ denotes equilibrium quantities and $\delta$ is the prefix of fluctuating fields. 

As anticipated in Sec.~\ref{sec:intro}, we assume that the characteristic nonlinear time, $\tau_{NL}$, satisfies the following ordering \cite{chen14,zonca14a,zonca14b}
\begin{equation}
1 \gg | \omega_0 \tau_{NL}|^{-1} \sim | \gamma_L/\omega_0 |  \gg \epsilon_\omega \equiv |\omega_0/\Omega| \;\; , \label{eq:timescaleorde}
\end{equation}
with $\Omega = e B_0/(m c)$ the cyclotron frequency, which generally holds for SAW/DAW excited by EPs in fusion plasmas \cite{zonca14e} but also applies to 
DWT in a variety of cases of practical interest \cite{chen00,chen01,chen04,zonca04b}. The ordering of Eq.~(\ref{eq:timescaleorde}) allows us to address
the nonlinear saturation of the considered instabilities, as well as to investigate post-saturation dynamics.

\subsection{Onset of nonlocal behaviors: resonance detuning and radial decoupling}
\label{sec:nonloc}

Based on Eq.~(\ref{eq:timescaleorde}) and on the low SAW fluctuation amplitudes (cf. Sec.~\ref{sec:intro}), it can be shown that resonant EP motion is slightly modified by fluctuations in one poloidal bounce/transit time. Therefore, fluctuation induced transport is a cumulative effect on resonant EP over many bounce/transit times, which can generally be secular, \idest, not bounded in time.
Detailed wave-particle interactions in axisymmetric toroidal magnetized plasmas are analyzed and discussed, 
under these assumptions,
in \ref{app:res2D}. There, 
it is shown that the effective action of a generic fluctuation field, $f(r,\theta,\zeta)$ (time dependences are left implicit), decomposed in Fourier harmonics, can be represented as
\begin{eqnarray}
f(r,\theta,\zeta) & = & \sum_{m,n \in \mathbb Z} e^{in\zeta-im\theta} f_{m,n}(r) \nonumber \\
& \rightarrow & \sum_{m,n,\ell \in \mathbb Z} e^{i \left( n \bar\omega_d + \ell \omega_b\right) \tau + i \Theta_{m,n,\ell} } {\cal P}_{m,n,\ell} \circ f_{m,n} (\bar r + \Delta r) 
\;\; , \label{eq:flift0}
\end{eqnarray}
where $\omega_b$ is the bounce/transit frequency for magnetically trapped/circulating particles and $\bar \omega_d$ is the toroidal precessional frequency, which are defined in Eqs.~(\ref{eq:omegab}) and~(\ref{eq:omegad}), respectively. Furthermore, $\tau$ is a time-like parameter that identifies the particle position along its integrable trajectory in the reference equilibrium, $\ell$ is the bounce harmonic, $\bar r$ is the bounce averaged radial coordinate, defined in Eq.~(\ref{eq:rparm}); and the projection operators ${\cal P}_{m,n,\ell}$ are defined by
Eq.~(\ref{eq:calpmnapp}). Meanwhile, $\Delta r$ and $\Theta_{m,n,\ell}$ are the nonlinear radial particle displacement and the nonlinear wave-particle phase shift, defined in Eqs.~(\ref{eq:deltar})  and~(\ref{eq:thetamnl}), respectively. Mathematically, Eq.~(\ref{eq:flift0}) is a lifting of $f(r,\theta,\zeta)$ to the phase space, using the time-like parameter $\tau$ to map the effective action of $f(r,\theta,\zeta)$ as the particle moves along its integrable equilibrium trajectory, identified by its constants of motion. Thus, Eq.~(\ref{eq:flift0}) reveals the dual nature of fluctuating fields in kinetic descriptions of collisionless plasmas: (i) the field observed in the laboratory frame, which enters in the evolution equation of mode structures; and (ii) the field effectively experienced by the particle in the particle-moving frame, which enters in the description of wave-particle resonances (cf. \ref{app:wpres}). 

For $\Theta_{m,n,\ell}=0$ and $\Delta r=0$, a fluctuation $\propto \exp (-i \omega_0 t)$ readily yields wave-particle resonance condition in the form of Eqs.~(\ref{eq:trapres}) and~(\ref{eq:circres}), respectively, for magnetically trapped and circulating particles. Meanwhile, in the presence of fluctuations, Eq.~(\ref{eq:flift0}) accounts for the cumulative bounce/transit averaged fluctuation effects, discriminating between ``resonance detuning'', $\propto \exp \left( i \Theta_{m,n,\ell} \right)$, and ``radial decoupling'' $\propto {\cal P}_{m,n,\ell} \circ f_{m,n} (\bar r + \Delta r)$.
These concepts, introduced theoretically in Ref. \cite{chen14} and further discussed below and in \ref{app:wpresNL}, are important for understanding and explaining nonlinear dynamics of SAW/DAW in nonuniform plasmas. 
In fact, based on the ``universal description of a nonlinear resonance''~\cite{chirikov79} as non-autonomous system with one degree of freedom, which yields a one-dimensional (1D) nonlinear pendulum Hamiltonian also referred to as ``standard Hamiltonian''~\cite{lichtenberg83}, resonance detuning is a general common feature of wave-particle interactions. On the contrary, radial decoupling exists only in nonuniform plasmas; and becomes important when $\Delta r$ is comparable with the perpendicular wavelength, $\lambda_\perp$, of perturbation structures $\propto {\cal P}_{m,n,\ell} \circ f_{m,n}$, which can be estimated as $\lambda_\perp \sim \epsilon_\Delta |nq'|^{-1}$, $\epsilon_\Delta < 1$ being a control parameter depending on the considered fluctuation
and the nonuniform plasma equilibrium, characterizing the mode transverse scale length \cite{chen14}.
Introducing the ``finite interaction length'', $\Delta r_L$, as the value of $\Delta r$ at which wave-particle resonance is lost due to plasma nonuniformity, Eq.~(\ref{eq:flift0}) shows that a quantitative and qualitative difference in nonlinear dynamics must be expected depending on the relative ordering of 
$\Delta r_L$ and $\lambda_\perp$. It is possible to show that $\Delta r_L \sim 3 |\gamma_L/\omega_0| \lambda_n^{-1} r$, with $\lambda_n = |n r q'|$ for circulating particles and $\lambda_n = 1$ for magnetically trapped particles (cf. \ref{app:finiteness}, where the corresponding ``finite interaction time'' is introduced as well). For $\Delta r_L \ll \lambda_\perp$, the plasma responds like a uniform system, and similarities can be drawn between EP-SAW interactions in burning plasmas of fusion interest and a 1D uniform beam-plasma system near marginality~\cite{berk90c}. 
This limit offers obvious advantages of adopting a simple 1D system for complex dynamics studies~\cite{berk90c,breizman11b}. 
However, as demonstrated first in numerical studies of Ref.~\cite{briguglio98} and further illustrated by recent numerical works \cite{wang12,zhang12}, important differences and novel behaviors of EP-SAW interactions in nonuniform plasmas appear for $\Delta r_L \gaeq \lambda_\perp$; \idest\ (cf. \ref{app:finiteness})
\begin{equation}
|\gamma_L/\omega_0| \gaeq \frac{\lambda_n}{3 |n r q'|} \epsilon_\Delta \;\; , \label{eq:trans}
\end{equation}
where $\epsilon_\Delta <1$ controls the perpendicular fluctuation scale length. Note that Eq.~(\ref{eq:trans}) depends on the type of resonance as well as system geometry and nonuniformity through the control parameters $\lambda_n$ and $\epsilon_\Delta$ \cite{zonca13,chen14}.
These numerical simulation studies,
illustrate the importance of mode structures in the nonlinear dynamics of SAW/DAW excited by EPs, consistent with Eq.~(\ref{eq:trans}) and theoretical predictions \cite{chen95,zonca96a}; and with recent experimental observations in DIII-D, interpreted and explained on the basis of dedicated  numerical simulations \cite{wang13}. Thus, Eq.~(\ref{eq:trans}) can be considered as a condition for the onset of nonlocal behaviors in SAW/DAW interactions with EPs; and as indication that the system behaves as a truly 3D nonuniform plasma, despite one isolated resonance can still be described as non-autonomous system with one degree of freedom \cite{zonca14e,chen14}. A systematic investigation of these issues is given in Ref.~\cite{briguglio14}, based on numerical simulation experiments analyzed with phase space diagnostics developed from Hamiltonian mapping techniques.

\subsection{Phase locking and non-adiabatic phase space dynamics}
\label{sec:lock}

Equation~(\ref{eq:trans}) suggests 
that onset of nonlocal behaviors and possibly novel features of SAW/DAW-EP interactions 
with respect to the beam-plasma system could emerge for sufficiently strong EP drive~\cite{zonca05}. This poses an interesting issue on whether EP dynamics can be considered perturbatively.
For example, the ``bump-on-tail paradigm'' \cite{chen14,chen14b,chen13} 
(cf. a recent review by Breizman and Sharapov~\cite{breizman11b}) 
considers perturbative EP response, which does not 
modify the thermal plasma dielectric behavior~\cite{berk90c}. 
Thus, in this paradigm, nonlinear dynamics is dominated by wave particle trapping, whose frequency $\omega_B$ (cf. \ref{app:finiteness}) provides the shortest characteristic temporal scale, $\tau_{NL}^{-1}$. Consequently, there then exists a conserved phase space action  connected with wave particle trapping; and nonlinear evolution of EP driven AEs becomes 
that of phase space holes and clumps, defined as regions with, respectively, lack of density and excess of density w.r.t.\
the surrounding phase space. 
These physics have been extensively investigated since the pioneering work by Bernstein, Greene and Kruskal (BGK)~\cite{bernstein57}; and used for analyzing nonlinear behaviors  of 1D uniform Vlasov plasmas~\cite{berk70,dupree70,dupree72,dupree82,berman83,tetreault83}, including sources and collisional dissipation~\cite{berk90c,berk90a,berk90b}. If the frequency of hole-clump pairs slowly evolves in time \cite{berk96,berk97,breizman97,berk99,wang12b,wang13b},
it happens adiabatically at a rate $|\dot \omega| \ll \omega_B^2$, set by balancing the rate of energy extraction of the moving holes/clumps in the phase space with the fixed background dissipation\footnote{Note that the original adiabatic theory of Ref.~\cite{berk99} has been recently extended in Refs.~\cite{wang12b,wang13b}, but still remains local in the sense discussed in Sec.~\ref{sec:nonloc} and hereafter. More detailed discussions of this point are given in Refs.~\cite{zonca14e,chen14}.}. 
EP transport in velocity space occurs in ``buckets'' \cite{breizman11a}, similar to ``bucket'' EP transport in real space introduced in Ref. \cite{mynick94}. However, unless diffusive transport process are considered due to many overlapping resonances, the EP transport in real space predicted by the ``bump-on-tail paradigm'' is implicitly local, since it assumes $\Delta r_L \ll \lambda_\perp$ (cf. Sec. \ref{sec:nonloc}). These transport processes are also similar to ``autoresonance'' \cite{meerson90,fajans01}, by which a nonlinear pendulum can be driven to large
amplitude, evolving in time to instantaneously match its frequency with that of an external
drive with sufficiently slow downward frequency sweeping.

It was noted by
Friedland \etal~\cite{friedland06} that spontaneous (``autoresonant'') nonlinear evolution of phase space structures becomes poorly controllable, when the underlying fluctuations are plasma instabilities rather than being imposed externally. In this respect, non-perturbative EP behaviors become particularly crucial for AEs;  when the following condition
\begin{equation}
|\Delta \omega_{EP}| \ll |\Delta \omega_{SAW}| \label{eq:perturbEP}
\end{equation}
is no longer satisfied \cite{zonca14a,zonca14b}. Here, $\Delta \omega_{EP}$ is the EP induced complex frequency shift of AEs; and $\Delta \omega_{SAW}$ is the frequency mismatch between AEs and the closest frequency of the SAW continuous spectrum computed at the peak AE amplitude. When Eq.~(\ref{eq:perturbEP}) is progressively broken, AE mode structures are also increasingly affected by EP dynamic response; consistent with theoretical analyses \cite{zonca14a,zonca14b,chen95,zonca96a} and numerical simulation results \cite{wang12,briguglio98,wang13,deng10,deng12b} as well as experimental observations (cf., \exgra, Refs. \cite{wang13,tobias11}).
Meanwhile, EP response is always non-perturbative for EPMs \cite{chen94}, excited within the SAW continuous spectrum as discrete fluctuations at the frequency
maximizing wave-EP power exchange above the threshold condition set by continuum damping \cite{chen14,zonca14a,zonca14b,zonca00a}. 

Clearly,  for non-perturbative EP dynamics and significant transports over the domain where fluctuations are localized, the interplay between mode evolution and EP radial redistributions has crucial impact on nonlinear dynamics. Two factors are fundamental for the nonlinear mode evolution: (i) the nonlinear frequency shift determined by EP redistributions, consistent with mode dispersion relation; and (ii) the change in growth rate, taking into account mode structure distortions and EP transports. Both effects are constrained, self-consistently, by the mode dispersive properties. For a given nonlinear frequency shift, there always exists a group of resonant EPs, with given phase w.r.t.\ the wave, that satisfies 
\begin{equation}
|\tau_{NL} \dot \Theta_{m,n,l}|, |\tau_{NL}^2 \ddot \Theta_{m,n,l}| \ll 1 \;\; ; \label{eq:phaselock} 
\end{equation}
\idest, ``phase locking''; as shown explicitly for, respectively, circulating and magnetically trapped particles by Eqs.~(\ref{eq:phaselockcirc}) and~(\ref{eq:phaselocktrap}) (cf. \ref{app:finiteness}). These are the resonant EPs that instantaneously maximize wave-particle power exchange \cite{white83,chen84,zonca13,chen14,zonca05,chen14b,zonca14d,briguglio14}, as they minimize resonance detuning. 
At the same time, frequency chirping connected with this process is non-adiabatic $|\dot \omega|\laeq \omega_B^2$, as demonstrated by Eqs.~(\ref{eq:omegaB2}), (\ref{eq:phaselockcirc}) and~(\ref{eq:phaselocktrap}). This means that, with non-perturbative EP response,  phase locking is signature of non-adiabatic frequency chirping and non-local EP redistributions; and {\sl vice versa} \cite{chen14}. Phase locking further  
extends (but not indefinitely) the finite interaction length and the corresponding finite interaction time (cf. Sec. \ref{sec:nonloc} and \ref{app:finiteness}) by a factor $\epsilon_{\dot\omega}^{-1}>1$. 
Here, $\epsilon_{\dot\omega}$ is defined by Eq.~(\ref{eq:epsilonomegadot}) and depends 
on the mode dispersive properties. It denotes the effectiveness of phase locking; and, in general, $\epsilon_{\dot\omega}<1$ requires the EP dynamics being non-perturbative. 
Adiabatic chirping ($|\dot \omega|\ll \omega_B^2$) or fixed frequency modes are characterized by $\epsilon_{\dot\omega}\simeq 1$. 
Taking into account 
the extended finite interaction length, Eq.~(\ref{eq:trans}) can be rewritten as 
\begin{equation}
|\gamma_L/\omega_0| \gaeq \frac{\lambda_n}{3 |n r q'|} \epsilon_\Delta \epsilon_{\dot\omega} \;\; . \label{eq:transext}
\end{equation}
Equation~(\ref{eq:transext}) indicates 
the growth rate threshold for radial mode structures to importantly affect nonlinear dynamics and the transition to non-local EP behaviors.
Detailed analysis, would, in general, require a full
3D kinetic description including realistic equilibrium geometry and nonuniformity \cite{zonca13,chen14}. 
Assuming typical plasma parameters and the fact that generally $\epsilon_\Delta \epsilon_{\dot \omega} \laeq 10^{-1}$, Eq.~(\ref{eq:transext}) yields the estimates that $|\gamma_L/\omega_0|\gaeq 10^{-3}$ for magnetically trapped particles, and $|\gamma_L/\omega_0|\gaeq 10^{-2}$ for circulating particles~\cite{chen07a,zonca13,chen14,zonca05,zonca06b}.

The onset of non-local EP behaviors, above the threshold given by Eq.~(\ref{eq:transext}), corresponds to nonlinear meso-scale dynamics involving mode structures \cite{zonca05,briguglio98,briguglio02,zonca02,vlad04,briguglio07}. We discuss these processes in Sec.~\ref{sec:EPMloc} by numerical simulation experiments, 
constructed {\em ad hoc} to illustrate phase locking and non-adiabatic frequency chirping. However, depending on the considered plasma equilibrium and the corresponding fluctuation spectrum;
and for increasing EP drive strength, phase locking and nonlinear meso-scale dynamics can extend to macro-scales. In this way, phase locked EPs  could be secularly transported outward by the ``mode particle pumping'' mechanism, originally introduced for explaining EP loss due to ``fishbone'' fluctuations~\cite{white83}. Meanwhile, self-consistent interplay between nonlinear mode dynamics and EP transports may give rise to EPM ``avalanches'', by which EPM wave packets are convectively amplified as they propagate radially, until they are quenched by spatial nonuniformity (cf. Sec.~\ref{sec:epmaval} and \ref{app:finiteness}).

\section{Nonlinear dynamics of localized EPMs}
\label{sec:EPMloc}

The important roles of plasma non-uniformities and finite radial mode structures for the nonlinear evolution of EP-driven SAW were first discussed by Briguglio \etal~\cite{briguglio98}, using hybrid MHD gyrokinetic numerical simulations~\cite{park92} applied to the analysis of low-$n$ toroidal AE (TAE) and EPM. 
Further insights into these physics were provided by hybrid MHD gyrokinetic simulations of nonlinear BAE dynamics~\cite{wang12}; employing 
high-resolution phase space numerical diagnostics techniques based on Hamiltonian mapping~\cite{white12}. Similar results have also been illustrated by nonlinear gyrokinetic simulations~\cite{zhang12}.
\begin{figure}
\centerline{\resizebox{0.8\linewidth}{!}{\includegraphics{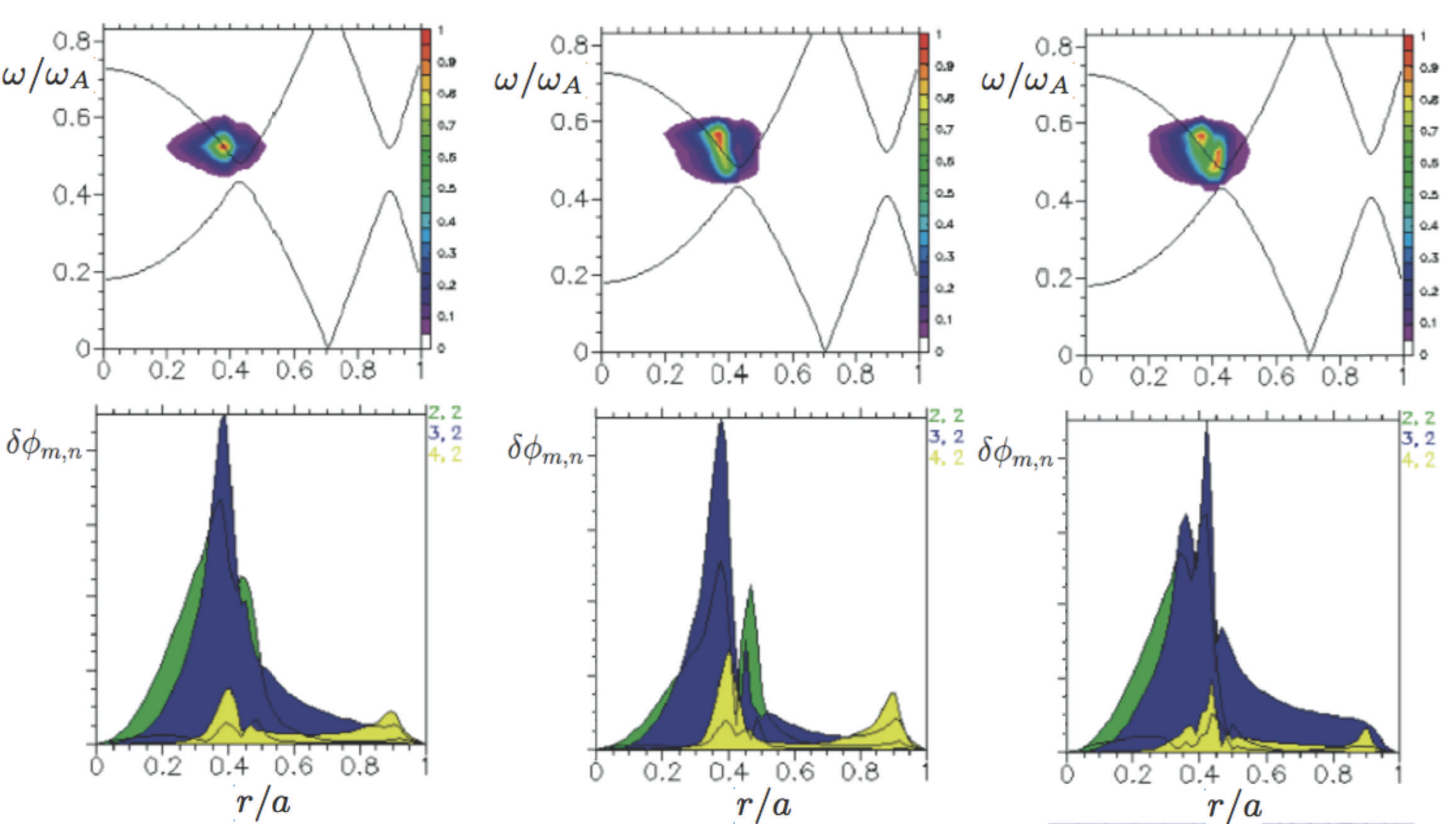}}}
\caption{Three snapshots of the $n=2$ EPM nonlinear evolution (adapted from the original Fig.~42 in Ref.~\cite{briguglio14}): $\omega_A t = 1200$ (left),  $\omega_A t = 1287$ (middle) and $\omega_A t = 1320$ (right), with $\omega_A=v_A/R_0$ computed at the magnetic axis. The top panels are mode intensity contour plots in the $(r/a,\omega/\omega_A)$ plane, with the $n=2$ SAW continuous spectrum shown by the thin black curve in background. Lower panels show the mode structures vs. $r/a$ of individual EPM poloidal harmonics $\delta \phi_{m,n}$: $m=2$ (green; 2nd dominant), $m=3$ (blue; dominant) and $m=4$ (yellow; subdominant). The $\delta \phi_{m,n}$ axis is normalized to the peak of the dominant harmonic. Reprinted with permission from Phys. Plasmas {\bf 21}, 112301 (2014). Copyright 2014, AIP Publishing LLC.}
\label{fig:epmloc}\end{figure}

In order to demonstrate non-perturbative EP behaviors and phase locking more clearly, Briguglio \etal~\cite{briguglio14} have carried out a ``numerical simulation experiment'' with the XHMGC code~\cite{briguglio95,wang11}, investigating the nonlinear dynamics of radially localized EPM near the TAE frequency gap. 
With the aim of controlling the EPM radial localization and inhibiting the onset of EPM convective amplifications (cf. Sec.~\ref{sec:epmaval})~\cite{zonca05,briguglio02,vlad04}, Ref.~\cite{briguglio14} selected a sufficiently flat equilibrium $q$ profile,  $q(r) = 1.1 + 0.8 (r/a)^2$. Adjacent toroidal
frequency gaps in the SAW continuous spectrum were, thus, so far apart that nonlocal couplings were minimized, as shown in Fig.~\ref{fig:epmloc} (upper panels). For a 
tokamak equilibrium with major/minor radii ratio $R_0/a = 10$, Fig.~\ref{fig:epmloc} (upper panels) illustrates the $n=2$ SAW continuous spectrum as thin black curve in the background of the intensity contour plots of an $n=2$ EPM at different times: $\omega_A t = 1200$ (right before mode saturation),  $\omega_A t = 1287$ (during saturation phase) and $\omega_A t = 1320$ (at the end of the EPM burst), with $\omega_A=v_A/R_0$ computed at the magnetic axis and $v_A$ the Alfv\'en speed. The thermal plasma density, $n_i (r) = n_{i0} q^2(0)/q^2(r)$, is chosen such that the SAW frequency gaps are aligned, while the EP guiding-center distribution function (cf. Sec.~\ref{sec:goveq}) is a Maxwellian, $\bar F_{0 EP} = (2\pi v_{EP})^{-3/2} n_{E}(r) \exp [- E/(m_{EP} v_{EP}^2)]$, with $n_{EP}(r) = n_{EP0} \exp [ - \psi^2(r)/\psi^2(a) ]$,  $\psi(r)$ is the poloidal magnetic flux function introduced by Eq.~(\ref{eq:B0}), $n_{EP0}/n_{i0} = 1.75 \times 10^{-3}$, $v_{EP}/(a\Omega_{EP})=0.01$ and $v_{EP}/v_A=1$ (cf. Ref. \cite{briguglio14} for further numerical simulation details).

\begin{figure}
\centerline{\resizebox{0.8\linewidth}{!}{\includegraphics{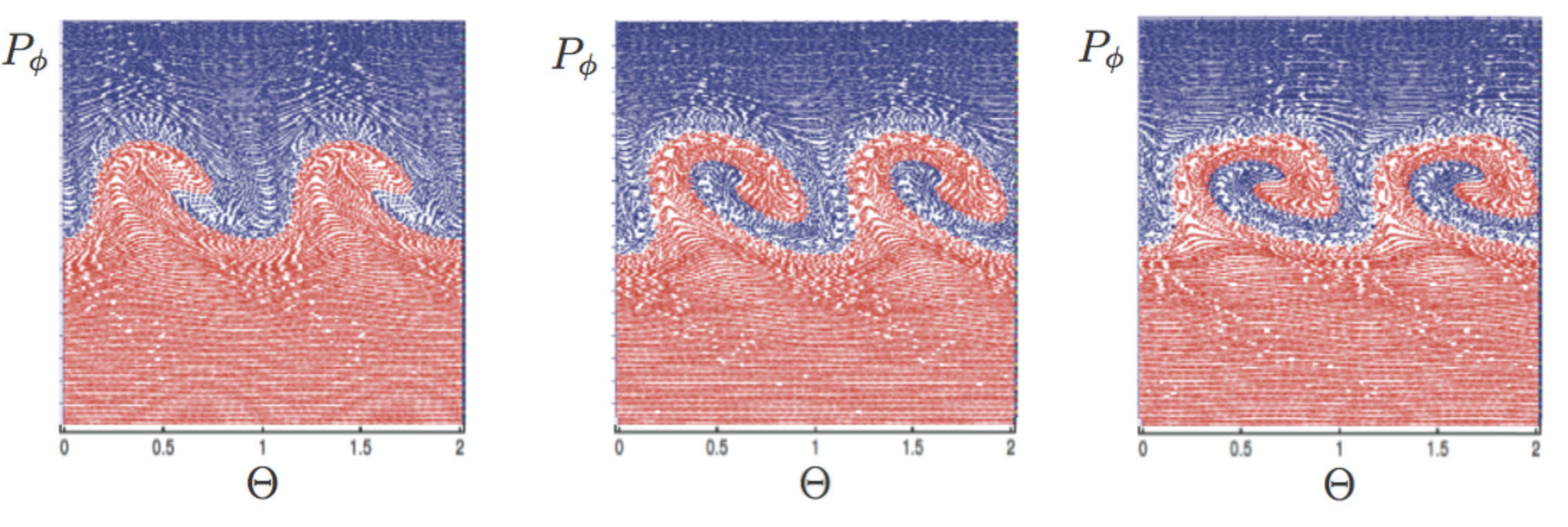}}}
\caption{Kinetic Poincar\'e plots \cite{briguglio14} illustrating the dynamic evolution of EP PSZS for the same snapshots as in Fig.~\ref{fig:epmloc}. Normalization of phase space variables are is discussed in the main text.}
\label{fig:kinepm}
\end{figure}

Kinetic Poincar\'e plots~\cite{briguglio14} in Fig.~\ref{fig:kinepm} show the dynamic evolution of EP PSZS for the same snapshots as in Fig.~\ref{fig:epmloc}.
Particle marker color denotes the initial  $P_\phi$, normalized to $m_{EP} v_{EP} a$, being larger (blue) or smaller (red) than the reference resonance value. The wave particle phase, $\Theta$, meanwhile, is normalized to $2\pi$ and refers to $\Theta_{m,n,\ell}$ (cf. Sec.~\ref{sec:nonloc}) computed for the dominant Fourier and bounce harmonics, $(m,n,\ell) = (3,2,1)$ \cite{briguglio14}, consistent with the transit resonance condition, Eq.~(\ref{eq:circres}). The $P_\phi$ axis is in a.u.\ and the $\Theta$ axis shows the 
range $0\leq \Theta_{3,2,1} \leq 4\pi$ in order to better visualize EP PSZS.
At $\omega_A t = 1200$ in Fig.~\ref{fig:epmloc}(left), the mode is approaching saturation 
when EPs have completed 1/4 oscillation in the instantaneous fluctuation-induced potential well,
as confirmed by the 
corresponding kinetic Poincar\'e plot in Fig.~\ref{fig:kinepm}; and the mode drive is significantly reduced at the original (linear) resonance location. In the further evolution of the mode and after mode saturation, 
the wave-particle power exchange becomes most negative (damping) at the original resonance location, when resonant EPs have completed 1/2 oscillation in the instantaneous fluctuation-induced potential well, as shown in Fig.~\ref{fig:kinepm} for $\omega_A t = 1287$. At the same time, due to the distortion of the EP distribution function, steeper gradient regions are formed at the outer limits of the PSZS, as these are the positions where particles tend to accumulate due to radial decoupling and diminishing fluctuation amplitude. This corresponds to an instantaneous strengthening of the mode drive at higher and lower frequencies, with respect to that of the original EPM (damped at this time), which ``locally forces'' wave-packets within the SAW continuous spectrum in the $(r/a,\omega/\omega_A)$ plane along the black lines in the background of the top panels of Fig.~\ref{fig:epmloc}. Thus, the mode structure and frequency tend to split, as shown in both intensity contour plots and in mode structure plots of Fig.~\ref{fig:epmloc}. Note that both mode saturation as well as mode structure and frequency splitting are non-adiabatic processes, since they occur on a time scale shorter than $2\pi \omega_B^{-1}$, as demonstrated in Fig.~\ref{fig:kinepm}.
Furthermore, phase locking plays an important role, since the radial dependence of SAW wave-packet frequency and of resonance condition, Eq.~(\ref{eq:circres}), is essentially the same; and, thus, Eq.~(\ref{eq:phaselockcirc}), the phase locking condition, is readily satisfied. 
Consistent with the analysis of Sec.~\ref{sec:lock}, non-adiabatic frequency chirping and phase locking are in one-to-one correspondence with non-perturbative EP behaviors. In fact, non-perturbative EP PSZS cause the two distinct dominant Fourier harmonics to shift radially and vary their relative weight; and 
yield the ``rabbit-ear'' structure of the $(3,2)$ mode, visible in the bottom right panel of Fig.~\ref{fig:epmloc}, that would be otherwise not justifiable.

\begin{figure}
\centerline{\resizebox{0.54\linewidth}{!}{\includegraphics{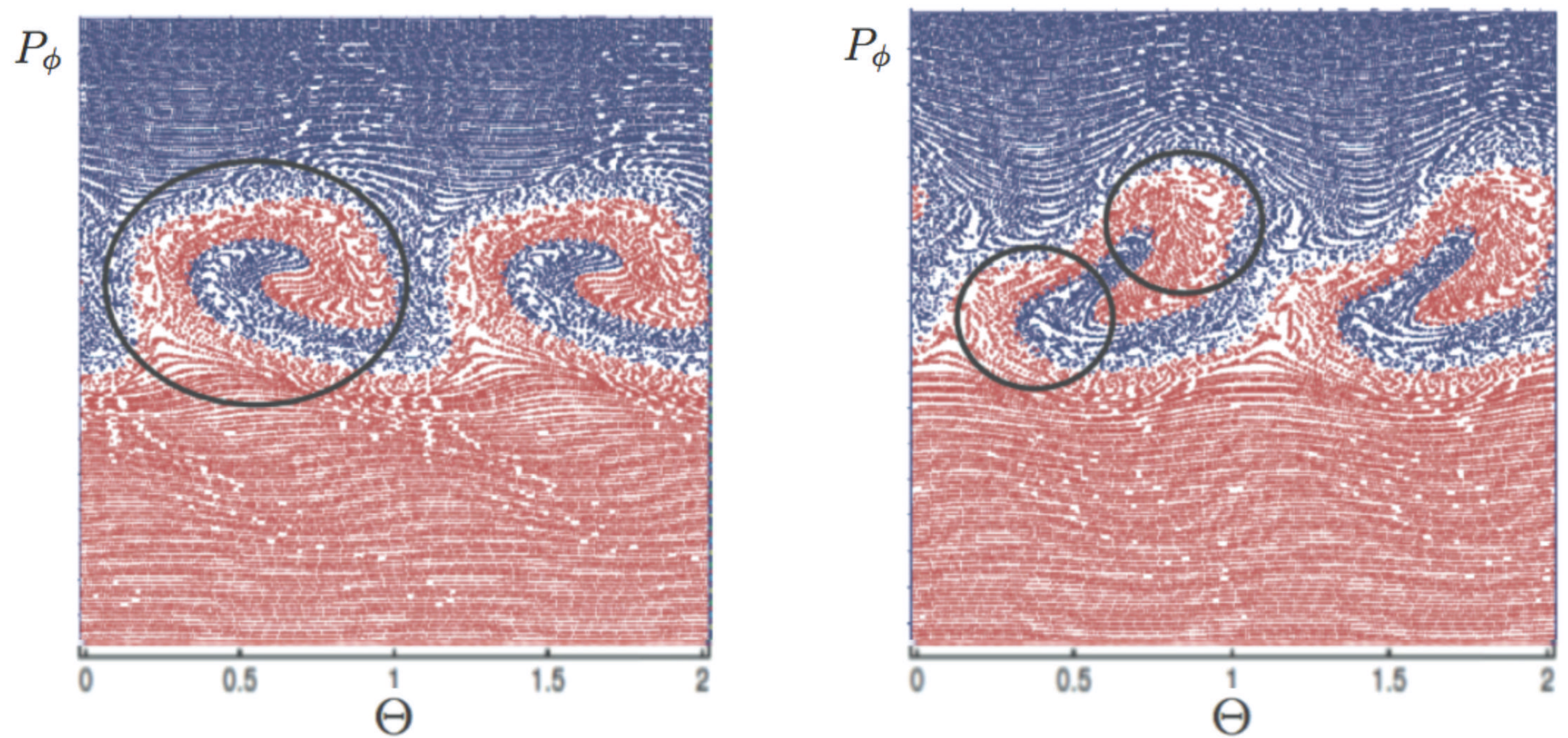}}}
\caption{Kinetic Poincar\'e plots illustrating non-adiabatic transition from single (left, $\omega_A t = 1320$) to double (right, $\omega_A t = 1353$) resonance structures in the EP phase space. Time normalization is the same as in Fig.~\ref{fig:epmloc}, while normalization of phase space variables are the same as in Fig.~\ref{fig:kinepm}.}
\label{fig:kinepm2} \end{figure}

In the further nonlinear evolution, mode structure and frequency splitting become more evident in Fig.~\ref{fig:epmloc} for $\omega_A t = 1320$ (right), although the same features are still not visible in Fig.~\ref{fig:kinepm}. As expected, mode structure splitting appears as double-resonance in the EP PSZS with some delay, as illustrated in 
Fig.~\ref{fig:kinepm2} after $\simeq 3/4$ oscillation of resonant EP in the instantaneous fluctuation-induced potential well, due to the time required for EPs to effectively
respond to the modified mode structures.  After this phase, EP PSZS as well as corresponding mode structures and frequencies tend to merge again, yielding a second EPM burst~\cite{briguglio14}. As said above, this is due to the {\em ad hoc} setup of the present numerical simulation experiment, aimed at inhibiting the onset of EPM convective amplifications (cf. Sec.~\ref{sec:epmaval})~\cite{zonca05,briguglio02,vlad04}. Detailed investigations of these dynamics require an in depth discussion, which will be reported elsewhere. Here, we emphasize again that
both intensity contour plots and mode structure distortions are non-perturbative and evidently interlinked self-consistently with the non-adiabatic evolution of EP PSZS ($|\dot\omega|\sim \omega_B^2$). A crucial role is played by the ``radial singular'' structures characterizing the SAW continuous spectrum \cite{zonca14e,chen14,zonca14a,zonca14b}.
In fact, while EP response to the regular EPM structure determines the mode frequency \cite{chen94,zonca14a,zonca14b} and phase locking (cf.\ also Sec.~\ref{sec:epmaval}), the mode structure is peaked (``singular'') at the radial position
where the SAW continuous spectrum is resonantly excited. Thus, consistent with EP radial redistributions as they are transported nonlinearly, SAW wave-packets readily respond to instantaneous local forcing and maximize
wave-particle power exchange.
This is consistent with observations from numerical simulation results, reporting that SAW continuous spectrum is crucial in the nonlinear mode dynamics and frequency chirping~\cite{briguglio98,briguglio02,vlad04,briguglio07,vlad99,vlad06,vlad09,bierwage11,bierwage12} (cf.\ also \ref{app:finiteness}).

\section{Governing equations of phase space zonal structures}
\label{sec:goveq}

As prevalent SAW/DAW and DWT fluctuations are in the low-frequency range, $|\omega_0| \ll |\Omega|$, 
kinetic description is based on the nonlinear gyrokinetic equation. Following Ref. \cite{frieman82}, the perturbed particle distribution function $\delta f$ can be written as
\begin{equation}
\delta f = e^{- \bm \rho \cdot \bm \nabla}  \left[ \delta g - \frac{e}{m} \frac{1}{B_0} \frac{\partial \bar F_0}{\partial \mu} \left \langle \delta L_g \right \rangle \right] + \frac{e}{m} 
\left[  \frac{\partial \bar F_0}{\partial \cal E} \delta \phi + \frac{1}{B_0} \frac{\partial \bar F_0}{\partial \mu} \delta L \right]\;\; . \label{eq:gypullback0}
\end{equation}
Here, adiabatic and non-adiabatic ($\delta g$) particle responses are separated, $e^{- \bm \rho \cdot \bm \nabla}$ is the {\em pull-back} transformation from guiding-center to particle coordinates \cite{brizard07}, $\bm \rho \equiv \Omega^{-1} \bm b \times \bm v$, $\bar F_0$ is the equilibrium guiding-center particle distribution function, 
\begin{equation}
\delta L_g = \delta \phi_g - \frac{v_\parallel}{c} \delta A_{\parallel g} = e^{\bm \rho \cdot \bm \nabla} \delta L  = e^{\bm \rho \cdot \bm \nabla} \left( \delta \phi - \frac{v_\parallel}{c}  \delta A_\parallel \right)\;\; , \label{eq:dpsigc}
\end{equation}
$\left\langle \cdots \right \rangle$ denotes gyrophase averaging, ${\cal E} = v^2/2$ is the energy per unit mass, and $\mu$ is the magnetic moment adiabatic invariant (cf. \ref{app:res2D}). 
Meanwhile, $\delta g$ is obtained from the Frieman-Chen nonlinear gyrokinetic equation \cite{frieman82}
\begin{eqnarray}
& & \left( \frac{\partial}{\partial t} + v_\parallel \nabla_\parallel + \bm v_d \cdot \bm \nabla_\perp + \delta \dot{\!\bar{\bm X}}_\perp \cdot \bm \nabla_\perp \right) \delta g  = \nonumber \\
& & \hspace*{4em} - \left( \frac{e}{m} \frac{\partial}{\partial t} \left \langle \delta L_g \right \rangle  \frac{\partial \bar F_0}{\partial \cal E}  + \frac{c}{B_0} \bm b \times \bm \nabla \left \langle \delta L_g \right \rangle \cdot \bm \nabla \bar F_0 \right) \, \, ,
\label{eq:fcnlgke}
\end{eqnarray}
where $\bm v_d$ is the magnetic drift velocity
\begin{equation}
\bm v_d = \frac{\bm b}{\Omega} \times \left( \mu \bm \nabla B_0 + \bm \kappa v_\parallel^2 \right) \simeq  \frac{\left( \mu B_0 + v_\parallel^2 \right)}{\Omega}  \bm b \times \bm \kappa  \;\; , \label{eq:magdrift}
\end{equation}
$\bm \kappa = \bm b \cdot \bm \nabla \bm b$ is the magnetic field curvature and $\bm \nabla B_0 \simeq \bm \kappa B_0$ in the low-$\beta$ limit. Furthermore, the fluctuation induced particle drift is
\begin{equation}
\delta \dot{\!\bar{\bm X}}_\perp = \frac{c}{B_0} \bm b \times \bm \nabla \left \langle \delta L_g \right \rangle + \frac{v_\parallel}{B_0} \bm \kappa \left\langle \delta A_{\parallel g} \right \rangle = \frac{c}{B_0} \bm b \times \bm \nabla \left \langle \delta \phi_g \right \rangle + v_\parallel \frac{\left\langle \delta \bm B_{\perp g} \right \rangle}{B_0} \, , \label{eq:pertvel}
\end{equation}
with $\left\langle \delta \bm B_{\perp g} \right \rangle = \bm \nabla \times \bm b \left\langle \delta A_{\parallel g} \right \rangle$. 
Note that Eq.~(\ref{eq:gypullback0}) is valid up to ${\cal O}(\epsilon_\delta)$ and that the relationship between $\delta g$ and  the fluctuating gyrocenter distribution function $\delta \bar F$ is given by $\delta  \bar F  = \delta g +  (e/m) \partial_{\cal E} \bar F_0 \left \langle \delta L_g \right \rangle$ \cite{brizard07}.
This approximation is justified within the characteristic (linear/nonlinear) time scale ordering of Eq.~(\ref{eq:timescaleorde}) (cf. \ref{app:wpresNL} for further discussions).

Consistent with Sec.~\ref{sec:intro}, the ``principal series'' of secular terms obtained by formal perturbation 
expansion in powers of fluctuating fields is dominated by PSZS \cite{vanhove55,prigogine62,balescu63}, which can be summed up and yields the solution of the corresponding Dyson equation \cite{altshul65,altshul66} (cf. Sec.~\ref{sec:dyson}). Thus, we formally separate $(m,n,\ell)=(0,0,0)$ terms in Eq.~(\ref{eq:flift0}) from $n\neq 0$ responses. PSZS, denoted hereafter by the subscript $z$, are then described by the distribution function $\delta f_z$, decomposed according to Eq.~(\ref{eq:gypullback0}); \idest, 
\begin{eqnarray}
\delta f_z & = & \sum_m \left\{ {\cal P}_{m,0,0} \circ \left[ J_0(\lambda)  \delta g \right]_{m,0} \right\} - \left[ J_0(\lambda)  \left( \frac{e}{m} \frac{1}{B_0} \frac{\partial \bar F_0}{\partial \mu} \left \langle \delta L_g \right \rangle \right)\right]_{0,0} \nonumber \\
& & + \frac{e}{m}
\left[  \frac{\partial \bar F_0}{\partial \cal E} \delta \phi + \frac{1}{B_0} \frac{\partial \bar F_0}{\partial \mu} \delta L \right]_{0,0} \;\; . \label{eq:dfz}
\end{eqnarray}
Here, the double subscript on fields stands for $(m,n)$, $J_0(\lambda) = \left\langle e^{\bm \rho \cdot \bm \nabla} \right\rangle$ accounts for finite Larmor radius effects and $\lambda = k_\perp v_\perp/\Omega$. Assuming $|k_\parallel|\ll | \bm k_\perp|$, the PSZS non-adiabatic response 
can be derived from Eqs.~(\ref{eq:flift0}) and~(\ref{eq:fcnlgke}); and 
is given by~\cite{chen14,zonca05}
\begin{eqnarray}
\frac{\partial \delta g_z}{\partial t}  & = &  -  {\cal P}_{0,0,0} \circ  \left( \frac{e}{m} \frac{\partial}{\partial t} \left \langle \delta L_g \right \rangle_z  \frac{\partial \bar F_0}{\partial \cal E} \right)_{0,0}
\nonumber \\ & & 
+ i \sum_m {\cal P}_{m,0,0} \circ  \frac{c}{d\psi/dr} \frac{\partial}{\partial r} \sum_{n\neq 0}  n \left( \delta g_n \left \langle \delta L_g \right \rangle_{-n}\right)_{m,0} 
\;\; , \label{eq:dgz}
\end{eqnarray}
where $\psi$ is the poloidal magnetic flux, $\bm B_0$ is represented as in Eq.~(\ref{eq:B0}), while the first and second terms on the right hand side (RHS) represent, respectively, the effects of ZS (zonal flows/fields) and the corrugation of radial profiles. 
Note that the formally nonlinear terms $\propto n$ on the RHS are dominant for $n\neq 0$ and $|k_\parallel|\ll | \bm k_\perp|$ modes,  
consistent with the standard gyrokinetic equation orderings~\cite{frieman82,brizard07}. However, if the nonlinear interactions between PSZS and ZS themselves are considered, 
Eq.~(\ref{eq:dgz}) should be suitably modified and an additional nonlinear term should be added on the RHS (cf. Refs. \cite{chen14b,qiu14d}). 
This is beyond the scope of the present paper, focusing on the nonlinear interaction of PSZS with $n\neq 0$ SAW/DAW; and is reported here only for the sake of completeness. 

Following the same procedure, the evolution equation for $\delta g_n$ can be derived from Eqs.~(\ref{eq:flift0}) and~(\ref{eq:fcnlgke}), yielding
\begin{eqnarray}
& & \left( \frac{\partial}{\partial t}  -  \frac{i n c}{d\psi/dr} \frac{\partial \left\langle  \delta L_g  \right \rangle_z}{\partial r}  + v_\parallel \nabla_\parallel + \bm v_d \cdot \bm \nabla_\perp \right) \delta g_n =
\nonumber \\
& & \hspace*{2cm} = i \frac{e}{m} \left( Q \bar F_0  - \frac{n B_0}{\Omega d\psi/dr} \frac{\partial}{\partial r} {\cal P}_{0,0,0}  \circ \delta g_z  \right) \left\langle \delta L_g \right\rangle_n \;\; . \label{eq:dgn} \\
& & Q\bar F_0 =  \frac{\partial \bar F_0}{\partial {\cal E}} i \frac{\partial}{\partial t} + \frac{\bm b \times \bm \nabla \bar F_0}{\Omega} \cdot ( - i \bm \nabla) \;\; . \label{eq:QF0}
\end{eqnarray}
On the left hand side (LHS) of Eq.~(\ref{eq:dgn}), it is interesting to note the effect of ZS in enhancing resonance detuning. Meanwhile, on the RHS, PSZS modify the mode 
dynamics via corrugation of resonant particle radial profiles.  
The lack of symmetry between effects of spatial and velocity space gradients of PSZS is only 
due to having dropped the $\propto \delta \dot{\cal E} \partial_{\cal E} \delta g_n$ term on the LHS of Eq.~(\ref{eq:fcnlgke}), with 
\begin{equation}\delta\dot{\cal E}= - \frac{e}{m} \bm v_d \cdot \bm \nabla_\perp \left\langle \delta L_g \right\rangle \;\; , \label{eq:deltadotcalE}
\end{equation}
consistent with the standard gyrokinetic equation orderings~\cite{frieman82,brizard07}. 
Similar to Eq.~(\ref{eq:dgz}), this term must be restored when considering nonlinear interactions between PSZS and ZS themselves \cite{chen14b,qiu14d}. 

Noting Eq.~(\ref{eq:kamcons}), fluctuation induced radial and energy excursion are ordered as $\Delta r/r \sim (\omega_{*EP}/\omega_0) \Delta {\cal E}/{\cal E}$~\cite{chen14,chen88b}, with $\omega_{*EP}$ the EP diamagnetic frequency and $|\omega_{*EP}/\omega_0|\gg 1$ for EP driven SAW/DAW in fusion plasmas (cf. \ref{app:finiteness}).  
Taking into account
the definition of the $Q\bar F_0$ operator in Eq.~(\ref{eq:QF0}),
Eq.~(\ref{eq:dgn}) can then be rewritten as
\begin{eqnarray}
& & \left( \frac{\partial}{\partial t}  -  \frac{i n c}{d\psi/dr} \frac{\partial \left\langle  \delta L_g  \right \rangle_z}{\partial r}  + v_\parallel \nabla_\parallel + \bm v_d \cdot \bm \nabla_\perp \right) \delta g_n =
- \frac{i n c}{d\psi/dr} \frac{\partial F_0}{\partial r}  \left\langle \delta L_g \right\rangle_n 
\; . \nonumber \\ & & \label{eq:dgn1}
\end{eqnarray}
Here, we have introduced 
\begin{equation}
F_0 \equiv \bar F_0 + {\cal P}_{0,0,0}  \circ \delta g_z \;\; ; \label{eq:F0def}
\end{equation} 
and have consistently neglected ${\cal O}(\omega_0/\omega_{*EP})$ terms on the RHS \cite{chen14}.
Using Eq.~(\ref{eq:F0def}), Eq.~(\ref{eq:dgz}) can be cast as
\begin{eqnarray}
\frac{\partial F_0}{\partial t}  & = &  -  {\cal P}_{0,0,0}^2 \circ  \left( \frac{e}{m} \frac{\partial}{\partial t} \left \langle \delta L_g \right \rangle_z  \frac{\partial \bar F_0}{\partial \cal E} \right)_{0,0}
\nonumber \\ & & 
+ i {\cal P}_{0,0,0} \circ \sum_m {\cal P}_{m,0,0} \circ  \frac{c}{d\psi/dr} \frac{\partial}{\partial r} \sum_{n\neq 0}  n \left( \delta g_n \left \langle \delta L_g \right \rangle_{-n}\right)_{m,0} 
\;\; . \label{eq:dgz1}
\end{eqnarray}
On the RHS, the dominant contribution in the nonlinear dynamics of EP PSZS is fluctuation induced resonant particle scattering. In Eq.~(\ref{eq:dgz1}), however, the contribution of 
other collisional~\cite{berk90c} and non-collisional~\cite{bergkvist04,bergkvist05,fu10,lang11} scattering processes may be readily included by corresponding scattering operators. Collision induced scattering, introduced in Ref. \cite{berk90c}, was the first one to be considered; but micro-turbulence induced diffusion has been demonstrated to be more relevant in typical ITER conditions~\cite{fu10,lang11}. Radio-frequency wave induced scattering~\cite{bergkvist04} has also been shown to dominate over collisional scattering in conditions of practical interest in present day experiments~\cite{bergkvist05}. 
These additional physics affect the nonlinear evolution of $F_{0EP}$ via self-consistent interplay of SAW/DAW and non-perturbative EP responses
on typical time scales longer than those considered here, $\tau_{NL} \sim \gamma_{L}^{-1}$; and, thus, will be neglected. 

Equations~(\ref{eq:dgn1}) and~(\ref{eq:dgz1}) completely describe PSZS nonlinear dynamics, once they are closed by evolution equations for SAW/DAW, $\left\langle  \delta L_g  \right \rangle_n$, and ZS, $\left\langle  \delta L_g  \right \rangle_z$. They also suggest that nonlinear dynamics of ZS and PSZS may be viewed as generators of neighboring nonlinear equilibria, which generally vary on the same time scale of the underlying fluctuation spectrum \cite{chen07b}. Here,  leaving time dependences implicit and using
the Poisson summation formula, we follow Ref. \cite{lu12} and write the spatiotemporal structures of SAW/DAW as \cite{zonca14e}
\begin{eqnarray}
\delta \phi_n (r,\theta,\zeta) & = & 2\pi \sum_{\ell \in \mathbb{Z}} e^{in\zeta-inq(\theta-2\pi\ell)} \delta \hat \phi_n (r,\theta-2\pi\ell)  \nonumber \\
& = & \sum_{m \in \mathbb{Z}} e^{in\zeta-i m\theta} \int e^{i(m-nq)\vartheta} \delta \hat \phi_n (r,\vartheta)  d\vartheta  \;\; ,  \label{eq:msd}
\end{eqnarray}
with the same decomposition assumed for $\delta A_{\parallel n}$. Meanwhile, separating parallel mode structure, $\delta \hat \phi_{0n}$, and radial envelope, $A_n(r)$, 
we let \cite{zonca14e}
\begin{equation}
\delta \hat \phi_n (r,\vartheta) = A_n(r) \delta \hat \phi_{0n} (r,\vartheta) \;\; . \label{eq:parenv}
\end{equation}
Finally, 
we adopt the theoretical framework of the general fishbone like dispersion relation (GFLDR) \cite{zonca14a,zonca14b},
that allows to write SAW/DAW dynamics equations as
\begin{equation}
\left[ i \Lambda_n -  \left( \delta \bar W_f + \delta \bar W_k \right)_n \right] A_n(r) = D_n(r,\theta_k,\omega) A_n(r) = 0 \;\; , \label{eq:fishlikeball0}
\end{equation}
which should be intended as the $n$-th row of a matrix equation for the complex amplitudes $(\ldots, A_n, \ldots)$ involving nonlinear mode couplings \cite{chen14,zonca06b}.
Equation~(\ref{eq:fishlikeball0}) is obtained by projection of nonlinear gyrokinetic quasineutrality condition and vorticity
equation on the parallel mode structures $\delta \hat \phi_{0n} (r,\vartheta)$; and assumes the eikonal form $A_n(r) \sim \exp i \int nq' \theta_k(r) dr$~\cite{dewar81,dewar82}.
Thus, $D_n(r,\theta_k,\omega)$ plays the role of a local WKB nonlinear dispersion function, but it can also be written for global fluctuations \cite{zonca14a,zonca14b}. 
General expressions of $\Lambda_n$, $\delta \bar W_{fn}$ and $\delta \bar W_{kn}$ are derived in Refs. \cite{zonca14a,zonca14b} and given in \ref{app:GFLDR} without derivation for readers' convenience: $\Lambda_n$ represents a generalized ``inertia'' 
accounting for the structures of the SAW continuous spectrum; while $\delta \bar W_{fn}$ and $\delta \bar W_{kn}$ can be interpreted as fluid and kinetic ``potential energies'' \cite{chen84}. 
Adopting the time scale ordering of Eq.~(\ref{eq:timescaleorde}), the 
spatiotemporal evolution of SAW wave packets can be described expanding the solutions of Eq.~(\ref{eq:fishlikeball0}) about the linear characteristics 
\begin{equation}
D_n^L(r,\theta_{k0}(r),\omega_0) = 0 \;\; , \label{eq:thetak0}
\end{equation}
where $D_n^L$ is the linearized dispersion function introduced in Eq.~(\ref{eq:fishlikeball0}). Thus, letting $A_n(r) = \exp ( - i \omega_0 t) A_{n0}(r,t)$, the  
spatiotemporal evolution equation 
for  $A_{n0}(r,t)$ becomes
\cite{zonca14a,lu12}
\begin{eqnarray}
& &\frac{\partial D_n^L}{\partial\omega_0}\left(i\frac{\partial}{\partial t}\right)A_{n0}+\frac{\partial
D_n^L}{\partial\theta_{k0}}\left(-\frac{i}{nq'}\frac{\partial}{\partial r}-\theta_{k0}\right)A_{n0} +\frac{1}{2}\frac{\partial^2D_n^L}{\partial\theta_{k0}^2} \nonumber \\ & & \hspace*{2em} 
\times \left[\left(-\frac{i}{nq'}\frac{\partial}{\partial r}-\theta_{k0}\right)^2 A_{n0} - \frac{i}{nq'}\frac{\partial\theta_{k0}}{\partial r} A_{n0} \right] = S_n (r,t)\;\; .  \label{eq:envelope0}
\end{eqnarray}
The $S_n(r,t)$ term on the RHS represents a generic source term, including external forcing, $S_n^{\rm ext}(r,t)$, and/or nonlinear interactions, denoted by the superscript $NL$ and extracted from the definition of $D_n$ in Eq.~(\ref{eq:fishlikeball0}) \cite{chen14,zonca14a,lu12}
\begin{equation}
S_n(r,t) = S_n^{\rm ext}(r,t) - e^{ i \omega_0 t} \left[ i \Lambda_n^{NL} -  \left( \delta \bar W_f^{NL} + \delta \bar W_k^{NL} \right)_n \right]  e^{- i \omega_0 t} A_{n0} \;\; . \label{eq:source}
\end{equation}
Equation~(\ref{eq:envelope0}) also holds as evolution equation for global modes, provided that $A_{n0}$ is interpreted as mode amplitude at a fixed 
radial position and $D_n$ is suitably redefined as a global dispersion function operator \cite{zonca14a}, with $\partial_r D_n = \partial_{\theta_k} D_n = \partial^2_{\theta_k} D_n = 0$. 
By letting $A_{n0} (r,t) \rightarrow A_z(r,t) \equiv e^{i \omega_z t} \delta \phi_z$, with $\omega_z$ the ZS complex frequency, Eq.~(\ref{eq:envelope0}) also describes  zonal flow dynamics, while the zonal field evolution,
assuming massless fluid electron response,
is given by~\cite{chen14}
\begin{equation}
\frac{\partial}{\partial t} \delta A_{\parallel z} = \left( \frac{c}{B_0} \bm b \times \bm \nabla \delta A_\parallel \cdot \bm \nabla \delta \psi \right)_z
\;\; , \label{eq:daz}
\end{equation}
with $\bm b \cdot \bm \nabla \delta \psi \equiv - c^{-1} \partial_t \delta A_\parallel$ for $k_\parallel \neq 0$ modes.

In summary, Eqs.~(\ref{eq:dgn1}) and~(\ref{eq:dgz1}), closed by Eqs.~(\ref{eq:envelope0}) and~(\ref{eq:daz}), describe the self-consistent evolution of ZS and PSZS in the presence of SAW/DAW and/or DWT, based on the time scale ordering of Eq.~(\ref{eq:timescaleorde}). Despite their relatively simple form, they have been investigated, so far, only in simplified limiting cases;
\idest, either neglecting the nonlinear effect of wave-particle resonances \cite{chen07b,chen00,chen01,chen13,guo09,chen12,kosuga12}, or the effects of ZS \cite{zonca05,zonca06b,zonca00b,zonca07b}. In particular, Eqs.~(\ref{eq:dgn1}) and~(\ref{eq:dgz1}) can describe the onset of nonlocal behaviors in SAW/DAW nonlinear dynamics and ensuing EP transport, discussed in Sec.~\ref{sec:nonloc}; and the extension of secular EP transport from meso- to macro-scales due to phase locking, described in Sec.~\ref{sec:lock}.
It was noted in Sec.~\ref{sec:theory}, that the onset of nonlocal behaviors is the signature that, even for an isolated resonance described as a 1D non-autonomous system, the
nonuniform plasma dynamics is fully 3D. It is, however, particularly simple and instructive to investigate this problem for precessional resonance only; \idest, Eq.~(\ref{eq:trapres}) with $\ell = 0$. 
In such a case, in fact, assuming $\omega_0 \sim n \bar\omega_d \ll \omega_b$, EP nonlinear dynamics preserves both $\mu$ and $J$ (cf. \ref{app:res2D}). Thus, the problem is that of a 1D non-autonomous nonuniform system,
which can be readily compared with the case of a 1D non-autonomous uniform plasma; \exgra, the ``bump-on-tail paradigm'' (cf. Sec.~\ref{sec:lock}).  

\section{Dyson equation and the ``fishbone paradigm''}
\label{sec:dyson}

In order to simplify the present analysis, we neglect ZS and restrict Eqs.~(\ref{eq:dgn1}) and~(\ref{eq:dgz1}) to precessional resonance with
magnetically trapped EPs, while neglecting finite magnetic drift orbit width effects and assuming $\omega_0 \sim n \bar\omega_d \ll \omega_b$. In this limit, as anticipated 
at the end of Sec.~\ref{sec:goveq}, we reduce the problem to analyzing a 1D non-autonomous nonuniform system, where onset of nonlocal behaviors is exclusively due to bounce averaged nonlinear EP
dynamics. This is the simplest possible case of PSZS evolution, self-consistently interacting with SAW/DAW excited by EPs in fusion plasmas,
and has been dubbed as ``fishbone paradigm'' \cite{chen14,chen14b,chen13}. In fact, 
this is also the case considered originally in the analysis of EP transport and nonlinear burst cycle of resonantly excited internal kink modes \cite{white83,chen84},
after their observation as ``fishbone'' oscillations in the 
Poloidal Divertor Experiment (PDX)~\cite{mcguire83}. The peculiar feature of the ``fishbone paradigm'', emphasized in Refs.~\cite{chen14,chen14b,chen13}, is the self-consistent interplay of mode structures and EP transport, reflecting equilibrium geometry and plasma nonuniformity. In the following, we also show that the ``fishbone paradigm''  recovers the 
``bump-on-tail paradigm'' in the (local) uniform plasma limit~\cite{chen14,chen13}.

\subsection{Renormalization of phase space zonal structures}
\label{sec:renorm}

\begin{figure}
\centerline{\includegraphics[width=0.5\linewidth]{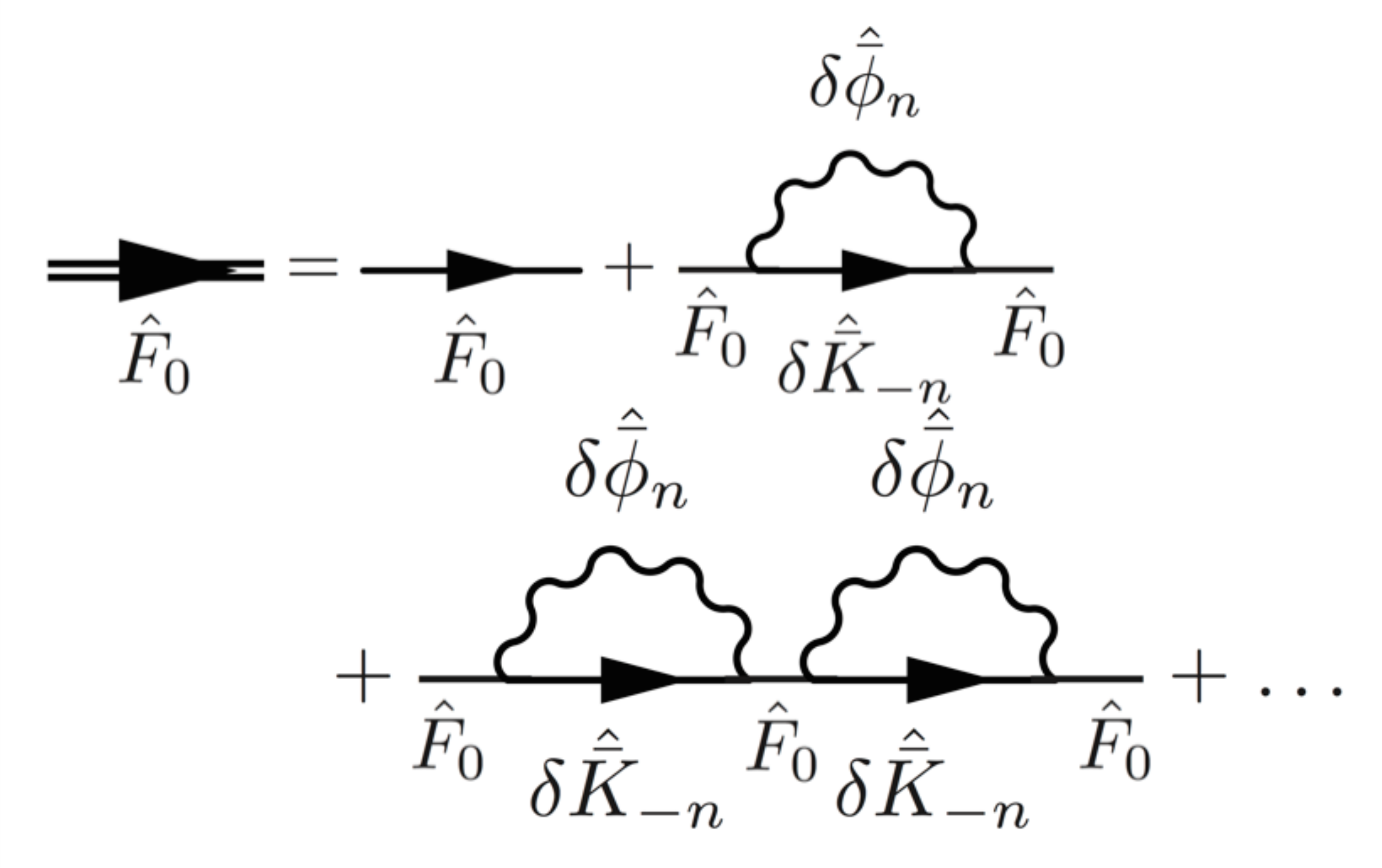}}
\caption{Schematic view of the Dyson series describing EP transport 
due to emission and reabsorption of (toroidal equilibrium) symmetry breaking perturbations \cite{zonca14e}.}
\label{fig:dyson} \end{figure}

Noting Eq.~(\ref{eq:flift0}), and due to the frequency ordering $\omega_0 \sim n \bar\omega_d \ll \omega_b$ as well as to the negligible magnetic drift orbit width limit, we can rewrite Eq.~(\ref{eq:dgz1}) in terms of 
\begin{equation}
\delta \bar \phi_n = e^{i n (\zeta - q \theta)} \sum_m {\cal P}_{m,n,0} \circ \delta \phi_{m,n} = e^{- i n  q \theta} \overline{e^{inq\theta} \delta \phi_n} \;\; , \label{eq:barphin}
\end{equation}
with $\overline{(\ldots)} = (2\pi)^{-1} \omega_b \oint (\ldots) d\theta/\dot \theta$ denoting magnetic drift bounce averaging (cf. \ref{app:res2D}); and the corresponding EP response 
\begin{equation}
\delta \bar g_n = e^{i n (\zeta - q \theta)} \sum_m {\cal P}_{m,n,0} \circ \delta g_{m,n} = e^{- i n  q \theta} \overline{e^{inq\theta} \delta g_n}  \;\; . \label{eq:dgbaven}
\end{equation}
Thus, recalling that the bounce averaged parallel velocity vanishes for magnetically trapped particles, we have
\begin{eqnarray}
\frac{\partial F_0}{\partial t} & = & i  \sum_n  \frac{nc}{d\psi/dr} \frac{\partial}{\partial r}  \left( \delta \bar g_n \delta \bar \phi_{-n} - \delta \bar g_{-n} \delta \bar \phi_n \right)  \label{eq:dgzonal3} \\
& = &   \sum_n  \frac{nc}{d\psi/dr} \frac{\partial}{\partial r} \left[ \frac{nc}{d\psi/dr} \frac{\partial F_0} {\partial r} \frac{\partial}{\partial t}  \left| \frac{\delta \bar \phi_n}{\omega_n} \right|^2 
+ i \left( \delta \bar K_n \delta \bar \phi_{-n} - \delta \bar K_{-n} \delta \bar \phi_n \right) \right] \nonumber \; . 
\end{eqnarray}
Here,  we have used the definition 
\begin{equation}
\delta g \equiv \delta K + i (e/m) Q\bar F_0 \partial_t^{-1} \left\langle  \delta \psi_g \right \rangle \;\;  \label{eq:dgdKrel}
\end{equation}
and the ideal MHD condition $\delta E_\parallel =0$; \idest, $\delta \phi = \delta \psi$, defined below Eq.~(\ref{eq:daz}), to explicitly rewrite the RHS of Eq.~(\ref{eq:dgzonal3}), formally separating reversible from irreversible processes connected with wave-particle interactions, which dominate the nonlinear dynamics in Eq.~(\ref{eq:dgzonal3}).  In fact, the contribution of reversible processes is 
${\cal O}(\gamma_L \tau_{NL}^{-1}/\omega_0^2)\sim {\cal O}(\gamma_L^2/\omega_0^2)$ 
with respect to irreversible ones, consistent with Eqs.~(\ref{eq:dyson3}) and~(\ref{eq:dyson4}) below. 
Hence, they will be neglected in the following.  
Adopting the notation
\begin{equation}
\hat F_0(\omega) = (2\pi)^{-1} \int_0^\infty e^{i \omega t} F_0(t) dt  \label{eq:laplacedef}
\end{equation}
for the Fourier-Laplace transform, Eq.~(\ref{eq:dgzonal3}) can be solved as
\begin{eqnarray}
\hat F_0(\omega) & = & \frac{i}{\omega} {\rm St} \hat F_0(\omega) +  \frac{i}{\omega} \hat S_0(\omega) + \frac{i}{2\pi\omega} \bar F_0(0) + \frac{nc}{\omega(d\psi/dr)} \nonumber \\ & & 
\times  \frac{\partial}{\partial r} \int_{-\infty}^\infty \left[ \delta \hat {\bar \phi}_n (y) \delta \hat {\bar K}_{-n} (\omega - y) -  \delta \hat {\bar \phi}_{-n} (y)  \delta \hat {\bar K}_n (\omega - y) \right] dy  \;\; . \label{eq:dyson2}
\end{eqnarray}
Here, we have straightforwardly included the effect of collisions, formally denoted by ${\rm St} \hat F_0(\omega)$, and of an external source term, $\hat S_0(\omega)$. Furthermore $\bar F_0(0)$ stands for the initial value of $F_0$ at $t=0$, while the repeated $n$ subscript assumes implicit summation. Note, also, that
we have explicitly indicated only the dependences on $\omega$ (and $y$ , as dummy integration frequency variable) for the sake of notation clarity. 
Meanwhile, for EP precessional resonance, it is possible to show \cite{chen14} 
\begin{equation}
\delta \hat{\bar K}_n (\omega) =  \frac{e}{m} \int_{-\infty}^{+\infty} \frac{\hat\omega_{dn}}{y} \frac{Q_{n,y} \hat F_0 (\omega - y)}{n\bar\omega_{d} - \omega} \delta \hat{\bar \phi}_n (y) d y \;\; , \label{eq:ldkprec}
\end{equation}
where the subscripts in $Q_{n,y}\hat F_0$ denote mode number and frequency at which the operator defined by Eq.~(\ref{eq:QF0}) must be evaluated; and, consistent with Eqs.~(\ref{eq:barphin}) and~(\ref{eq:dgbaven}), we have introduced the definition
\begin{equation}
e^{- inq\theta} \overline{e^{inq\theta} \omega_d \delta \phi_n} \equiv \hat\omega_{dn} \delta \hat{\bar \phi}_n \;\; . \label{eq:hatomegad}
\end{equation}
It is straightforward to verify that Eq.~(\ref{eq:ldkprec}) gives back  
the linear limit for $\hat F_0(\omega) = (2\pi \omega)^{-1} i \bar F_0(0)$.
By substitution of Eq.~(\ref{eq:ldkprec}) into Eq.~(\ref{eq:dyson2}), one readily obtains
\begin{eqnarray}
\hat F_0(\omega) & = & \frac{i}{\omega} {\rm St} \hat F_0 (\omega) + \frac{i}{\omega} \hat S_0(\omega) + \frac{i}{2\pi \omega} \bar F_0(0)  
+ \frac{e}{m} \frac{nc}{\omega(d\psi/dr)} \nonumber
\\ & & \times \frac{\partial}{\partial r} \iint_{-\infty}^\infty \left[ \delta \hat {\bar \phi}_n (y) \frac{\hat\omega_{d-n}}{y'} \frac{Q_{-n,y'} \hat F_0 (\omega - y - y')}{- n \bar\omega_{d} + y - \omega} \delta \hat{\bar \phi}_{-n} (y') \right.  \nonumber \\
& & -  \left. \delta \hat {\bar \phi}_{-n} (y) \frac{\hat\omega_{dn}}{y'} \frac{Q_{n,y'} \hat F_0 (\omega - y - y')}{n \bar\omega_{d} + y - \omega} \delta \hat{\bar \phi}_n (y') \right] dy dy' 
\;\; .  \label{eq:dyson3} \end{eqnarray}
This equation is the analogue of the Dyson equation in quantum field theory (cf., \exgra, Ref. \cite{kaku93}) 
describing EP transport 
due to emission and reabsorption of (toroidal equilibrium) symmetry breaking perturbations on all possible ``loops'' \cite{zonca14e},  
schematically shown in Fig.~\ref{fig:dyson}.
Its solution provides the renormalized expression of $\hat F_0$, and, thereby, of $F_0$, taking into account
nonuniform toroidal plasmas effects with the addition of sources and collisions.
The nonlinear term in Eq.~(\ref{eq:dyson3}) can be further simplified considering that
\begin{equation}
Q_{n,\omega}\hat F_0 \simeq - \frac{nB_0}{\Omega d\psi/dr} \frac{\partial \hat F_0}{\partial r} \label{eq:QF0lim}
\end{equation}
for $|\omega_{*EP}/\omega|\gg1$, typical of SAW/DAW excited by EP in fusion plasmas (cf. Sec.~\ref{sec:goveq}).

\subsection{The uniform plasma limit and the bump-on-tail paradigm}
\label{sec:uniform}

The uniform plasma limit of Eq.~(\ref{eq:dyson3}) illuminates the connection of ``fishbone'' and ``bump-on-tail'' paradigms (cf. Secs.~\ref{sec:lock} and~\ref{sec:goveq}). It is obtained postulating constant $\delta \hat{\bar \phi}_n (\omega)$ fluctuations and neglecting spatial dependences of $\hat F_0$ by mapping them into corresponding velocity space variations. 
By direct inspection of Eqs.~(\ref{eq:dgn}) and~(\ref{eq:QF0}), and by their comparison with the nonlinear Vlasov equation for a 1D uniform plasma, the configuration (radial) to
velocity space mapping is obtained by letting
\begin{equation}
- \frac{nB_0}{d\psi/dr} \frac{\partial}{\partial r} \leftrightarrow k_0 \frac{\partial}{\partial v} \;\; . \label{eq:botequiv}
\end{equation}
Here, $k_0$ is the reference wave vector and $\omega_0 = k_0 v_0$ is the resonance condition to be used instead of the precessional resonance of Eq.~(\ref{eq:trapres}) with $\ell=0$. Thus, Eq.~(\ref{eq:botequiv}) is completed by the mapping of resonant denominators
\begin{equation}
n\bar\omega_{d} - \omega_0 \simeq n \bar\omega_{d0} \frac{(r-r_0)}{L_{d0}}  \leftrightarrow  k_0 (v-v_0) = k_0 u \;\; , \label{eq:precresmap}
\end{equation}
where $L_{d0}$ is the characteristic length of variation of $\bar\omega_{d}$\footnote{Note that Eq.~(\ref{eq:botequiv}) implies that directions of incrementing $v$ corresponds to decreasing $r$ and 
{\sl vice versa}. However, $\bar\omega_{d}$ is also a generally decreasing function of $r$.} and $u\equiv (v-v_0)$. 

Introducing a source term $Q(v)$ and a simple Krook collision operator, $-\nu(v) f(v)$, the Vlasov equation for a 1D uniform plasma equilibrium is written as
\begin{equation}
d_t f(v) = Q(v) - \nu(v) f(v) \;\; . \label{eq:vlasovkrook}
\end{equation}
The configuration to velocity space mapping of Eqs.~(\ref{eq:botequiv}) and~(\ref{eq:precresmap}), thus, allows to rewrite Eqs.~(\ref{eq:dyson3}) and~(\ref{eq:QF0lim}) as
\begin{eqnarray}
(-i\omega + \nu ) \delta \hat f_0 (\omega) & = & i \frac{e^2 k_0^2}{m^2} \frac{\partial}{\partial u} \iint_{-\infty}^\infty \left[ \delta \hat {\phi}_{k_0} (y) \frac{ - \partial_u \hat F_0 (\omega - y - y')}{y - k_0 u - \omega - i \nu} \delta \hat{\phi}_{-k_0} (y') \right.  \nonumber \\
& & -  \left. \delta \hat {\phi}_{-k_0} (y) \frac{\partial_u \hat F_0 (\omega - y - y')}{ y + k_0 u - \omega - i \nu} \delta \hat{\phi}_{k_0} (y') \right] dy dy' \;\; , \label{eq:dysonbot}
\end{eqnarray}
when expressed for the nonlinear deviation $\delta \hat f_0 (\omega)$ of the particle distribution function from the equilibrium (initial) value $F_0(0) = Q(v)/\nu(v)$.
Following Ref. \cite{altshul65,altshul66}, it is possible to show that, in the case of many waves characterized by a broad frequency spectrum with overlapping resonances, Eq.~(\ref{eq:dysonbot}) reduces to the quasilinear theory of a weakly turbulent plasma~\cite{drummond62,vedenov61c}. In the opposite limit of a narrow frequency spectrum of nearly periodic nonlinear fluctuations (cf. Sec.~\ref{sec:periodicD}), Eq.~(\ref{eq:dysonbot}) describes the oscillations of particles that are trapped in the wave, which, however, do not decay in time as expected as consequence of phase mixing. Therefore, the scalar field $\delta \phi_{k_0}$ oscillations are also predicted to continue indefinitely rather than to fade away, as in actual physical conditions~\cite{oneil65}, due to the fact that the processes described in Fig.~\ref{fig:dyson} do not account for $k_0$-harmonics generation produced by spatial bunching~\cite{oneil71}. 

Equation~(\ref{eq:dysonbot}) can also be used to analyze the adiabatic frequency chirping ($|\dot \omega| \ll \omega_B^2$) of phase space hole-clump pairs
due to the balance of EP power extraction by PSZS with a fixed background dissipation~\cite{berk96,berk97,breizman97,berk99,berk97b}. 
In particular, two limiting cases have been investigated in the literature: (i) the $\omega_B t \ll 1$ limit~\cite{berk96,berk97,breizman97}, where a recursive solution can be adopted near marginal stability;
and (ii) the $\omega_B t \gg 1$ limit~\cite{berk99,berk97b}, where the lowest order solution corresponds to a constant distribution function inside the wave-particle trapping separatrix (hole/clump), which is slowly shifting in velocity space (chirping). 
The recursive solution of Refs.~\cite{berk96,berk97,breizman97}, obtained for $\omega_B t \ll 1$,
corresponds to taking $\hat F_0 (\omega - y - y') = i (2\pi)^{-1} F_0(0) (\omega - y - y')^{-1}$ in Eq.~(\ref{eq:dysonbot}); \idest, to considering only the first loop in the Dyson series, schematically shown in Fig~\ref{fig:dyson}. Moving to the $t$-representation, the recursive solution of Eq.~(\ref{eq:dysonbot}) is then obtained as
\begin{eqnarray}
\left (\frac{\partial}{\partial t} + \nu \right) \delta f_0  & = & i \frac{e^2 k_0^2}{m^2} \frac{\partial}{\partial u} \iint_{-\infty}^\infty e^{-i(y+y')t} \left[ \delta \hat {\phi}_{k_0} (y) \frac{\partial_u F_0 (0)}{y' + k_0 u + i \nu} \delta \hat{\phi}_{-k_0} (y') \right.  \nonumber \\
& & + \left. \delta \hat {\phi}_{-k_0} (y) \frac{\partial_u F_0 (0)}{ y' - k_0 u +  i \nu} \delta \hat{\phi}_{k_0} (y') \right] dy dy' \;\; , \label{eq:bot0}
\end{eqnarray}
which is readily cast as
\begin{eqnarray}
\left (\frac{\partial}{\partial t} + \nu \right) \delta f_0  = \frac{\omega_B^2 (t)}{4k_0^2} \frac{\partial}{\partial u} \int_{0}^t \left[ e^{-(\nu + i k_0 u)(t-t')} + c.c. \right]  \omega_B^2(t') \frac{\partial F_0( 0)}{\partial u}  dt' \, . \label{eq:bot1}
\end{eqnarray}
This equation coincides with the evolution equation of PSZS given by Ref.~\cite{berk96}, noting that we have introduced $\omega_B^4 \equiv 4 (e/m)^2 k_0^4 |\delta \phi_{k_0}|^2$, in order to preserve the same normalizations of Fourier amplitudes used in the original work. The resulting PSZS dynamics can be shown to admit fixed point solutions, nonlinear oscillations
as well as finite time singularities~\cite{berk96,berk97,breizman97} (cf. Ref.~\cite{breizman11b} for a recent review). This latter case, obtained for sufficiently low collisionality,
corresponds to breaking down of the truncation in the perturbation expansion of the Dyson series~\cite{chen14}; and coincides with the formation of hole/clump pairs, demonstrated in numerical simulations~\cite{vann03,vann05,lesur09,lesur12}. The analytic description of adiabatic chirping of hole/clump pairs~\cite{berk99,berk97b}, valid for $\omega_B t \gg 1$  and for isolated PSZS with frequency separation much larger than $\omega_B$, shows that 
the EP distribution function is flattened (in a coarse-grain sense~\cite{sagdeev69}) inside the wave-particle trapping region. Meanwhile, the energy transfer from EP to PSZS during the chirping process is balanced by fixed background dissipation; that is, the condition under which fluctuations are maintained near marginal stability during the adiabatic evolution of hole/clump structures. These features are embedded in the general solution of Eq.~(\ref{eq:dysonbot}), which, as noted above, can address wave-particle trapping as the shortest time scale nonlinear dynamics~\cite{altshul65,altshul66}. Thus, Eq.~(\ref{eq:dysonbot}) can reproduce
both $\omega_B t \ll 1$~\cite{berk96,berk97,breizman97}
and $\omega_B t \gg 1$ limits~\cite{berk99,berk97b}; and addresses the general $\omega_B t \sim 1$ case, where formation of PSZS is expected~\cite{chen14} (cf. Sec.~\ref{sec:EPMloc}).

\subsection{Dyson equation for nearly periodic fluctuations}
\label{sec:periodicD}

The same results that hold for Eq.~(\ref{eq:dysonbot}) with broad and narrow frequency spectrum fluctuations can be obtained for Eq.~(\ref{eq:dyson3}), given the mapping of 
Eqs.~(\ref{eq:botequiv}) and~(\ref{eq:precresmap}). Thus, one can demonstrate, respectively, quasilinear transport in the radial direction in the case of overlapping resonances, and
nonlinear oscillations due to wave-particle trapping in the presence of a nearly periodic large amplitude wave\footnote{A proof of this can be found, \exgra, in the Lectures on {\sl Nonlinear dynamics of phase-space zonal structures and energetic particle physics in fusion plasmas}, F. Zonca, Institute for Fusion Theory and Simulation, Zhejiang University, May 5-16, (2014); http://www.afs.enea.it/zonca/references/seminars/IFTS\_spring14.}. Further to this, Eq.~(\ref{eq:dyson3}) can address the onset of nonlocal behaviors in EP transport,
including effects of equilibrium geometries and plasma nonuniformity discussed in Sec.~\ref{sec:theory}, 
which are expected to become progressively more important for increasing EP drive and non-perturbative response \cite{chen14}. 
The present approach also shows that, even in the perturbative EP local limit and given the mapping of 
Eqs.~(\ref{eq:botequiv}) and~(\ref{eq:precresmap}), some features remain in the ``fishbone'' paradigm that make it qualitatively but not quantitatively the same as the ``bump-on-tail'' paradigm.
For example, the energy dependence of $\bar\omega_d$ and $\hat \omega_{dn}$ give a different weighting of the particle phase space
contributing to the resonant mode drive with respect to the uniform plasma case.

Equation~(\ref{eq:dyson3}), when applied to many modes, provides a framework to explore the transition of EP transports through stochasticity threshold with all the necessary physics ingredients for a realistic comparison with experimental observations and for applications to the dense SAW/DAW fluctuation spectrum typical of fusion plasmas, since it addresses resonance detuning and radial decoupling in wave-particle interactions on the same footing.
Equation~(\ref{eq:dyson3}), meanwhile, specialized to nearly periodic fluctuations, can be adopted for nonlinear analyses of ``fishbone'' oscillations as well as EPMs (cf. Sec.~\ref{sec:epmaval}).

Given a nearly periodic fluctuation $\delta \bar\phi_{n} (t)$ (spatial dependences are left implicit for the sake of notation simplicity), defined in Eq.~(\ref{eq:barphin}) and with oscillation frequency centered about
$\omega_0(\tau) = \omega_{0r}(\tau) + i \gamma_0(\tau)$ (slowly varying in time), we can write $\delta \bar \phi_{n} (t) \equiv \lim_{\tau \rightarrow t } \delta \bar \varphi_{n} (\tau) \exp \left( - i \omega_{0}(\tau) t \right)$, with $\delta \bar \varphi_{n} (\tau) \equiv  \delta \bar \phi_{n0} \exp \left[ -i \int_0^\tau \omega_{0}(t') dt' + i \omega_{0}(\tau) \tau \right]$. Thus, as shown in \ref{app:Laplace}, $\delta \bar \phi_{n} (t)$ admits the Laplace transforms 
\begin{equation}
\delta \hat {\bar \phi}_n (\omega) = \frac{i}{2\pi} \frac{\delta \bar\varphi_{n} (r,\tau) }{\omega - \omega_0 (\tau)} \;\; ,  \;\;\;\;\; {\rm and} \;\;\;\;\;
\delta \hat {\bar \phi}_{-n}(\omega) = \frac{i}{2\pi} \frac{\delta \bar\varphi_{-n} (r,\tau) }{\omega + \omega^*_0 (\tau)} \;\; , \label{eq:monochr1}
\end{equation}
provided that $|t - \tau| \laeq \tau_{NL} \sim |\gamma_{0}|^{-1}$; and $|\dot\omega_{0r}|\ll |\gamma_0\omega_{0r}|$, $|\dot\gamma_0|\ll |\gamma_0^2|$. Therefore, Eq.~(\ref{eq:monochr1}) applies in general to adiabatic as well as non-adiabatic frequency
chirping modes. When substituted back into Eq.~(\ref{eq:dyson3}), Eq.~(\ref{eq:monochr1}) yields the following
form of the Dyson equation
\begin{eqnarray}
\hat F_0(\omega) & = & \frac{i}{\omega} {\rm St} \hat F_0 (\omega) + \frac{i}{\omega} \hat S_0(\omega) + \frac{i}{2\pi \omega} \bar F_0(0) +  
\frac{e}{m} \frac{nc}{\omega(d\psi/dr)} \nonumber \\
& & \times \frac{\partial}{\partial r} \left\{ \left[ \frac{Q^*_{n,\omega_0(\tau)}}{\omega_0^*(\tau)} \frac{\hat F_0 \left(\omega - 2 i \gamma_0 (\tau) \right)}{\omega - \omega_0(\tau) + n \bar\omega_{d}} + \frac{Q_{n,\omega_0(\tau)}}{\omega_0(\tau)} \right.\right. \nonumber \\ & & \times \left. \left.  \frac{\hat F_0 \left(\omega - 2 i \gamma_0 (\tau) \right)}{\omega + \omega_0^*(\tau) - n \bar\omega_{d}} \right] \hat\omega_{dn} \left| \delta \bar \varphi_{n}(r,\tau)\right|^2 \right\} \;\; , \label{eq:dyson4}
\end{eqnarray}
where $Q_{n,\omega_0(\tau)} \hat F_0$ can be approximated by Eq.~(\ref{eq:QF0lim}). 

Equation~(\ref{eq:dyson4}) is the renormalized form of PSZS generated by nearly periodic fluctuations, which allows investigating the self-consistent nonlinear evolution of SAW/DAW and EP PSZS when ``closed'' by Eqs.~(\ref{eq:dgn1}) (dropping the ZS shearing effect $\propto \partial_r \langle \delta L_g \rangle_z$) and~(\ref{eq:envelope0}).
A crucial element, here, is the different argument of $\hat F_0$ on the LHS and RHS
of Eq.~(\ref{eq:dyson4}), since it bears the information on the self-consistent interplay of nonlinear mode dynamics and EP transport. 
Assuming $|\omega|\gg|\gamma_0| \rightarrow 0$; \idest, assuming fixed amplitude fluctuations, allows to solve Eq.~(\ref{eq:dyson4}) in the uniform radial limit
and to recover nonlinear oscillations due to wave-particle trapping, discussed above in this section. In general, however, we have $|\omega|\sim|\gamma_0|$, and Eq.~(\ref{eq:dyson4})
describes much richer dynamics, among which those to be discussed in Sec.~\ref{sec:epmaval}.

\section{EPM convective amplification and avalanches}
\label{sec:epmaval}  

We analyze the nonlinear dynamics of a nearly periodic EPM burst using Eqs.~(\ref{eq:dgn1}), (\ref{eq:envelope0}) and~(\ref{eq:dyson4}). In particular, we
consider a radially localized EPM wave-packet with perpendicular wavelength $\lambda_\perp \ll L_{EP}$, with $L_{EP}$ the characteristic 
EP pressure scale length. Furthermore, the mode frequency is assumed to be near the lower accumulation point of the TAE frequency gap in the SAW continuum (cf.\ also Sec.~\ref{sec:EPMloc}). 
Thus, $\theta_{k0}=0$ and $\partial D_n^L/\partial \theta_{k0}=0$ in Eq.~(\ref{eq:envelope0})~\cite{zonca14a,zonca14b,chen95,zonca96a,zonca00a}; and  
\begin{equation}
\Lambda_n^L \equiv \Lambda_T = (1/2)[(\omega^2-\omega_\ell^2)/(\omega_u^2-\omega^2)]^{1/2} \;\; .  \label{eq:LambdaT}
\end{equation}
Here, the subscript $T$ stands for ``toroidal''; and $\omega_\ell/\omega_u$ are the lower/upper SAW continuum accumulation point frequency. 
Using a simple $(s,\alpha)$ model for high aspect ratio ($R_0/a\gg 1$) tokamak equilibria with circular shifted magnetic flux surfaces~\cite{connor78}, with $s\equiv rq'/q$ the magnetic shear and $\alpha \equiv - R_0 q^2 \beta'$ the ``ballooning'' dimensionless pressure gradient parameter, 
it is possible to show~\cite{zonca14a,zonca14b,zonca93a} 
\begin{equation}
\delta \bar W_{fn}^L =  \frac{|s|\pi}{8} \left( 1 + 2 \kappa(s) - \frac{\alpha}{\alpha_c} \right) \; ; \;\;\;\;\; {\rm and} \;\;\;\;\; \frac{1}{2}\frac{\partial^2D_n^L}{\partial\theta_{k0}^2} = \frac{|s|\pi}{8} \kappa(s)  \;\; , \label{eq:dwbarfepm}
\end{equation}
for $|s|,|\alpha| < 1$ 
in Eq.~(\ref{eq:envelope0}), with $\alpha_c = s^2/(1+ |s|)$ and $\kappa(s) \simeq (1/2) ( 1 + 1/|s|) \exp(-1/|s|)$.
Meanwhile, 
consistent with the ``fishbone'' paradigm (cf. Sec.~\ref{sec:dyson}), $\delta \bar W_{kn}$ can be written as~\cite{chen14,zonca14a,zonca14b}
\begin{eqnarray}
\delta \bar W_{kn}  & = & 
\int {\cal E} d {\cal E} d \lambda \sum_{v_\parallel/|v_\parallel| = \pm} \frac{\pi^2 q R_0}{c^2 k_\vartheta^2 |s|} \frac{e^2}{m} \left( \frac{\tau_b n^2 \bar\omega_{d}^2}{\omega_0 (\tau)} \right) 
\nonumber \\ & & \times \int_{-\infty}^{+\infty}  \frac{\omega + \omega_0 (\tau)}{n\bar\omega_{d} - \omega_0 (\tau) - \omega} e^{-i \omega t} Q_{n,\omega_0(\tau)} \hat F_{0} (\omega) d\omega \label{eq:dwbarprec3} \;\; ,
\end{eqnarray}
where $\lambda = \mu B_0/{\cal E}$, $\tau_b = 2\pi/\omega_b$, $k_\vartheta = - nq/r$, and we have further assumed that magnetically trapped EPs are characterized by harmonic motion between magnetic mirror points; \idest, they are deeply trapped, such that 
$\hat \omega_{dn} \simeq n\bar\omega_d$ in Eq.~(\ref{eq:hatomegad}). The linear contribution $\delta \bar W_{kn}^L$ is readily obtained from Eq.~(\ref{eq:dwbarprec3}) by
substitution of $\hat F_{0} (\omega) = (2\pi\omega)^{-1} i \bar F_0(0)$, as suggested by Eq.~(\ref{eq:dyson4}). In this case, we assume
an initial (equilibrium) EP isotropic slowing down distribution function of particles born at energy $E_F$. Thus, 
\begin{equation}
\bar F_0 (0) = \frac{3 P_{0EP}}{4\pi E_F} \frac{ {\rm H} (E_F/m_{EP} - {\cal E})}{(2{\cal E})^{3/2} + (2E_c/m_{EP})^{3/2}} \;\; ; \label{eq:F0slow}
\end{equation}
where ${\rm H}$ denotes the Heaviside step function and the normalization condition is chosen such that the EP energy density is $(3/2)P_{0EP}$ for $E_F \gg E_c$,
and EP energy is predominantly transferred to thermal electrons by collisional friction~\cite{stix72}  as it occurs for $\alpha$-particles in fusion plasmas.
We then obtain
\begin{equation}
\delta \bar W_{kn}^L (\omega_0) = \frac{3\pi (r/R_0)^{1/2}}{8\sqrt{2}|s|} \alpha_{EP} \left[1 + \frac{\omega_0}{\bar{\omega}_{dF}}
\ln \left( \frac{\bar{\omega}_{dF}}{\omega_0}-1\right) + i \pi \frac{\omega_0}{\bar{\omega}_{dF}} \right]  \;\; , \label{eq:dwkt0lin}
\end{equation} 
where $\alpha_{EP} = - 8\pi R_0 q^2 P_{0EP}'/B_0^2$ and $\bar{\omega}_{dF} = n\bar\omega_d ({\cal E} = E_F/m_{EP})$. Considering
a radially localized EP source, $\alpha_{EP} = \alpha_{0EP}$ $\exp [ - (r-r_0)^2/L_{EP}^2]$~\cite{zonca05}; and using Eqs.~(\ref{eq:LambdaT}), (\ref{eq:dwbarfepm}) and~(\ref{eq:dwkt0lin}),
it is possible to show
that the linear EPM wave-packet, obtained as solution of Eq.~(\ref{eq:envelope0}), has a characteristic radial envelope width $\sim |L_{EP}/k_\vartheta|^{1/2}$, with $\lambda_\perp \sim |k_\vartheta|^{-1} \ll  |L_{EP}/k_\vartheta|^{1/2} \ll L_{EP}$, consistent with the initial assumptions. Most important, however, is that resonant EPM-EP interactions play a double role: (i) they 
give mode drive in excess of the local threshold condition set by SAW continuum damping, $\propto \Lambda_T(\omega_{0r})$~\cite{chen94}; and (ii) they  
provide the radial potential well that allows EPM to be radially bounded and excited as absolute instability \cite{zonca14b,zonca00a}. These properties of resonant EPM-EP interactions are both crucial
to explain and understand the nonlinear dynamics of an EPM burst. In fact, they suggest that EPM mode structures may adapt to the nonlinearly modified EP source; and, {\sl vice versa}, that
self-consistent evolution of EP PSZS and EPM mode structures may take the form of avalanches~\cite{zonca14e,zonca05}. It is, thus, reasonable to
seek solutions representing the nearly periodic EPM burst as radially localized, traveling and convectively amplified wave-packets~\cite{chen14}.

It is possible to demonstrate that nonlinear dynamics of EP PSZS and EPM are dominated by $\delta \bar W_{kn}^{NL}$. In fact, EP effects $\Lambda_n^{NL}$ are negligible due to 
finite orbit averaging in the kinetic/inertial layer, while $\delta \bar W_{fn}^{NL}$ accounts for non-resonant EP responses \cite{chen14,zonca14a,zonca14b}. Thus, in the present
analysis we may assume $\Lambda_n^{NL} = \delta \bar W_{fn}^{NL} = 0$. We further set $S^{\rm ext}_n (r,t) = 0$ in Eq.~(\ref{eq:envelope0}); and neglect source and 
collision terms in Eq.~(\ref{eq:dyson4}), as our aim is to investigate the nonlinear
evolution of one single EPM burst excited by the initial unstable distribution given in Eq.~(\ref{eq:F0slow}). By direct substitution into Eq.~(\ref{eq:dwbarprec3}) of $\hat F_0(\omega)$ from Eq.~(\ref{eq:dyson4}), and keeping the leading order terms in the time scale ordering of Eq.~(\ref{eq:timescaleorde}), we obtain
\begin{eqnarray}
\delta \bar W_{kn}^{NL}  & \simeq & -
\int {\cal E} d {\cal E} d \lambda \sum_{v_\parallel/|v_\parallel| = \pm} \frac{\pi^2 q R_0}{c^2 k_\vartheta^2 |s|} \frac{e^2}{m} \int_{-\infty}^{+\infty}  \frac{\tau_b n^2 \bar\omega_{d}^2(k_\vartheta/\Omega)
e^{-i \omega t}}{(n\bar\omega_d - \omega_{0r}) - i (\gamma_0 - i \omega)}  
\nonumber \\ & \times & \frac{\partial^2}{\partial r^2} 
\left[ \frac{c^2 k_{\vartheta}^2}{B_0^2} \frac{(2i/\omega)(\gamma_0 + i \omega) \left| \delta \bar\varphi_n(r,\tau) \right|^2}{(n\bar\omega_d - \omega_{0r})^2+(\gamma_0 + i \omega)^2}  \frac{\partial}{\partial r} \hat F_0 (\omega - 2 i \gamma_0)  \right] d\omega \label{eq:dwkt0NL} \; .
\end{eqnarray}
Note that, here, we have taken into account that nonlinear EPM dynamics occurs on meso-scales, intermediate between $\lambda_\perp \sim |k_\vartheta|^{-1}$ and $L_{EP}$, to have $\partial_r$ commute with quantities that vary on the equilibrium length scale. 

In Eq.~(\ref{eq:dwkt0NL}), in addition to the resonant denominators, the integrand has two 
complex-$\omega$ dependences that predominantly contribute to the overall integral value: one is connected with the $\propto 1/\omega$ dependence; the other one is
due to the nearly singular/peaked structure of $\hat F_0 (\omega - 2 i \gamma_0)$ at $\omega = 2 i \gamma_0$. Due to the $\propto \exp ( - i \omega t)$ dependence of the Laplace integrand,
the contribution from $\omega\simeq 0$ is exponentially smaller than that from $\omega \simeq 2 i \gamma_0$ in the growth and saturation phases of the burst; and is, thus, negligible. 
This further demonstrates the importance of the self-consistent interplay of nonlinear mode dynamics and EP transport, 
already noted in Sec.~\ref{sec:periodicD}. Equation~(\ref{eq:dwkt0NL}) also illuminates the competition of phase locking and wave-particle trapping. In fact, 
noting that the radially localized EPM wave-packet travels at the group velocity $v_g$, there results a relationship between the wave-packet radial position and its time
evolving frequency. Thus, 
focusing on meso-scale radial structures of the integrand in Eq.~(\ref{eq:dwkt0NL}), at the leading order we can rewrite
\begin{eqnarray}
\frac{\partial_r^2 \left( \ldots \right)}{(n\bar\omega_d - \omega_{0r}) - i (\gamma_0 - i \omega)} & = & \frac{\partial^2}{\partial r^2} \left[ \frac{\left( \ldots \right)}{(n\bar\omega_d - \omega_{0r}) - i (\gamma_0 - i \omega)}   \right] \nonumber \\
& & + 2 \frac{(n\partial_r \bar\omega_d - \dot \omega_{0r}/v_g) - i \dot \gamma_0/v_g}{\left[(n\bar\omega_d - \omega_{0r}) - i (\gamma_0 - i \omega)\right]^2} \partial_r \left( \ldots \right)  \nonumber \\
& & - 2 \frac{\left[(n\partial_r \bar\omega_d - \dot \omega_{0r}/v_g) - i \dot \gamma_0/v_g\right]^2}{\left[(n\bar\omega_d - \omega_{0r}) - i (\gamma_0 - i \omega)\right]^3}  \left( \ldots \right)  \nonumber \\
& & + \frac{(n\partial_r^2 \bar\omega_d - \ddot \omega_{0r}/v_g^2) - i \ddot \gamma_0/v_g^2}{\left[(n\bar\omega_d - \omega_{0r}) - i (\gamma_0 - i \omega)\right]^2} \left( \ldots \right) 
\;\; . \label{eq:plcommute}
\end{eqnarray}
Noting that $|(n\bar\omega_d - \omega_{0r})|\sim |(\gamma_0 - i \omega)| \sim |k_r v_g| \sim \tau_{NL}^{-1}$; and consistent with the time scale ordering of Eq.~(\ref{eq:timescaleorde}) and 
discussed after Eq.~(\ref{eq:monochr1}), the dominant contributions on the RHS of Eq.~(\ref{eq:plcommute}) are given by the first three lines, where terms $\propto \dot \gamma_0$ and $\propto \ddot \gamma_0$ can be dropped. Meanwhile, given Eq.~(\ref{eq:dotthetatrap}) for $\ell =0$, the contribution of second and third line on the RHS are, respectively $\sim |\tau_{NL}^2 \ddot \Theta_{m,n,\ell=0}|$ and
$\sim |\tau_{NL}^4 \ddot \Theta_{m,n,\ell=0}|^2$ w.r.t.\ the first line. These terms are generally important, and, in the ``bump-on-tail'' paradigm or the uniform plasma limit
for SAW/DAW near marginal stability (cf. Sec.~\ref{sec:dyson}), account for wave-particle trapping~\cite{altshul65,altshul66}. However, when phase locking occurs, Eq.~(\ref{eq:phaselocktrap}), second and third line on the RHS of Eq.~(\ref{eq:plcommute}) become negligible, consistent with Eq.~(\ref{eq:phaselock}). Thus, phase locking {\em de facto} prevents wave-particle trapping to occur (cf. \ref{app:finiteness}). Note also that $|\tau_{NL}^2 \ddot \Theta_{m,n,\ell=0}| \sim \epsilon_{\dot\omega} \ll 1$ by definition, Eq.~(\ref{eq:epsilonomegadot}). Thus, on time scales shorter than $\epsilon_{\dot\omega}^{-1} \tau_{NL}$, Eq.~(\ref{eq:plcommute}) with only the first line on the RHS can be used to calculate $\delta \bar W_{kn}^{NL}$, which describes convective amplification of a nearly periodic phase locked EPM burst when substituted back into
Eq.~(\ref{eq:envelope0}). Phase locking, Eq.~(\ref{eq:phaselocktrap}), assumed here as a simplifying Ansatz, can be self-consistently verified {\sl a posteriori}, once the solution is obtained
as shown below. On longer time scales, wave-particle trapping and/or resonance detuning must be accounted for, along with non-resonant EP responses as well as collisions and other relevant phenomena, neglected here but discussed in Ref.~\cite{zonca14e}.

For the phase locked EPM, the velocity space integral of Eq.~(\ref{eq:dwkt0NL}) is dominated by resonant particles and by $\omega \simeq 2 i \gamma_0$, as discussed above. Thus,
noting that $n\bar\omega_d \simeq - k_\vartheta {\cal E}/(R_0\Omega)$ and $\tau_b \simeq 2\pi q R_0 (R_0/r)^{1/2} {\cal E}^{-1/2}$ for the simplified equilibrium considered in this section~\cite{zonca14b},  Eq.~(\ref{eq:dwkt0NL}) can be reduced to~\cite{chen14}\footnote{Note, here, that Eq.~(\ref{eq:dwkt1NL}) assumes ${\rm Res} \left(\hat F_0 (\omega)|_{n\bar\omega_d = \omega_{0r}}, 0
\right) = (2\pi i)^{-1} \oint_{\Gamma_r}   \hat F_0 (\omega; z) dz = 0$, with $z \equiv (n\bar\omega_d/\omega_{0r} - 1)$ and ${\Gamma_r}$ a vanishing circular path in the complex-$z$ plane centered at $z=0$. In the general case, an expression for $\delta \bar W_{kn}^{NL}$ can still be written but is slightly more complicated. For the application discussed here, Eq.~(\ref{eq:dwkt1NL}) is adequate.}
\begin{eqnarray}
\delta \bar W_{kn}^{NL}  & \simeq & - i \frac{8\pi^4 E_F}{|s|B_0^2}\frac{\omega_{0r}}{\bar{\omega}_{dF}} \left( \frac{r}{R_0} \right)^{1/2} q^2 R_0 \frac{\partial^2}{\partial r^2} 
\int_{-\infty}^{+\infty} \left[ \frac{c^2 k_{\vartheta}^2}{B_0^2}  \frac{\left| \delta \bar\phi_n(r,t) \right|^2}{(2\gamma_0 - i \omega)^2} \right. \nonumber \\
& & \times \left.  \frac{\partial}{\partial r} \left( {\cal E}^{3/2} \hat F_0 (\omega) \right)_{n\bar\omega_d = \omega_{0r}}\right] e^{-i\omega t} d\omega 
= - i \frac{8\pi^4 E_F}{|s|B_0^2}\frac{\omega_{0r}}{\bar{\omega}_{dF}} \left( \frac{r}{R_0} \right)^{1/2} \nonumber \\
& & \times q^2 R_0 \frac{\partial^2}{\partial r^2} \partial_t^{-2}
\left[ \frac{c^2 k_{\vartheta}^2}{B_0^2}  \left| \delta \bar\phi_n(r,t) \right|^2   \frac{\partial}{\partial r} \left( {\cal E}^{3/2}  F_0 (t) \right)_{n\bar\omega_d = \omega_{0r}}\right]  \label{eq:dwkt1NL} \; .
\end{eqnarray}
Furthermore, 
following Ref.~\cite{zonca05} and noting that the analysis is limited to times shorter than $\epsilon_{\dot\omega}^{-1} \tau_{NL}$, $F_0(t) \simeq \bar F_0(0)$  
with $\bar F_0(0)$ given by Eq.~(\ref{eq:F0slow}). Thus, Eq.~(\ref{eq:dwkt0NL}) can be further simplified and yields
\begin{equation}
\delta \bar W_{kn}^{NL}  \simeq  i \mathbb I{\rm m} \left[ \delta \bar W_{kn}^L (\omega_{0r}) \right] k_\vartheta^2  \rho^2_{EP} v_{EP}^2 \partial_t^{-2} \partial_r^2 \left| \bar A_{n0} (r,t)  \right|^2
 \label{eq:dwkt2NL} \; ,
\end{equation}
where $v_{EP}^2 = E_F/m_{EP}$, $\rho_{EP}^2 = v_{EP}^2/\Omega_{EP}^2$, $\mathbb I{\rm m} \left[ \delta \bar W_{kn}^L (\omega_{0r}) \right] $ is given by Eq.~(\ref{eq:dwkt0lin}), and we have introduced the normalized amplitude~\cite{chen14}
\begin{equation}
(e_{EP}/E_F)  \delta \bar \phi_n (r,t) = \bar A_n (r,t) = e^{-i\omega_{0r} t} \bar A_{n0} (r,t) \;\; . \label{eq:barandef}
\end{equation}
Collecting terms, the desired nonlinear evolution equation for $\bar A_{n0} (r,t)$, Eq.~(\ref{eq:envelope0}), can then be rewritten as
\begin{eqnarray}
& &\left[ i \left( \Lambda_T(\omega_{0r}) - \mathbb I{\rm m} \left[ \delta \bar W_{kn}^L (\omega_{0r}) \right] \right) - \left( \delta \bar W_{fn}^L + \mathbb R{\rm e} \left[ \delta \bar W_{kn}^L (\omega_{0r})\right] \right) \right] \bar A_{n0} \nonumber \\ & & \hspace*{2em} = 
\left[ \frac{\partial \mathbb R{\rm e} \left[ \delta \bar W_{kn}^L (\omega_{0r})\right]}{\partial \omega_{0r}} \left( i \frac{\partial}{\partial t} \right) + \frac{1}{2 k_\vartheta^2 s^2} \frac{\partial^2 D_n^L}{\partial \theta_{k0}^2}\frac{\partial^2}{\partial r^2} \right. \nonumber \\ & & \hspace*{4em}  \left. + i  \mathbb I{\rm m} \left[ \delta \bar W_{kn}^L (\omega_{0r}) \right] k_\vartheta^2  \rho^2_{EP} v_{EP}^2 \partial_t^{-2} \frac{\partial^2}{\partial r^2} \left| \bar A_{n0} \right|^2  \right] \bar A_{n0} \;\; , \label{eq:envelope1}
\end{eqnarray} 
where all contributions are defined in Eqs.~(\ref{eq:LambdaT}), (\ref{eq:dwbarfepm}) and~(\ref{eq:dwkt0lin}). 

\begin{figure}
\centerline{\resizebox{0.4\linewidth}{!}{\includegraphics{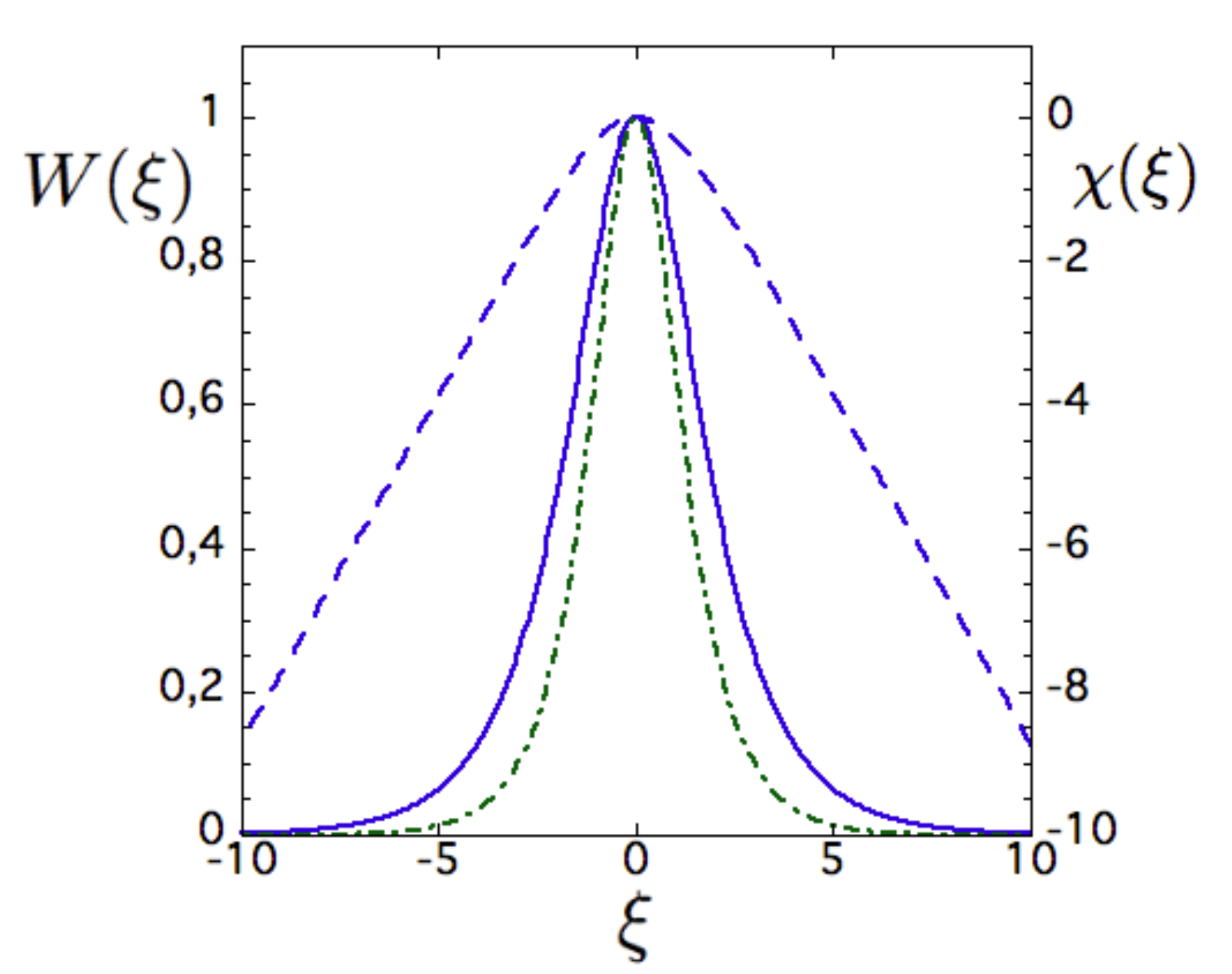}}}
\caption{Normalized amplitude, $W(\xi)$ (solid blue line), and phase, $\chi(\xi)$ (dashed blue line), of the nonlinear EPM wave-packet. The function ${\rm sech} (\xi)$ (green dash-dotted line) is also shown as reference.}
\label{fig:complexU}\end{figure}

Equation~(\ref{eq:envelope1}) readily recovers the linear limit~\cite{zonca14b,zonca00a}.
In the general case~\cite{chen14,zonca14a}, as suggested by the last term on the RHS, Eq.~(\ref{eq:envelope1})  is a nonlinear Schr\"odinger equation describing the evolution of the nearly periodic phase locked EPM burst in an ``active medium''. We can look for solutions in the convectively amplified self-similar form~\cite{chen14} 
\begin{equation}
\bar A_{n0} (r,t) = U(\xi) \exp \int^t \gamma_0(t') dt' = W(\xi) \exp \left(  i \chi (\xi) + \int_0^t \gamma_0 (t') dt' \right) \; , \label{eq:ampsoliton}
\end{equation}
with 
\begin{equation}
\xi \equiv k_{n0} \left(  r -  r_0 - \int_0^t v_g(t') d t' \right) \;\; , \label{eq:xidef}
\end{equation}
and $k_{n0}$ denoting the nonlinear radial wave vector. Noting that 
$\partial_r \bar A_{n0} = k_{n0}  \partial_\xi U (\xi) \exp \int^t \gamma_0(t') dt'$ and $\partial_t \bar A_{n0} =  ( \gamma_0 - k_{n0} v_g \partial_\xi )U (\xi) \exp \int^t \gamma_0(t') dt'$, there exist two regimes in the nonlinear EPM burst evolution: (i) $\gamma_0 \gaeq | k_{n0} v_g|$, where the linear EPM is increasingly modified by the
nonlinear interplay with EP transport as the fluctuation intensity increases; and (ii) $\gamma_0 < | k_{n0} v_g|$, where the nonlinear EPM-EP interaction dominates the dynamics. In this second case, it is possible to solve Eq.~(\ref{eq:envelope1})
in a simple analytical form. In fact, 
balancing the nonlinear term with the linear dispersiveness in Eq.~(\ref{eq:envelope1}) yields 
\begin{equation}
v_g = \lambda_g \hat v_{EP}^{E\times B} \;\; , \;\;\;\;\; {\rm and} \;\;\;\;\; k_{n0}^2 = \frac{k_\vartheta^2}{\lambda_g^2} \frac{s^2 \mathbb I{\rm m} \left[ \delta \bar W_{kn}^L (\omega_{0r}) \right]}{\partial^2 D_n^L/\partial \theta_{k0}^2} \;\; , \label{eq:vgkn0}
\end{equation}
with $\hat v_{EP}^{E\times B} =  (-k_\vartheta c/B_0) \max [ \delta \bar \phi_n(r,t)]$ the radial $\bm E \times \bm B$ velocity at the fluctuation peak, $\lambda_g$ an ${\cal O}(1)$ control parameter to be determined; and
\begin{equation}
\partial_\xi^2 U = \lambda_0 U - 2 i U |U|^2 \;\;  ,
\label{eq:complexU}
\end{equation}
where $\lambda_0 \simeq - 0.47 + i 1.33$ corresponds to the ground state of the complex nonlinear oscillator described by Eq.~(\ref{eq:complexU}), whose numerical solution is
shown in Fig.~\ref{fig:complexU}, separating amplitude and phase, $U(\xi) = W(\xi) e^{i \chi(\xi)}$.
With this solution, Eq.~(\ref{eq:envelope1}) becomes the ``nonlinear'' EPM dispersion relation
\begin{equation}
D_n^{L\ell}(\omega_0) - \frac{\lambda_0}{2 \lambda_g^2} \mathbb I{\rm m} \left[ \delta \bar W_{kn}^L (\omega_{0r}) \right] = 0  \;\; , \label{eq:EPMnldisp}
\end{equation} 
where
\begin{eqnarray}
D_n^{L\ell}(\omega_0) & = & i \left( \Lambda_T(\omega_{0r}) - \mathbb I{\rm m} \left[ \delta \bar W_{kn}^L (\omega_{0r}) \right] \right) - \left( \delta \bar W_{fn}^L + \mathbb R{\rm e} \left[ \delta \bar W_{kn}^L (\omega_{0r})\right] \right)\nonumber \\ & & - \left(  \omega_{0r} \frac{\partial \mathbb R{\rm e} \left[ \delta \bar W_{kn}^L (\omega_{0r})\right]}{\partial \omega_{0r}} \right) i \frac{\gamma_0}{\omega_{0r}} 
\end{eqnarray}
is the linearized ``local'' EPM dispersion function \cite{chen94}, obtained neglecting the linear dispersiveness $\propto \partial^2 D_n^L/\partial \theta_{k0}^2$ \cite{zonca14b}.

Equations~(\ref{eq:vgkn0}) and~(\ref{eq:EPMnldisp}) describe a one-parameter family, $\lambda_g$, of possible EPM wave-packets that are convectively amplified as they radially propagate with a group velocity $\propto \hat v_{EP}^{E\times B}$. The dominant nonlinear mode is obtained for 
\begin{equation}
\frac{d\gamma_0}{d \lambda_g^2} = \frac{\partial\gamma_0}{\partial \lambda_g^2} + \frac{\partial\gamma_0}{\partial\omega_{0r}} \frac{d\omega_{0r}}{d\lambda_g^2} = 0 \;\; ;
\label{eq:dominantEPM}
\end{equation}
\idest, for maximized mode growth and wave-EP power transfer \cite{chen14}. For typical tokamak plasma parameters, and the simplified model plasma equilibrium adopted here, one generally finds $\lambda_g \simeq 0.5 \div 0.6$\footnote{It is possible to verify that $d\gamma_0/d\lambda_g^2 >0$ for $\lambda_g^2 \rightarrow 0$, and $d\gamma_0/d\lambda_g^2 <0$ for $\lambda_g^2 \rightarrow \infty$.}. 
Thus, the EPM wave-packet propagates radially at a substantial fraction of the peak EP  $\bm E \times \bm B$ velocity and, as a result, resonant EP motion is predominantly ballistic in the radial direction,
similar to the ``mode particle pumping'' mechanism introduced in Ref.~\cite{white83} for explaining EP secular loss due to ``fishbones''. 
Equation~(\ref{eq:EPMnldisp}) applies at the instantaneous radial position $r_1 (t) = r_0 + \int_0^t v_g(t') dt'$ of the EPM wave-packet.  
The dependence on $r_1$, in the present model equilibrium, is explicit in terms $\propto \alpha_{EP}$ and $\propto \bar \omega_{dF}$; and implicit in terms depending on $\omega_0$. 
Due to the $\propto  \ln (\bar\omega_{dF}/\omega_0 - 1)$ term and $\omega_0/\bar\omega_{dF}$ dependences of $\delta \bar W_{kn}^L$ in Eq.~(\ref{eq:dwkt0lin}), the real EPM frequency is given by
\begin{equation}
\omega_{0r}/\bar\omega_{dF} (r_1) \simeq {\rm const} \;\; , \label{eq:omega0r}
\end{equation}
at the leading oder, so that the phase locking condition, Eqs.~(\ref{eq:phaselock}) and~(\ref{eq:phaselocktrap}), is readily satisfied, consistent with the Ansatz and introductory discussions above in this section.
Meanwhile, since $\mathbb I{\rm m}\lambda_0 > 0$, the nonlinear EPM-EP interplay results into a strengthening of the mode drive w.r.t.\ the linear value $\propto \mathbb I{\rm m} \left[ \delta \bar W_{kn}^L (\omega_{0r})\right]$. Thus, the convective amplification of the phase locked EPM is an ``avalanche''~\cite{zonca05}; and wave-packet amplification can continue until quenched by radial nonuniformities; \idest, the combined effect of decreasing values of linear drive $\mathbb I{\rm m} \left[ \delta \bar W_{kn}^L (\omega_{0r})\right]\propto \alpha_{EP}$ and increasing values of SAW continuum damping $\propto \Lambda_T(\omega_{0r})$ (cf. Secs.~\ref{sec:lock} and~\ref{app:finiteness}; and Refs. \cite{zonca14e,zonca13,chen14}). Hence, as consequence of phase locking, meso-scale nonlinear dynamics of the EPM burst as well as EP transport are extended to the macro-scales.

The self-consistent interplay of nonlinear EPM radial structure evolution with EP transport suggests that this is a non-adiabatic ``autoresonance'' effect (cf. Sec.~\ref{sec:lock}), caused by the non-perturbative EP response and the ensuing kinetic EPM dispersiveness. Plasma instability and non-perturbative EP behavior are the crucial differences with respect to adiabatic ``autoresonance'', where fluctuations are driven/controlled externally and adiabatic frequency sweeping is imposed~\cite{meerson90,fajans01}. 
During the amplification process and non-adiabatic frequency chirping, the relevant nonlinear time scale is $\tau_{NL} \sim|\partial_t^{-1}| \sim|A_{n0}|^{-1}$, consistent with Eq.~(\ref{eq:vgkn0}). Furthermore,  since $v_g = \lambda_g \hat v_{EP}^{E\times B}$, the EPM intensity is proportional to the square of the traveled distance, similar to the propagation of a FEL short optical pulse  in the superradiant regime~\cite{bonifacio90}. Similar to superradiance is also the resonant EPM-EP interaction, which takes place when the wave-EP phase is amplifying due to the phase locking condition. By the time residual resonance detuning changes the wave-EP phase to damping, the resonance condition is lost by radial decoupling. Thus, EPMs can be considered as examples of ``autoresonance'' and ``superradiance'' in fusion plasmas~\cite{chen14,zonca13b}. Other interesting analogies with neighboring research areas are further discussed in Ref.~\cite{chen14}. For example, Eq.~(\ref{eq:complexU})
shows interesting analogies with $\partial_\xi^2 U = U - 2 U^3$ yielding $U = {\rm sech} (\xi)$; \idest, the equation of motion of a nonlinear oscillator in the so-called ``Sagdeev potential'' $V = ( - U^2 + U^4)/2$, which can be obtained, \exgra, in the analysis of Ion Temperature Gradient turbulence spreading via soliton formation~\cite{guo09}, but also 
from the Gross-Pitaevsky equation, describing the ground state of a quantum system of identical bosons~\cite{gross61,pitaevsky61};  and from the investigation of the envelope of modulated water wave groups~\cite{zakharov68}. However, Eq.~(\ref{eq:complexU}) maintains its peculiarities, mostly due to its complex nature coming from the self-consistent interplay of EPM radial structure and EP transport.

\section{Summary and Discussions} 
\label{sec:summary}

Phase space zonal structures play fundamental roles in EP transport induced by SAW/DAW instabilities with $|\gamma_L/\omega_0|\ll 1$ in burning fusion plasmas. 
In this work, we have derived general equations that can be used as starting point in analytic as well as numerical investigations of 
the self-consistent interplay of SAW/DAW instabilities and PSZS in the presence of a non-perturbative EP population in fusion plasmas
with realistic equilibrium geometry and plasma nonuniformities. As particular case of the general equations, we have derived a reduced
description that may be adopted for nearly periodic fluctuations within the ``fishbone'' paradigm; and applied it to the analytic
study of convective EPM amplification as solitons in active media. Phase locking, in this case, extends the meso-scale dynamics
of EPM ``avalanches'' and causes secular EP redistribution on the macro-scales. 
These processes have interesting analogies with 
``autoresonance'' in nonlinear dynamics and ``superradiance'' in FEL operations. 

The strength
of the instability drive controls the mechanism by which nonlinear saturation is reached in nonuniform plasma equilibria and realistic magnetic field geometries. The reduced 1D uniform plasma
description, adopted in the ``bump-on-tail'' paradigm, applies near marginal stability as long as the nonlinear particle displacement at saturation is significantly smaller than
the radial mode width. As a result, mode saturation occurs via wave-particle trapping and resonance detuning. 
For sufficiently strong EP drive, $|\gamma_L/\omega_0|>|\gamma_L/\omega_0|_c$ ($|\gamma_L/\omega_0|_c\laeq 10^{-2}$ for typical parameters), 
the fluctuation induced EP displacement becomes comparable to the radial mode width, and radial decoupling of EPs from mode structures is increasingly more important in the nonlinear
saturation dynamics, until it eventually dominates over resonance detuning. This transition is demonstrated in numerical simulation experiments of meso-scale dynamics of EPMs; and is indicative of the plasma behavior as a 3D system, with given equilibrium geometry and plasma nonuniformity.
Restricting the analysis to precessional resonance with magnetically trapped particles only, the general case can be reduced to a 1D nonuniform plasma, adopted in the ``fishbone'' paradigm, which recovers the ``bump-on-tail'' paradigm in the uniform plasma limit; and allows to understand the onset of non-local behaviors, such as self-consistent interplay of mode structures and EP transport.

The strength of instability drive or, more precisely, the perturbative vs. non-perturbative EP behavior, also controls the frequency dynamics of PSZS. Very near SAW/DAW marginal stability, such that the 
``bump-on-tail'' paradigm can be applied, wave-particle trapping is the fastest nonlinear time scale and, for sufficiently weak collisions, PSZS may evolve into hole/clump pairs with adiabatically sweeping frequency ($|\dot\omega|\ll\omega_B^2$), which are maintained near marginal stability by balancing energy extraction from EP phase space with background dissipation. Meanwhile, for sufficiently strong EP drive that the ``fishbone'' paradigm must be adopted and EP response is non-perturbative, kinetic SAW/DAW dispersiveness, self-consistently modified by EP transport, causes non-adiabatic frequency evolution ($|\dot\omega|\sim\omega_B^2$). Phase locked resonant EPs dominate nonlinear dynamics and, by reduction of resonance detuning, nonlinear interplay of secular EP transport with radial mode structures becomes the most important nonlinear physics. 

The present theoretical framework, conceived and developed for application to SAW/DAW instabilities in burning fusion plasmas
with a non-perturbative EP population, is characterized by the peculiar system geometry and equilibrium non-uniformity. However,
it also sheds light on physics aspects that may be of relevance to neighboring fields of research, such as space plasma physics,
nonlinear dynamics, condensed matter and accelerator physics.

\section*{Acknowledgments}

This work was supported by European UnionÕs Horizon 2020 research and innovation program under grant agreement number 633053 as Enabling Research Project CfP-WP14-ER-01/ENEA\_Frascati-01; and by US DoE, ITER-CN, and NSFC No. 11235009 grants.

\appendix 

\section{Wave particle interactions in axisymmetric toroidal systems.}
\label{app:res2D}

For gyrokinetic theory in axisymmetric toroidal systems, considered in this work, two obvious pairs of action angle coordinates for charged particle motions are $(m\mu, \alpha)$, with $\mu = v_\perp^2/(2 B_0) + \ldots$ the magnetic moment (see, \exgra, Refs. \cite{brizard07,cary09}) and $\alpha$ the gyrophase; and $(P_\phi, \phi)$, with $P_\phi$ the canonical toroidal angular momentum and $\phi$  the toroidal angle\footnote{A recent review of coordinates systems and their connection with the description of the guiding-center particle motion is given by Cary and Brizard~\cite{cary09}.}.
Here, perpendicular ($\perp$) and parallel ($\parallel$) directions are defined with respect to the unit vector $\bm b = \bm B_0/B_0$; and the axisymmetric equilibrium magnetic field configuration is assumed in the form
\begin{equation}
\bm B_0 = F(\psi) \bm \nabla \phi + \bm \nabla \phi \times \bm \nabla \psi \;\; ; \label{eq:B0}
\end{equation}
$2\pi\psi$ denoting the poloidal magnetic flux within the magnetic surface labeled by $\psi$. Consistent with the main text, we adopt field aligned toroidal flux coordinates $(r,\theta,\zeta)$, with $r=r(\psi)$ a radial-like flux variable, the poloidal angle $\theta$ is chosen such that the Jacobian ${\cal J} = (\bm \nabla \psi \times \bm \nabla \theta \cdot \bm \nabla \zeta)^{-1}$ satisfies the condition
${\cal J} B_0^2$ being a magnetic flux function \cite{boozer81,boozer82}; and the general toroidal angle $\zeta \equiv \phi - \nu(\psi,\theta)$, $\nu(\psi,\theta)$ being a suitable periodic function of $\theta$ \cite{white89b}, such that the safety factor $q$,
\begin{equation}
q = \frac{\bm B_0 \cdot \bm \nabla \zeta}{\bm B_0 \cdot \bm \nabla \theta} = q(\psi) \;\; , \label{eq:straight}
\end{equation}
is a function of $\psi$. 
At the leading order, $P_\phi$ is given by 
\begin{equation}
P_\phi = \frac{e}{c} \left( F(\psi) \frac{v_\parallel}{\Omega} - \psi \right)  \;\; ; \label{eq:pphiapp}
\end{equation}
with $\Omega = e B_0/(mc)$ the cyclotron frequency. 
The third pair of action angle coordinates is $(J,\theta_c)$, with $J$ the ``second invariant'' and $\theta_c$ the respective conjugate canonical angle
\begin{equation}
J = \oint v_\parallel d l \;\; , \;\;\;\;\;
\theta_c = \omega_b \int_0^\theta d\theta'/\dot \theta' \;\; . \label{eq:jetaapp}
\end{equation}
Here,  integrals are taken along the particle orbits in the equilibrium magnetic field, $d l$ denoting the arc-length element along it; and we have introduced the unified notation $\omega_b$ for bounce and transit frequency of trapped and circulating particles, respectively,
\begin{equation}
\omega_b(\mu,J,P_\phi) = \frac{2\pi}{\oint d\theta/\dot \theta} \;\; . \label{eq:omegab}
\end{equation} 
Given $H_0 = m (v_\parallel^2 + \mu B_0 )$ and the definitions of $J$ and $\omega_b$,
\begin{displaymath}
\frac{\partial J}{\partial H_0} = \frac{2\pi}{\omega_b} \;\; \leftrightarrow \;\; \omega_b = 2\pi \frac{\partial H_0}{\partial J} \;\;.
\end{displaymath}
Therefore, $J$ can be readily computed given the reference magnetic equilibrium
\begin{displaymath}
J = J(\mu,P_\phi,H_0) \;\; \leftrightarrow \;\; H_0 = H_0(\mu,J,P_\phi) \;\; . 
\end{displaymath}

Equation~(\ref{eq:jetaapp}) yields $\theta_c = \omega_b \tau$, with $\tau$ a time-like parameter tracking the particle position along the closed poloidal orbit of period $2\pi/\omega_b$. Thus, for a particle with given constants of motion,
\begin{equation}
r = \bar r + \tilde \rho_c(\theta_c) \;\; , \label{eq:rparm} \\
\end{equation}
where $\bar r$ and $\tilde \rho_c(\theta_c)$ are parameterized by $(\mu,J,P_\phi)$ and the initial particle radial position\footnote{The initial particle poloidal angle can be reabsorbed by suitable time shift in the time-like parameter $\tau$; \idest, in $\theta_c$.}; and ``tilde'' denotes a generic periodic function in $\theta_c$ with zero average, which can be computed from the particle equations of motion in the equilibrium $\bm B_0$.
Similarly, it is readily demonstrated that, for circulating particles (\idest; particles whose $v_\parallel$ does not change sign)
\begin{equation}
\theta = \theta_c + \tilde \Theta_c(\theta_c) \;\; , \label{eq:thetaparmcirc}
\end{equation}
while, for magnetically trapped particles (\idest; particles whose $v_\parallel$ changes sign twice along a closed orbit in the $(r,\theta)$ plane),
\begin{equation}
\theta = \tilde \Theta_c(\theta_c) \;\; . \label{eq:thetaparmtrap}
\end{equation}
Note, again, that $\tilde \Theta_c$ is a periodic function of $\theta_c$ with zero average, which is parameterized by $(\mu,J,P_\phi)$. Thus it takes different values in Eqs.~(\ref{eq:thetaparmcirc}) and~(\ref{eq:thetaparmtrap}).
At last, considering Eqs.~(\ref{eq:straight}) and the definition of $\zeta\equiv \phi - \nu(\psi,\theta)$, 
the parametric representation of $\zeta$ along the particle orbit in the equilibrium $\bm B_0$ is
\begin{equation}
\zeta = \bar{\omega}_d \tau + \bar q \theta + \tilde \Xi_c(\theta_c) \;\; . \label{eq:zetaparm}
\end{equation}
Here, $\theta$ is given by Eqs.~(\ref{eq:thetaparmcirc}) or~(\ref{eq:thetaparmtrap}), $\bar\omega_d(\mu,J,P_\phi)$ is the toroidal precession frequency and 
\begin{equation}
\bar q (\mu,J,P_\phi) = \frac{\oint q d\theta}{\oint d\theta} = \frac{\oint q \dot \theta d\theta_c}{\oint \dot \theta d\theta_c}  \;\; . \label{eq:barqdef}
\end{equation}
Note that, even though $q$ does not depend on $\theta$, it depends on the particle position because of the particle radial displacement.
Meanwhile, from Eq.~(\ref{eq:zetaparm}) and~(\ref{eq:barqdef}), 
one readily derives the following definition for the toroidal precession frequency
\begin{equation}
\bar\omega_d (\mu,J,P_\phi) =  \frac{\omega_b}{2\pi} \oint \left( \dot \zeta - q \dot \theta \right) \frac{d\theta}{\dot \theta} \;\; . \label{eq:omegad}
\end{equation}

\subsection{Wave particle resonance condition}
\label{app:wpres}

Using Eqs.~(\ref{eq:rparm}) to~(\ref{eq:zetaparm}),  
the Fourier decomposition 
\begin{equation}
f(r,\theta,\zeta) = \sum_{m,n \in \mathbb{Z}} e^{i n \zeta - i m \theta} f_{m,n} (r) \label{eq:fourierdec}
\end{equation}
of a scalar function $f(r,\theta,\zeta)$, describing a generic fluctuating field for which time dependence has been left implicit, can be projected along the unperturbed
particle motion in the equilibrium $\bm B_0$ using the time-like parameter $\tau = \theta_c/\omega_b$; 
expressing $(r,\theta,\zeta)$ in terms of the particle position along its unperturbed trajectory. More specifically, we have
\begin{equation}
f(r,\theta,\zeta) = \sum_{m,n,\ell \in \mathbb{Z} } e^{i \left( n \bar\omega_d + \ell \omega_b\right) \tau }  {\cal P}_{m,n,\ell} \circ f_{m,n} (\bar r) \;\; . \label{eq:kindec1app}
\end{equation}
Here, we have introduced the projection operators along the particle motion in the equilibrium $\bm B_0$, defined as
\begin{eqnarray}
{\cal P}_{m,n,\ell} \circ f_{m,n} (\bar r) & \equiv & \frac{\lambda_{m,n}}{2\pi} \oint  \exp \left\{ i n \tilde \Xi_c (\theta_c) + i \left[ n \bar q (\bar r) 
\right. \right. \nonumber \\ & & \left. \left. - m  \right] \tilde \Theta_c( \theta_c )\right\}  f_{m,n} ( \bar r + \tilde \rho_c(\theta_c) ) e^{-i \ell \theta_c} d \theta_c
\;\; . \label{eq:calpmnapp}
\end{eqnarray}
Furthermore, $\lambda_{m,n} = 1$ for magnetically trapped particles, while, for circulating particles,
\begin{equation}
\lambda_{m,n} = \exp \left[i \left( n \bar q (\bar r ) - m \right)  \omega_b \tau \right] \;\; . \label{eq:lambdamnapp}
\end{equation}
Note that Eq.~(\ref{eq:kindec1app}) represents the value of $f(r,\theta,\zeta)$ effectively experienced by the particle along its orbit, parameterized by the actions $(\mu,J,P_\phi)$, or equivalently by $(\mu,P_\phi,H_0)$ (see above), and by the time-like variable $\tau = \theta_c/\omega_b$. The integer $\ell \in \mathbb{Z}$ stands for the ``bounce harmonic'', similarly to $m,n \in \mathbb{Z}$ denoting poloidal and toroidal Fourier harmonics. Equation~(\ref{eq:kindec1app}) is completely defined, together with the functions $\tilde\Theta_c(\theta_c)$, $\tilde \Xi_c(\theta_c)$ and $\tilde \rho_c(\theta_c)$ that are obtained from the integration of particle motion, once the reference equilibrium $\bm B_0$ is assigned.
This mode structure decomposition corresponds to a ``Lagrangian'' description of the fluctuation while moving in the reference frame of the considered particle; \idest, it is
a lifting of a generic scalar field to the particle phase space \cite{chen14}. Meanwhile, Eq.~(\ref{eq:fourierdec}) is the usual mode structure decomposition 
of the fluctuating field in the laboratory frame. These two descriptions are the basis of the dual nature of fluctuation structures within a kinetic analysis of wave particle interactions: one with emphasis on the ``effective'' fluctuation strength experienced by the particles; and another one, reflecting the fields measured by the ``observer''.  Of these two descriptions, the former enters in the resonant particle dynamic response and, thus, is crucially important for wave-particle power exchange, resonant excitation and instability drive; and particle transports \cite{chen99}. The latter, meanwhile, is 
what enters in the evolution equation(s) for the mode structures.

As noted above, time dependences in $f(r,\theta,\zeta)$ are left implicit to simplify notation. Once a monochromatic wave $f(r,\theta,\zeta) \propto \exp (- i \omega_0 t)$ is assumed, the resonance condition; \idest, the stationarity of wave-particle phase in the particle moving frame where $\tau \leftrightarrow t$, is readily derived from Eq.~(\ref{eq:kindec1app}) and yields
\begin{equation}
\omega_0 = \omega_0 (\mu,J,P_\phi) = n\bar \omega_d + \ell \omega_b \label{eq:trapres}
\end{equation}
for magnetically trapped particles; while, for circulating particles,
\begin{equation}
\omega_0 = \omega_0 (\mu,J,P_\phi) = n\bar \omega_d + \ell \omega_b + \left( n \bar q (\bar r ) - m \right)  \omega_b \;\; . \label{eq:circres} 
\end{equation}
Here, we remind that $\omega_b$ stands for both bounce and transit frequencies. We also note that $\bar \omega_d$ is typically negligible for circulating particles, except for particles that are close to the trapped-to-circulating boundary in the action space.

\subsection{The effect of fluctuations}
\label{app:wpresNL}

When considering the effect of fluctuations, for every bounce/transit; \idest, a $2\pi$ increment in the angle $\theta_c$, a shift will be accumulated in the wave-particle phase (``resonance detuning''), because of radial nonuniformities and the dependence of $\bar\omega_d$ and $\omega_b$ on $J,P_\phi$ that are not conserved anymore in the nonlinear regime ($\mu$ is conserved for low frequency fluctuations, analyzed here, for which $|\omega_0|\ll |\Omega|$). Moreover, the wave-particle phase in the mode structure decomposition in action-angle variables will also be shifted, since $\tilde\Theta_c(\theta_c)$, $\tilde \Xi_c(\theta_c)$ and $\tilde \rho_c(\theta_c)$ are not simple periodic functions of $\theta_c$ any longer, as in the given plasma equilibrium. Consistent with the time scale ordering assumed in this work and, in particular, with the condition $1 \gg |\omega_0 \tau_{NL}|^{-1} \sim |\gamma_L/\omega_0| \gg \epsilon_\omega \equiv |\omega_0/\Omega|$ (cf. Sec.~\ref{sec:theory}), we assume
that for every bounce/transit the effect of the nonlinear dynamics is small compared with the oscillations of the equilibrium particle trajectory. 
However, the cumulative effect of the nonlinear dynamics on many bounce/transit times can be large and even connected with a secular process; \idest, not bounded in time. 

The perturbed equations of motion, which are needed to extend Eq.~(\ref{eq:kindec1app}) to handle the effect of small but finite amplitude fluctuations, can be readily obtained from the
expression of the gyrocenter Hamiltonian up to order $\sim \epsilon_\delta \equiv |\delta \bm B_\perp/B_0| \ll 1$~\cite{brizard07}; yielding the expressions for $\delta \dot r$ ($\delta \dot \psi$), $\delta \dot \theta$ and $\delta \dot \zeta$, as well as for $\dot H_0$, $\dot P_\phi$ and $\dot J$. Here, $(\delta r, \delta \theta, \delta \zeta)$ denote the change in particle position due to the small but finite amplitude fluctuations, which are also the cause of $\delta H_0$, $\delta P_\phi$ and $\delta J$ and finite $\dot H_0$, $\dot P_\phi$ and $\dot J$. All these perturbed quantities are considered here to be formally linear in the fluctuation fields. 
Within this level of approximation of gyrokinetic description of particle motions, the conserved invariants are $\mu$ and the extended phase space Hamiltonian $K = H_0 + e \left\langle \delta \phi_g - v_\parallel \delta A_{\parallel g}/c \right\rangle - \bar H$ (cf., \exgra, \cite{lichtenberg83,lichtenberg10}), whose conservation can be readily cast as
\begin{equation}
\frac{\partial}{\partial t} \dot \Pi_\phi + \frac{\partial}{\partial \zeta} \dot{\bar H} = 0 \;\; . \label{eq:kamcons}
\end{equation}
Here, $\bar H$ is the the gyrocenter Hamiltonian up to order $\sim \epsilon_\delta$~\cite{brizard07}
\begin{equation}
\bar H = H_0 + e \left\langle \delta \phi_g \right\rangle - e \frac{v_\parallel}{c} \left\langle \delta A_{\parallel g} \right\rangle \;\; ; \label{eq:hbaro1}
\end{equation}
and $\Pi_\phi$ is the extension of the canonical toroidal angular momentum, Eq.~(\ref{eq:pphiapp}), in the presence of fluctuations and for $\bm B_0$ expressed as in Eq.~(\ref{eq:B0}), 
\begin{equation}
\Pi_\phi = P_\phi + \frac{e}{c} \frac{F(\psi)}{B_0} \left\langle \delta A_{\parallel g} \right\rangle \;\; . \label{eq:pphipertapp}
\end{equation}
Furthermore, it is readily shown that, for trapped particles, we have
\begin{equation} 
\theta = \tilde\Theta_c(\theta_c) + \Delta \theta \label{eq:thetatrapnl} \;\; ,
\end{equation}
while, for circulating particles,
\begin{eqnarray}
\theta & = & \omega_b \tau + \tilde \Theta_c(\theta_c) + \Delta \theta \nonumber \\ & &  + \frac{\partial \omega_b}{\partial P_\phi} \int_0^\tau \delta P_\phi d \tau'  +  \frac{\partial \omega_b}{\partial J} \int_0^\tau \delta J d \tau'  \label{eq:thetacircnl} \;\; ,
\end{eqnarray}
with
\begin{equation}
\Delta \theta =  \int_0^\tau \delta \dot \theta \frac{d\theta}{\dot \theta} \label{eq:deltatheta} \;\; .
\end{equation}
Note that the difference between Eqs.~(\ref{eq:thetatrapnl}) and~(\ref{eq:thetacircnl}) stems from the different nature of unperturbed orbits, respectively described by Eqs.~(\ref{eq:thetaparmtrap}) and~(\ref{eq:thetaparmcirc}), and from the varied expression of $\theta_c \rightarrow \theta_c + \delta \theta_c$ ($\delta \theta_c = \int_0^\tau \delta \omega_b d\tau'$) for circulating particles in the presence of fluctuations. 
Similarly, the radial motion is described as
\begin{eqnarray}
r & = &  \bar r + \tilde \rho_c(\theta_c) + \Delta r  \label{eq:rhoctnl} \;\; , \\ 
\Delta r & = & \int_0^\tau \delta \dot r \frac{d\theta}{\dot \theta} \label{eq:deltar} \;\; ,
\end{eqnarray} 
while $\zeta$ is given by
\begin{eqnarray}
\zeta & = & \bar\omega_d \tau + \bar q \theta + \tilde \Xi_c(\theta_c)  + \Delta \zeta \nonumber \\
& & + \frac{\partial \bar \omega_d}{\partial P_\phi} \int_0^\tau \delta P_\phi d \tau'  +  \frac{\partial \bar \omega_d}{\partial J} \int_0^\tau \delta J d \tau' \label{eq:xictnl} \\ & &  + \bar q \left( \frac{\partial \omega_b}{\partial P_\phi} \int_0^\tau \delta P_\phi d \tau'  +  \frac{\partial \omega_b}{\partial J} \int_0^\tau \delta J d \tau' \right) + \omega_b \frac{d \bar q}{d \bar r} \int_0^\tau \delta r  d \tau' \nonumber \;\; , \\
\Delta \zeta & = &  \int_0^\tau \delta \dot \zeta \frac{d\theta}{\dot \theta} \label{eq:deltaxi} \; \; . 
\end{eqnarray}
Here, the terms involving $\omega_b$ and its derivatives with respect to $P_\phi$ and $J$ are to be considered for circulating particles only.
In fact, these terms are originated by the effect of fluctuations on the term $\propto \bar q \theta_c = \int_0^\tau \omega_b \bar q d \tau'$, due to the unperturbed circulating particle motion, Eq.~(\ref{eq:thetaparmcirc}), which has no correspondence in the unperturbed trapped particle motion, Eq.~(\ref{eq:thetaparmtrap}). In particular, terms $\propto \partial \omega_b/\partial P_\phi , \partial \omega_b/\partial J$ have the same origin of those included in Eq.~(\ref{eq:thetacircnl}) for circulating particles and excluded in Eq.~(\ref{eq:thetatrapnl}) for magnetically trapped particles. Similarly, the term
involving $\sim \omega_b (d \bar q/d \bar r)$ stems from $\int_0^\tau \omega_b \delta \bar q d \tau'$ for circulating particles only.

With the above results, 
we can extend the mode structure decomposition in action-angle variables, Eq.~(\ref{eq:kindec1app}), to include 
nonlinear particle orbit distortions due to fluctuations: 
\begin{equation} 
f(r,\theta,\zeta) = \sum_{m,n,\ell \in \mathbb Z}  \lambda_{m,n}^{NL} e^{i \left( n \bar\omega_d + \ell \omega_b\right) \tau + i \Theta_{m,n,\ell}^{NL} }  {\cal P}_{m,n,\ell} \circ  e^{\Delta r \partial_{\bar r}} f_{m,n}  (\bar r) \;\; , \label{eq:kindec0nl}
\end{equation}
where ${\cal P}_{m,n,\ell} \circ e^{\Delta r \partial_{\bar r}} f_{m,n}$ is defined as in Eq.~(\ref{eq:calpmnapp}), 
$\lambda_{m,n}^{NL}$ is the
nonlinear extension of Eq.~(\ref{eq:lambdamnapp}), \idest, $\lambda_{m,n}^{NL}=1$ for trapped particles, while for circulating particles
\begin{eqnarray}
\lambda_{m,n}^{NL} & = & \exp \left[i \left( n \bar q (\bar r ) - m \right) \left(\frac{\partial \omega_b}{\partial P_\phi} \int_0^\tau \delta P_\phi d \tau'  \right. \right. \nonumber 
\\ & & \left. \left. +  \frac{\partial \omega_b}{\partial J} \int_0^\tau \delta J d \tau' \right)
+ i n \omega_b \frac{d \bar q}{d \bar r} \int_0^\tau \delta r d \tau' \right] \label{eq:lambdanmnl} \;\; ; 
\end{eqnarray} 
and the definition of $\Theta_{m,n,\ell}^{NL}$ is
\begin{eqnarray}
\Theta_{m,n,\ell}^{NL}  & = & n \Delta \zeta - m  \Delta \theta + n \left( \frac{\partial \bar \omega_d}{\partial P_\phi} \int_0^\tau \delta P_\phi d \tau'  +  \frac{\partial \bar \omega_d}{\partial J} \int_0^\tau \delta J d \tau' \right) 
 \nonumber \\
& & + \ell \left(\frac{\partial \omega_b}{\partial P_\phi} \int_0^\tau \delta P_\phi d \tau'  +  \frac{\partial \omega_b}{\partial J} \int_0^\tau \delta J d \tau' \right)  \;\; .  \label{eq:thetanmnl}
\end{eqnarray}
Equation~(\ref{eq:kindec0nl}) can be rewritten in a more compact form, introducing the definition 
\begin{equation}
\Theta_{m,n,\ell} \equiv \Theta_{m,n,\ell}^{NL} - i \ln \lambda_{m,n}^{NL} - \int_0^{\tau} \Delta \omega (t) dt \;\; , \label{eq:thetamnl}
\end{equation} 
so that  the nonlinear frequency shift $\Delta \omega (t) = \omega(t) - \omega_0$ for a nearly monochromatic wave
(cf. Sec.~\ref{sec:periodicD}) is explicitly taken into account, leaving implicit only the time dependence of the reference linear instability. Note that the concept of nearly monochromatic wave still applies for frequency chirping modes as long as $|\dot \omega| = | \Delta \dot \omega| \ll |\gamma_L \omega_0|$; \idest, for both adiabatic ($|\dot \omega| \ll \omega_B^2$) and non-adiabatic ($|\dot \omega| \laeq \omega_B^2$) frequency sweeping, $\omega_B$ denoting the wave-particle trapping frequency. Meanwhile, we have, at the lowest order, 
\begin{equation}
{\cal P}_{m,n,\ell} \circ e^{\Delta r \partial_{r}}  f_{m,n} (\bar r) \simeq {\cal P}_{m,n,\ell} \circ f_{m,n} (\bar r + \Delta r) \;\; ; \label{eq:nlbounceav}
\end{equation}
with $\Delta r$ being the bounce-averaged nonlinear radial particle displacement, accumulated over many bounces/transits ($2\pi$ increments in $\theta_c$). For each bounce/transit, the bounce-averaging
involved in the definition of ${\cal P}_{m,n,\ell} \circ f_{m,n} (\bar r)$, Eq.~(\ref{eq:calpmnapp}), is carried out at the ``effective'' $\bar r$ location shifted by the accumulated bounce-averaged nonlinear radial particle displacement. 
Therefore, Eq.~(\ref{eq:kindec0nl}) can be finally cast as
\begin{equation}
f(r,\theta,\zeta) = \sum_{m,n,\ell \in \mathbb Z} e^{i \left( n \bar\omega_d + \ell \omega_b\right) \tau + i \Theta_{m,n,\ell} }  {\cal P}_{m,n,\ell} \circ f_{m,n} (\bar r + \Delta r)  \;\; . \label{eq:kindec1nlapp}
\end{equation}
Thus, the dominant effect of nonlinear physics on wave-particle resonances enters via the nonlinear phase shift $ \Theta_{m,n,\ell}$ (``resonance detuning'') and/or the ${\cal P}_{m,n,\ell} \circ f_{m,n} (\bar r + \Delta r)$ functions (``radial decoupling''), which regulate the strength of wave-particle resonances due to the radial mode structure. In both cases, these processes can be understood as cumulative effects of bounce-averaged responses on linear particle motions. This viewpoint is very useful when describing wave-particle interactions by Hamiltonian techniques; \exgra, by ``kinetic Poincar\'e plots'' introduced recently by White~\cite{white12}. Meanwhile, a rather detailed analysis, based on implementation of Hamiltonian phase space diagnostics, aimed at investigating the relative importance of  ``resonance detuning'' vs. ``radial decoupling'' in numerical simulations of nonlinear dynamics of SAW excited by EPs, is given in Ref. \cite{briguglio14}.

\subsection{Finite interaction length and finite interaction time}
\label{app:finiteness}

By direct inspection of Eqs.~(\ref{eq:lambdanmnl}) and~(\ref{eq:thetanmnl}), it is readily recognized that resonance detuning is drastically different for circulating particle and trapped particle resonances. In fact, from EP equations of motion and Eq.~(\ref{eq:kamcons}) one generally has $\Delta r/r \sim \Delta P_\phi/P_\phi \sim (\omega_{*EP}/\omega_0) \Delta H_0/H_0$~\cite{chen14,chen88b}, with $\omega_{*EP}$ the EP diamagnetic frequency. Thus, not only radial decoupling but also resonance detuning for EPs is expected to be dominated by the nonlinear radial displacement, since $|\omega_{*E}/\omega_0|\gg 1$
for SAW excited by EPs in fusion plasmas. 
Let us also consider high-$n$ toroidal mode number modes, as expected in ITER~\cite{zonca14a,zonca14b}, and moderate/low bounce harmonics; \idest, $\ell = {\cal O}(1)$.
Specializing, for simplicity, to the case of a toroidal equilibrium with major radius $R_0$ and shifted circular magnetic surfaces, we readily obtain
\begin{equation}
\dot \Theta_{m,n,l}  \simeq  n \frac{d \bar q}{d \bar r} \omega_t \Delta r - \Delta \omega\;\;  , \label{eq:dotthetacirc}
\end{equation}
for circulating particles, having denoted the transit frequency as $\omega_t$ and neglected $\bar \omega_d$; whereas, for magnetically trapped particle resonance,
\begin{equation}
\dot \Theta_{m,n,l} \simeq \left( n \frac{\partial \bar \omega_d}{\partial \bar r}  + \ell \frac{\partial \bar \omega_b}{\partial \bar r} \right)  \Delta r  - \Delta \omega \;\; . \label{eq:dotthetatrap}
\end{equation}
Meanwhile, for both circulating and magnetically trapped particles, the equation for $\Delta r$ is
\begin{eqnarray}
\Delta \dot r  & \simeq & i \frac{c}{B_0} \sum_{m,n,\ell \in \mathbb Z} \frac{n \bar q}{\bar r} e^{i \left( n \bar\omega_d + \ell \omega_b\right) \tau + i \Theta_{m,n,\ell} }  \nonumber \\ 
& & \times {\cal P}_{m,n,\ell} \circ \left\langle \frac{\bar r q}{r \bar q} \left[1 - \frac{v_\parallel}{q R_0\omega} (nq-m) \right] \delta \phi_g  \right\rangle_{m,n}  (\bar r + \Delta r)
\;\; , \label{eq:deltardot1}
\end{eqnarray}
having used $\delta A_{\parallel m,n} \simeq (nq-m)/(qR_0) (c/\omega) \delta \phi_{m,n}$; \idest, $\delta E_\parallel \simeq 0$ at the lowest order for SAW.  
For a resonance given by Eqs.~(\ref{eq:trapres}) or~(\ref{eq:circres}), Eqs.~(\ref{eq:dotthetacirc}) to~(\ref{eq:deltardot1}) may describe an isolated resonance as well as resonance overlap in the case of many modes. It is also worthwhile emphasizing that, for the same toroidal mode number $n$, different poloidal mode numbers $m$ are generally coherent, since they belong to the same (generally nonlinear) mode structure. From Eqs.~(\ref{eq:dotthetacirc}) to~(\ref{eq:deltardot1}) with $\Delta \omega =0$, the wave-particle trapping frequency $\omega_B$ is readily estimated as
\begin{equation}
\omega_B^2 \simeq  \lambda_n \left| \frac{\omega}{r} \delta \dot{\!\bar{\bm X}}_\perp \right| \simeq \lambda_n \left| \frac{\omega}{r} \frac{nq}{r} \frac{c}{B_0} \delta \phi \right| \;\; , \label{eq:omegaB2}
\end{equation}
with $\lambda_n = |n r q'|$ for circulating, $\lambda_n = 1$ for magnetically trapped particles; and $\delta \dot{\!\bar{\bm X}}_\perp$ the fluctuating $\bm E \times \bm B$ guiding-center drift velocity~\cite{brizard07,cary09}. In the derivation of Eq.~(\ref{eq:omegaB2}), the estimate $d_{tt} \Delta r \sim \omega_B^2 \Delta r$ has been adopted, and resonant particles have been assumed; \idest,
Eq.~(\ref{eq:trapres}) for magnetically trapped and Eq.~(\ref{eq:circres}) for circulating particles, respectively. Given $\omega_B$, the separatrix width in particle phase space, $\Delta r_{sx}$, is estimated as $\Delta r_{sx} \simeq |\delta \dot{\!\bar{\bm X}}_\perp/\omega_B|$. Thus, $\Delta r_{sx}$ and $\omega_B$ are of order $\sim \epsilon_\delta^{1/2}$.

Equations~(\ref{eq:dotthetacirc}) and~(\ref{eq:dotthetatrap}) can be used for estimating the finite interaction length and finite interaction time for resonant particles; \idest, the typical spatial and time scales required for particles to effectively loose the resonance condition in a non-uniform system. 
For $\Delta \omega = 0$;  
\idest, no frequency chirping, 
the finite interaction time for resonant particles defines the characteristic nonlinear time scale and is given by 
\begin{equation}
\dot \Theta_{m,n,\ell}  \tau_{NL} \sim 1 \;\; ; \;\;\;\;\; {\rm and} \;\;\;\;\;  \tau_{NL} \sim 1/(3 \gamma_L) \;\; . \label{eq:taunl}
\end{equation}
Note that the $\gamma_L^{-1}$ enters via the anticipated saturation process (either resonance detuning or radial decoupling; or both, with varying relative weight); \idest, saturation occurs when
$\tau_{NL} \sim 1/(3 \gamma_L)$. 
Here, the factor $3$ is a typical value adopted from analyses of the beam-plasma system \cite{levin72a,levin72b}, as well as numerical simulations of Alfv\'en Eigenmodes (AE) in toroidal plasmas \cite{wu95}.  Meanwhile, neglecting the radial mode structure, the finite interaction length, $\Delta r_L$, is obtained from Eq.~(\ref{eq:dotthetacirc}) as
\begin{equation}
3\gamma_L \sim n q' \omega \Delta r_L \;\; \label{eq:deltarlcirc0}
\end{equation}
for circulating particles; while, from Eq.~(\ref{eq:dotthetatrap}),
\begin{equation}
3\gamma_L \sim \omega (\Delta r_L/r) \label{eq:deltarltrap0}
\end{equation}
for magnetically trapped particles; where characteristic radial scale lengths of $\omega_t$, $\bar \omega_d$ and $\omega_b$ are taken to be $\sim r$. 
From these estimates, we readily recognize that the finite interaction length, $\Delta r_L$, for circulating resonant particles is typically smaller than the characteristic width, $\sim |nq'|^{-1}$ \cite{zonca14a,zonca14b,lu12}, of one poloidal harmonic, $\delta \phi_{m,n}$; whereas it may be comparable or larger than $\sim |nq'|^{-1}$ for magnetically trapped resonant particles if $(\gamma_L/\omega) \gaeq |3 nrq'|^{-1}$. For high-$n$ modes as expected in ITER \cite{zonca14a,zonca14b}, this condition is $(\gamma_L/\omega) \gaeq 10^{-2}$.
Thus, circulating particles transport due to isolated resonances is predicted to be local, causing minor radial profile relaxation, while transport is expected to be mostly diffusive in the presence of many modes \cite{chen14}. Meanwhile, magnetically trapped EP transports is intrinsically non-local; \idest, characterized by meso-scales, larger than $|nq'|^{-1}$ and shorter than the equilibrium scale length, and convective (ballistic) processes \cite{chen14}. 
Different trapped particle vs. circulating particle behaviors are very general, as they are connected with the properties of Eqs.~(\ref{eq:dotthetacirc}) and~(\ref{eq:dotthetatrap}), and are expected to play important role whenever resonant particle transport is significant. In fact, both convective and diffusive electron motion have been reported in gyrokinetic numerical simulations of collisionless trapped electron mode turbulence~\cite{xiao11}. Ballistic and (super/sub) diffusive EP behaviors have also been demonstrated in the presence of interchange turbulence in simple magnetized toroidal plasmas \cite{gustafson12a,gustafson12b}; and, more recently, diffusive circulating particle transport vs. ballistic magnetically trapped particle behaviors induced by electrostatic drift wave turbulence have been discussed in Ref. \cite{feng13}.

Somewhat different behaviors are expected for low-$n$ modes, routinely excited in present day experiments. Here, convective (ballistic) supra-thermal particle transport is observed typically 
in connection with significant non-adiabatic [$|\dot \omega| \laeq \omega_B^2$; cf. discussions following Eq.~(\ref{eq:thetamnl})] frequency sweeping as, \exgra, in NSTX experiments with neutral beam injection~\cite{fredrickson09,podesta11,podesta12}.
Secular radial motions of resonant particles were recognized by White \etal \cite{white83} to be a key mechanism, dubbed ``mode particle pumping'', for explaining fishbone 
nonlinear dynamics~\cite{chen84} and related particle losses. This phenomenology is readily explained by Eqs.~(\ref{eq:dotthetacirc}) to~(\ref{eq:deltardot1}), which have the same structure 
of the original equations derived in Ref.~\cite{white83}; and predict secular (ballistic) particle motions if the $\dot \Theta_{m,n,\ell} = 0$ resonance condition is preserved as the particles are 
transported out of the system. In fact, Eqs.~(\ref{eq:deltarlcirc0}) and~(\ref{eq:deltarltrap0}) give ``mode particle pumping'' by ``phase locking'' for circulating and magnetically trapped particles when, respectively~\cite{white83,zonca00c},
\begin{equation}
\Delta \dot \omega \simeq n \frac{d \bar q}{d \bar r} \omega_t \Delta \dot r \;\; , \label{eq:phaselockcirc} 
\end{equation}
and
\begin{equation}
\Delta \dot \omega \simeq \left( n \frac{\partial \bar \omega_d}{\partial \bar r}  + \ell \frac{\partial \bar \omega_b}{\partial \bar r} \right)  \Delta \dot r \;\;  .  \label{eq:phaselocktrap}
\end{equation} 
It is worthwhile noting that, because of Eqs.~(\ref{eq:phaselockcirc}) and~(\ref{eq:phaselocktrap}) together with Eqs.~(\ref{eq:deltardot1}) and~(\ref{eq:omegaB2}), 
``phase locking'' 
and secular (ballistic) particle transport always imply non-adiabatic frequency sweeping, $\Delta \dot \omega \sim \omega_B^2$. 
Another important feature of the phase locking condition and $\Delta \dot r$ given by Eq.~(\ref{eq:deltardot1}) is that the frequency sweeping rate is proportional to the mode amplitude; \idest, it is fastest when the mode is strongest. This is observed in typical experimental conditions, when resonant particle transport is ballistic (see, \exgra, Refs. \cite{heidbrink08,podesta11}), and in numerical simulations of, \exgra, nonlinear EPM evolutions~\cite{briguglio98,briguglio02,zonca02,vlad04,briguglio07,vlad99} and, more recently, nonlinear fishbone dynamics~\cite{vlad13,fu06b,vlad12}. 

The effect of phase locking is to extend the finite interaction lengths, given by 
Eqs.~(\ref{eq:phaselockcirc}) and~(\ref{eq:phaselocktrap}) in the $\Delta \omega = 0$ no-frequency sweeping limit, to, respectively
\begin{equation}
3\gamma_L  \sim  n q' \omega \epsilon_{\dot\omega} \Delta r_L \;\; , \label{eq:deltarlcirc1} 
\end{equation}
and
\begin{equation}
3\gamma_L \sim \omega \epsilon_{\dot\omega} (\Delta r_L/r) \;\; . \label{eq:deltarltrap1}
\end{equation}
Here, $\epsilon_{\dot\omega}$ is defined as
\begin{equation}
\epsilon_{\dot\omega} \equiv \left| \frac{\dot \Theta_{m,n,\ell}(\Delta \omega)}{\dot \Theta_{m,n,\ell}(\Delta \omega=0)}\right|\;\; , \label{eq:epsilonomegadot}
\end{equation}
where numerator and denominator are computed from Eq.~(\ref{eq:thetamnl}), respectively, at the actual value of the nonlinear frequency shift (frequency chirping) and at $\Delta \omega=0$ 
(no chirping). The value of $\epsilon_{\dot\omega}$ denotes the effectiveness of phase locking; it depends 
on the wave dispersive properties \cite{chen14} and, in general, $\epsilon_{\dot\omega}<1$ requires the EP dynamics being non-perturbative. 
A similar argument applies to the finite interaction time, which is also extended by $\sim \epsilon_{\dot\omega}^{-1}$. Because of the extended interaction length, circulating resonant particles can also  be displaced by a significant fraction of $\sim |nq'|^{-1}$ before resonance detuning sets in. Thus, in the presence of  significant non-adiabatic frequency sweeping, circulating EP transports are also expected to occur as convective processes even for long-wavelength modes. This has, indeed, been observed experimentally, \exgra, in recent NSTX experiments with neutral beam injection~\cite{fredrickson09,podesta11}; and in numerical simulations~\cite{briguglio98,briguglio02,briguglio07,vlad99}, where it is shown that the structure of the SAW continuous spectrum is crucial in the nonlinear mode dynamics and frequency chirping~\cite{briguglio98,briguglio02,vlad04,briguglio07,vlad99,vlad06,vlad09,bierwage11,bierwage12}. 

The secular resonant particle motion, predicted by Eqs.~(\ref{eq:phaselockcirc}) and~(\ref{eq:phaselocktrap}) is restricted to a group of particles that have similar initial wave-particle phase 
and can be maintained as long as the mode frequency can be swept preserving phase locking on the one hand and, on the other hand, satisfying the mode dispersion relation. 
When phase locked resonant particles eventually loose the resonance condition after the finite interaction time/length,  
they may be 
substituted by others; and the process can continue until quenched by equilibrium nonuniformity (cf. Secs.~\ref{sec:lock} and~\ref{sec:epmaval}; and Refs. \cite{zonca14e,zonca13,chen14}). In fact, non-adiabatic frequency sweeping prevents the {\em de facto} wave-particle trapping to occur and particle can enter and leave the resonant region during the nonlinear evolution,
which maximizes wave-particle power exchange as well as mode growth \cite{chen14} (cf. Sec.~\ref{sec:epmaval}).

\section{The general fishbone like dispersion relation}
\label{app:GFLDR}

Here,  expressions for $\Lambda_n$, $\delta \bar W_{fn}$ and $\delta \bar W_{kn}$ to be used in Eq.~(\ref{eq:fishlikeball0}) are given, without derivation, for the readers' convenience. Detailed derivations of these
expressions can be found in Refs. \cite{chen14,zonca14a,zonca14b}.
The generalized ``inertia'', $\Lambda_n$, accounting for the radially local structures of the SAW continuous spectrum,§ is given by
\begin{equation}
i \Lambda_n \equiv  \frac{1}{2} \left( \delta \hat \Psi_{-n0^+}^\dag \delta \hat \Psi_{n0^+} \right)^{-1}
\left[ \delta \hat \Psi_{-n}^\dag (\vartheta) \partial_\vartheta \delta \hat \Psi_n (\vartheta) \right]_{\vartheta \rightarrow 0^-}^{\vartheta \rightarrow 0^+}
\;\; , \label{eq:biglambdadef}
\end{equation}
where $\delta \hat \Psi_{n0^+} = \lim_{\vartheta \rightarrow 0^+} 
\delta \hat \Psi_n(\vartheta)$, and $\delta \hat \Psi_n(\vartheta)$ is obtained
from the solution of the short radial scale limit ($k_\perp^2 \simeq k_r^2$) of the gyrokinetic vorticity equation
written as \cite{chen14,zonca14a,zonca14b}
\begin{eqnarray}
& & \left( \frac{\partial^2}{\partial \vartheta^2} - \frac{\partial_\vartheta^2 \hat \kappa_\perp}{\hat \kappa_\perp} \right) \delta \hat \Psi_n - 
\frac{{\cal J}^2 B_0^2}{v_A^2}  \frac{\partial}{\partial t} \left[ \frac{\partial}{\partial t} + i \omega_{*pi} \right. \nonumber \\
& & \hspace*{2em} \left. - \frac{3}{4} k_\vartheta^2 \rho_i^2 \hat \kappa_\perp^2 \left( \frac{\partial}{\partial t} + i \omega_{*pi} + i \omega_{*Ti} \right)  \right] \delta \hat \Phi_n \nonumber \\
& & \hspace*{2em} - \lim_{\hat \kappa_\perp \rightarrow \infty} \frac{4\pi {\cal J}^2 B_0}{c k_\vartheta^2 \hat \kappa_\perp} \bm b \times \bm \kappa \cdot \bm \nabla 
\sum \left\langle m \left( \mu B_0 + v_\parallel^2 \right) J_0 \frac{\partial}{\partial t} \delta \hat g_n \right\rangle_v  
\nonumber \\ & & \hspace*{2em} + \left[ {\rm NL} \, {\rm TERMS} \right] = 0  \label{eq:vortixball0}
\;\; .
\end{eqnarray}
Here, we have adopted Eq.~(\ref{eq:msd}) for representing spatiotemporal structures of SAW/DAW, $\bm{k}_\perp = - i \bm \nabla_\perp$, $\kappa_\perp^2 \equiv \hat \kappa_\perp^2 k_\vartheta^2$, $k_\vartheta \equiv - n q/r$, $\delta \hat \Psi_n = \hat \kappa_\perp \delta \hat \psi_n$, $\delta \hat \Phi_n = \hat \kappa_\perp \delta \hat \phi_n$, $\rho_i = (T_i/m_i)^{1/2}/\Omega_i$ is the thermal ion Larmor radius, ${\cal J}$ is the Jacobian [cf. \ref{app:res2D}, following  Eq.~(\ref{eq:B0})], summation is extended to all particle species, and we have introduced the thermal ion diamagnetic frequencies 
\begin{eqnarray}
\omega_{*pi} & = & \omega_{*ni} + \omega_{*Ti} \;\; , \nonumber \\ 
\omega_{*ni} & = & \left(\frac{T_0c}{e n_0 B_0} \right)_i (\bm{b}  \times  \bm \nabla n_{0i} ) \cdot \bm{k}_\perp \;\; ,  \nonumber \\
\omega_{*Ti} & = &  \left(\frac{c}{e B_0} \right)_i ( \bm{b} \times \bm \nabla T_{0i} ) \cdot \bm{k}_\perp \;\; . \label{eq:diamfreq}
\end{eqnarray}
Furthermore, for the sake of simplicity, we have assumed the long wavelength limit $k_\perp^2 \rho_i^2 \ll 1$, except in the magnetic curvature coupling term, where $J_0$ is maintained for brevity. Interested readers can find a discussion of the more general $k_\perp \rho_i \sim 1$ case in 
Refs. \cite{chen14,zonca14a,zonca14b}. Note that the $\sim \delta \hat g_n$ term generally includes nonlinear particle response via $\delta \hat g_n$ itself. Finally, formally nonlinear terms, denoted as $\left[ {\rm NL} \, {\rm TERMS} \right]$, are given by
\begin{eqnarray}
\left[ {\rm NL} \, {\rm TERMS} \right] & = & \lim_{\hat \kappa_\perp \rightarrow \infty} \frac{{\cal J}^2 B_0^2}{k_\vartheta^2 \hat \kappa_\perp v_A^2}  \frac{\partial^2}{\partial t^2} {\cal P}_{Bn} \left( \frac{\delta \varrho_m}{\varrho_{m0}} \nabla_\perp^2 \delta \phi \right) \;\; , \label{eq:nlterms} \\ 
& &  - \lim_{\hat \kappa_\perp \rightarrow \infty} \frac{{\cal J}^2 B_0}{c k_\vartheta^2 \hat \kappa_\perp} {\cal P}_{Bn} \left( \bm b \times \bm \nabla \delta A_\parallel \cdot \bm \nabla \nabla_\perp^2 \delta A_\parallel \right) \nonumber \\
& & - \lim_{\hat \kappa_\perp \rightarrow \infty} \sum_{\neq e} \frac{2\pi e {\cal J}^2 B_0^2}{c k_\vartheta^2  \hat \kappa_\perp \Omega^2} {\cal P}_{Bn}  \left[  \bm b \times \bm \nabla \left( \nabla_\perp^2 \delta \phi \right)  \cdot \bm \nabla \left\langle \mu \delta g \right\rangle_v 
\right. \nonumber \\ & & \left.
- \bm b \times \bm \nabla  \delta \phi  \cdot \bm \nabla \left\langle \mu  \nabla_\perp^2\delta g \right\rangle_v
-  \nabla_\perp^2\left( \bm b \times \bm \nabla  \delta \phi  \cdot \bm \nabla \left\langle \mu \delta g \right\rangle_v \right) \right] \;\; . \nonumber
\end{eqnarray}
Here, $\delta \varrho_m$ stands for plasma mass density fluctuation about the equilibrium value $\varrho_{m0}$, summation is extended to all particle species except thermal electrons and, in order to simplify notation, we have introduced the operators ${\cal P}_{Bn}$~\cite{zonca14a,zonca14b}, 
\begin{equation}
{\cal P}_{Bn} (r,\vartheta) \, : \, f(r,\theta,\zeta) \mapsto \hat f_n (r,\vartheta) \;\; ,   \label{eq:ballproj}
\end{equation}
which act on a generic function $f(r,\theta,\zeta)$ in $(r,\theta,\zeta)$ space, producing the corresponding $\hat f_n (r,\vartheta)$ component in $(r,\vartheta)$ space,
consistent with Eq.~(\ref{eq:msd}).
In Eq.~(\ref{eq:nlterms}), the first line represents the polarization current nonlinearity \cite{sagdeev69}, the second one accounts for the Maxwell stress, while the third line
is the kinetic expression representing both nonlinear diamagnetic response and Reynolds stress for $k_\perp^2 \rho_i^2 \ll 1$~\cite{chen14,zonca14a}.

The expressions of $\delta \bar W_{fn}$ and $\delta \bar W_{kn}$ are obtained by construction of a quadratic form starting from the gyrokinetic vorticity equation~\cite{chen14,zonca14a}, 
and are given by
\begin{eqnarray}
\hspace*{-1cm}\delta \bar W_{nf} + \delta \bar W_{nk} & = &  \left( \delta \hat \Phi_{-n0^+}^\dag \delta \hat \Phi_{n0^+}  \right)^{-1} \frac{1}{2} \int_{-\infty}^\infty \left[ \left( \frac{\partial}{\partial \vartheta} \delta \hat \Phi_{-n} \right)^\dag \left( \frac{\partial}{\partial \vartheta} \delta \hat \Phi_{n} \right) \right. \label{eq:dwbar} \\ & & + \frac{\partial_\vartheta^2 \hat \kappa_\perp}{\hat \kappa_\perp} \delta \hat \Phi_{-n}^\dag \delta \hat \Phi_{n}  + \delta \hat \Phi_{-n}^\dag
\frac{{\cal J}^2 B_0^2}{v_A^2}  \frac{\partial}{\partial t} \left( \frac{\partial}{\partial t} + i \omega_{*pi}  \right) \delta \hat \Phi_{n}  \nonumber \\
& & \left.  + \delta \hat \Phi_{-n}^\dag \frac{4\pi {\cal J}^2 B_0}{c k_\vartheta^2 \hat \kappa_\perp} \bm b \times \bm \kappa \cdot \bm \nabla 
\sum \left\langle m \left( \mu B_0 + v_\parallel^2 \right) J_0 \frac{\partial}{\partial t} \delta \hat g_n \right\rangle_v  \right] d \vartheta  \;\; . \nonumber
\end{eqnarray}
Here, we noted that $\delta \hat \Phi_n(\vartheta) \simeq\delta \hat \Psi_n(\vartheta)$ for regular SAW/DAW radial structures away from resonances with the SAW continuous spectrum. 
Furthermore,
the nonlinear contribution to $\delta \bar W_{nf} + \delta \bar W_{nk}$ is 
mainly due to the $\sim \delta \hat g_n$ term, while formally nonlinear terms in the gyrokinetic vorticity equation [cf. Eq.~(\ref{eq:nlterms})] predominantly contribute to $\Lambda_n$~\cite{chen14,zonca14a}.
The nonlinearity due to EP response in the $\sim \delta \hat g_n$ term is what is considered in this work for the analyses of EPM avalanches, carried out in Sec.~\ref{sec:epmaval} adopting the ``fishbone paradigm'' (cf. Sec.~\ref{sec:dyson}).
The expressions of $\delta \bar W_{fn}$ and $\delta \bar W_{kn}$ in Eq.~(\ref{eq:dwbar}) can be extracted, dividing $\delta \hat g_n$ in ``fluid''  and ``kinetic'' responses as specified in Eq.~(\ref{eq:dgdKrel}) \cite{chen84}. The expression
for $\delta \bar W_{kn}$ is then obtained from Eq.~(\ref{eq:dwbar}) using the ``kinetic'' response only in the $\sim \delta \hat g_n$ term. Note that the remaining $\delta \bar W_{fn}$ does not include the (linear) ``kink drive'' contribution, which is typically negligible for SAW/DAW and can be anyway readily included for application of the present theoretical framework to MHD modes \cite{chen14,zonca14a,zonca14b}.

\section{Laplace transform of nearly periodic fluctuations}
\label{app:Laplace}

In this Appendix, we justify the Laplace representation of Eq.~(\ref{eq:monochr1}) for nearly periodic fluctuations, and discuss
its validity in terms of relevant time scales. 

Introduce the slow and fast time variables $\tau$ and $t$, with $\tau \simeq t$ but $|\partial_t| \gg |\partial_\tau|$ and $d_t = \partial_t + \partial_\tau$. Furthermore,
given a nearly periodic fluctuation $\delta \bar\phi_{n} (t)$ (spatial dependences are left implicit for the sake of notation simplicity), defined in Eq.~(\ref{eq:barphin}) and with oscillation frequency centered about
$\omega_0(\tau) = \omega_{0r}(\tau) + i \gamma_0(\tau)$ (slowly varying in time), we let
\begin{eqnarray}
\delta \bar \phi_{n} (t) & \equiv & \lim_{\tau \rightarrow t } \delta \varphi_{n} (t,\tau) \equiv \lim_{\tau \rightarrow t } \delta \bar \varphi_{n} (\tau) \exp \left( - i \omega_{0}(\tau) t \right) \nonumber \\
\delta \bar \varphi_{n} (\tau) & \equiv  &\delta \bar \phi_{n0} \exp \left[ -i \int_0^\tau \omega_{0}(t') dt' + i \omega_{0}(\tau) \tau \right] \;\; . \label{eq:periodicphi}
\end{eqnarray}
Noting that
$$(\partial_t + \partial_\tau)  \delta \varphi_{n} (t,\tau) = \left[ - i \omega_{0} (\tau) + i \dot \omega_{0}(\tau) ( \tau - t ) \right]  \delta \varphi_{n} (t,\tau) \;\; , $$
Eq.~(\ref{eq:periodicphi}) admits the Laplace transforms 
\begin{equation}
\delta \hat {\bar \phi}_n (\omega) = \frac{i}{2\pi} \frac{\delta {\bar \varphi}_{n} (r,\tau) }{\omega - \omega_0 (\tau)} \;\; ,  \;\;\;\;\; {\rm and} \;\;\;\;\;
\delta \hat {\bar \phi}_{-n}(\omega) = \frac{i}{2\pi} \frac{\delta {\bar \varphi}_{-n} (r,\tau) }{\omega + \omega^*_0 (\tau)} \;\; , \label{eq:monochr1app}
\end{equation}
provided that $|t - \tau| \laeq \tau_{NL} \sim |\gamma_{0}|^{-1}$; and $|\dot\omega_{0r}|\ll |\gamma_0\omega_{0r}|$, $|\dot\gamma_0|\ll |\gamma_0^2|$. Thus,
considering the time scale ordering of Eq.~(\ref{eq:timescaleorde}), Eq.~(\ref{eq:monochr1app}) applies in general to adiabatic as well as non-adiabatic frequency
chirping modes; and is adopted in this work as Eq.~(\ref{eq:monochr1}).

\section*{References}

\begin{thebibliography}{100}
\expandafter\ifx\csname url\endcsname\relax
  \def\url#1{{\tt #1}}\fi
\expandafter\ifx\csname urlprefix\endcsname\relax\def\urlprefix{URL }\fi
\providecommand{\eprint}[2][]{\url{#2}}

\bibitem{chen07a}
Chen L and Zonca F 2007 {\em Nucl. Fusion\/} {\bf 47} S727

\bibitem{fasoli07}
Fasoli A, Gormezano C, Berk H~L, Breizman B~N, Briguglio S, Darrow D~S,
  Gorelenkov N~N, Heidbrink W~W, Jaun A, Konovalov S~V, Nazikian R, Noterdaeme
  J, Sharapov S~E, Shinohara K, Testa D, Tobita K, Todo Y, Vlad G and Zonca F
  2007 {\em Nucl. Fusion\/} {\bf 47} S264

\bibitem{heidbrink08}
Heidbrink W~W 2008 {\em Phys. Plasmas\/} {\bf 15} 055501

\bibitem{belikov68}
Belikov V~S, Kolesnichenko {\relax Ya I} and Oraevskij V~N 1968 {\em Zh. Eksp.
  Teor. Fiz.\/} {\bf 5} 2210

\bibitem{belikov69}
Belikov V~S, Kolesnichenko {\relax Ya I} and Oraevskij V~N 1969 {\em Sov. Phys.
  JETP\/} {\bf 28} 1172

\bibitem{rosenbluth75}
Rosenbluth M~N and Rutherford P~H 1975 {\em Phys. Rev. Lett.\/} {\bf 34} 1428

\bibitem{mikhailovskii75a}
Mikhailovskii A~B 1975 {\em Sov. Phys. JETP\/} {\bf 41} 890

\bibitem{mikhailovskii75b}
Mikhailovskii A~B 1975 {\em Zh. Eksp. Teor. Fiz.\/} {\bf 68} 1772

\bibitem{zonca99}
Zonca F, Chen L, Dong J~Q and Santoro R~A 1999 {\em Phys. Plasmas\/} {\bf 6}
  1917

\bibitem{nazikian06}
Nazikian R, Berk H~L, Budny R~V, Burrell K~H, Doyle E~J, Fonck R~J, Gorelenkov
  N~N, Holcomb C, Kramer G~J, Jayakumar R~J, {La Haye} R~J, McKee G~R, Makowski
  M~A, Peebles W~A, Rhodes T~L, Solomon W~M, Strait E~J, {Van Zeeland} M~A and
  Zeng L 2006 {\em Phys. Rev. Lett.\/} {\bf 96} 105006

\bibitem{hasegawa76a}
Hasegawa H and Chen L 1976 {\em Phys. Fluids\/} {\bf 19} 1924

\bibitem{zonca08b}
Zonca F and Chen L 2008 {Nonlinear Dynamics and Complex Behaviors in Magnetized
  Plasmas of Fusion Interest} {\em Frontiers in Modern Plasma Physics\/} vol
  CP1061 ed Shukla P~K, Eliasson B and Stenflo L (AIP) p~34

\bibitem{zonca08d}
Zonca F 2008 {\em Int. J. Mod. Phys. A\/} {\bf 23} 1165

\bibitem{qiu12}
Qiu Z, Zonca F and Chen L 2012 {\em Phys. Plasmas\/} {\bf 19} 082507

\bibitem{zonca14e}
Zonca F, Chen L, Briguglio S, Fogaccia G, Milovanov A~V, Qiu Z, Vlad G and Wang
  X 2014 {Energetic particles and multi-scale dynamics in fusion plasmas} 
  {\em Plasma Phys. Control. Fusion\/}  to be published

\bibitem{white83}
White R~B, Goldston R~J, McGuire K, Boozer A~H, Monticello D~A and Park W 1983
  {\em Phys. Fluids\/} {\bf 26} 2958

\bibitem{chen84}
Chen L, White R~B and Rosenbluth M~N 1984 {\em Phys. Rev. Lett.\/} {\bf 52}
  1122

\bibitem{chen99}
Chen L 1999 {\em J. Geophys. Res.\/} {\bf 104} 2421

\bibitem{white10a}
White R~B, Gorelenkov N, Heidbrink W~W and {\relax Van Zeeland} M~A 2010 {\em
  Phys. Plasmas\/} {\bf 17} 056107

\bibitem{white10b}
White R~B, Gorelenkov N, Heidbrink W~W and {\relax Van Zeeland} M~A 2010 {\em
  Plasma Phys. Control. Fusion\/} {\bf 52} 045012

\bibitem{vanhove55}
{\relax Van Hove} L 1955 {\em Physica\/} {\bf 21} 517

\bibitem{prigogine62}
Prigogine I 1962 {\em {Nonequilibrium Statistical Mechanics}\/} (New York:
  Interscience)

\bibitem{balescu63}
Balescu R 1963 {\em {Statistical Mechanics of Charged Particles}\/} (New York:
  Interscience)

\bibitem{altshul65}
{\relax Al'tshul'} L~M and Karpman V~I 1965 {\em Zh. Eksp. Teor. Fiz.\/} {\bf
  49} 515

\bibitem{altshul66}
{\relax Al'tshul'} L~M and Karpman V~I 1966 {\em Sov. Phys. JETP\/} {\bf 22}
  361

\bibitem{zonca13}
Zonca F, Briguglio S, Chen L, Fogaccia G, Vlad G and Wang X 2013 {Nonlinear
  dynamics of phase-space zonal structures and energetic particle physics} {\em
  Proceedings of the 6th IAEA Technical Meeting on ``Theory of Plasmas
  Instabilities''\/} (Vienna, Austria, May 27 - 29: IAEA, Vienna)

\bibitem{chen14}
Chen L and Zonca F 2014 {Physics of Alfv\'en waves and energetic particles in 
  burning plasmas} {\em Rev. Mod. Phys.\/}  submitted

\bibitem{hasegawa79}
Hasegawa A, {\relax Maclennan} C~G and Kodama Y 1979 {\em Phys. Fluids\/} {\bf
  22} 2122

\bibitem{lin98}
Lin Z, Hahm T~S, Lee W~W, Tang W~M and White R~B 1998 {\em Science\/} {\bf 281}
  1835

\bibitem{diamond05}
Diamond P~H, Itoh {\relax S-I}, Itoh K and Hahm T~S 2005 {\em Plasma Phys.
  Control. Fusion\/} {\bf 47} R35

\bibitem{itoh06}
Itoh K, Itoh {\relax S-I}, Diamond P~H, Hahm T~S, Fujisawa A, Tynan G~R, Yagi M
  and Nagashima Y 2006 {\em Phys. Plasmas\/} {\bf 13} 055502

\bibitem{taylor71}
Taylor J~B and {\relax McNamara} B 1971 {\em Phys. Fluids\/} {\bf 14} 1492

\bibitem{okuda73}
Okuda H and Dawson J~M 1973 {\em Phys. Fluids\/} {\bf 16} 408

\bibitem{shukla84}
Shukla P~K, Yu M~Y, Rahman H~U and Spatschek K~H 1984 {\em Phys. Rep.\/} {\bf
  105} 227

\bibitem{zonca05}
Zonca F, Briguglio S, Chen L, Fogaccia G and Vlad G 2005 {\em Nucl. Fusion\/}
  {\bf 45} 477

\bibitem{zonca06b}
Zonca F, Briguglio S, Chen L, Fogaccia G, Hahm T~S, Milovanov A~V and Vlad G
  2006 {\em Plasma Phys. Control. Fusion\/} {\bf 48} B15

\bibitem{chen07b}
Chen L and Zonca F 2007 {\em Nucl. Fusion\/} {\bf 47} 886

\bibitem{wang07b}
Wang W~X, Hahm T~S, Lee W~W, Rewoldt G, Manickam J and Tang W~M 2007 {\em Phys.
  Plasmas\/} {\bf 14} 072306

\bibitem{difpradalier10}
{\relax Dif-Pradalier} G, Diamond P~H, Grandgirard V, Sarazin Y, Abiteboul J,
  Garbet X, Ghendrih {\relax Ph}, Strugarek A, Ku S and Chang C~S 2010 {\em
  Phys. Rev. E\/} {\bf 82} 025401(R)

\bibitem{jolliet12}
Jolliet S and Idomura Y 2012 {\em Nucl. Fusion\/} {\bf 52} 023026

\bibitem{gurcan13}
{\relax G\"urcan} {\relax \"O D}, Vermare L, Hennequin P, Berionni V, Diamond
  P~H, {\relax Dif-Pradalier} G, Garbet X, Ghendrih P, Grandgirard V, McDevitt
  C~J, Morel P, Sarazin Y, Storelli A, Bourdelle C and {\relax the Tore Supra
  Team} 2013 {\em Nucl. Fusion\/} {\bf 53} 073029

\bibitem{bernstein57}
Bernstein I~B, Greene J~M and Kruskal M~D 1957 {\em Phys. Rev.\/} {\bf 108} 546

\bibitem{berk70}
Berk H~L, Nielson C~W and Roberts K~W 1970 {\em Phys. Fluids\/} {\bf 13} 980

\bibitem{dupree70}
Dupree T~H 1970 {\em Phys. Rev. Lett.\/} {\bf 25} 789

\bibitem{dupree72}
Dupree T~H 1972 {\em Phys. Fluids\/} {\bf 15} 334

\bibitem{dupree82}
Dupree T~H 1982 {\em Phys. Fluids\/} {\bf 25} 277

\bibitem{berman83}
Berman R~H, Tetreault D~J and Dupree T~H 1983 {\em Phys. Fluids\/} {\bf 26}
  2437

\bibitem{tetreault83}
Tetreault D~J 1983 {\em Phys. Fluids\/} {\bf 26} 3247

\bibitem{wang12}
Wang X, Briguglio S, Chen L, {\relax Di Troia} C, Fogaccia G, Vlad G and Zonca
  F 2012 {\em Phys. Rev. E\/} {\bf 86} 045401(R)

\bibitem{vlad13}
Vlad G, Briguglio S, Fogaccia G, Zonca F, Fusco V and Wang X 2013 {\em Nucl.
  Fusion\/} {\bf 53} 083008

\bibitem{chen14b}
Chen L and Zonca F 2014 {\em JPS Conf. Proc.\/} {\bf 1} 011001

\bibitem{zonca14d}
Zonca F and Chen L 2014 {\em AIP Conf. Proc.\/} {\bf 1580} 5

\bibitem{briguglio14}
Briguglio S, Wang X, Zonca F, Vlad G, Fogaccia G, {\relax Di Troia} C and Fusco
  V 2014 {\em Phys. Plasmas\/}  {\bf 21} 112301

\bibitem{meerson90}
Meerson B and Friedland L 1990 {\em Phys. Rev. A\/} {\bf 41} 5233

\bibitem{fajans01}
Fajans J and Friedland L 2001 {\em Am. J. Phys.\/} {\bf 69} 1096

\bibitem{chen94}
Chen L 1994 {\em Phys. Plasmas\/} {\bf 1} 1519

\bibitem{briguglio95}
Briguglio S, Vlad G, Zonca F and Kar C 1995 {\em Phys. Plasmas\/} {\bf 2} 3711

\bibitem{wang11}
Wang X, Briguglio S, Chen L, {Di Troia} C, Fogaccia G, Vlad G and Zonca F 2011
  {\em Phys. Plasmas\/} {\bf 18} 052504

\bibitem{frieman82}
Frieman E~A and Chen L 1982 {\em Phys. Fluids\/} {\bf 25} 502

\bibitem{brizard07}
Brizard A~J and Hahm T~S 2007 {\em Rev. Mod. Phys.\/} {\bf 79} 421

\bibitem{dicke54}
Dicke R~H 1954 {\em Phys. Rev.\/} {\bf 93} 99

\bibitem{bonifacio90}
Bonifacio R, {\relax De Salvo} L, Pierini P and Piovella N 1990 {\em Nucl.
  Instrum. Methods Phys. Res., Sect. A\/} {\bf 296} 358

\bibitem{zonca14a}
Zonca F and Chen L 2014 {\em Phys. Plasmas\/} {\bf 21} 072120

\bibitem{zonca14b}
Zonca F and Chen L 2014 {\em Phys. Plasmas\/} {\bf 21} 072121

\bibitem{chen00}
Chen L, Lin Z and White R~B 2000 {\em Phys. Plasmas\/} {\bf 7} 3129

\bibitem{chen01}
Chen L, Lin Z, White R~B and Zonca F 2001 {\em Nucl. Fusion\/} {\bf 41} 747

\bibitem{chen04}
Chen L, White R~B and Zonca F 2004 {\em Phys. Rev. Lett.\/} {\bf 92} 075004

\bibitem{zonca04b}
Zonca F, White R~B and Chen L 2004 {\em Phys. Plasmas\/} {\bf 11} 2488

\bibitem{chirikov79}
Chirikov B~V 1979 {\em Phys. Rep.\/} {\bf 52} 263

\bibitem{lichtenberg83}
Lichtenberg A~J and Lieberman M~A 1983 {\em Regular and Stochastic Motion\/}
  (Springer - Verlag)

\bibitem{berk90c}
Berk H~L and Breizman B~N 1990 {\em Phys. Fluids B\/} {\bf 2} 2246

\bibitem{breizman11b}
Breizman B~N and Sharapov S~E 2011 {\em Plasma Phys. Control. Fusion\/} {\bf
  53} 054001

\bibitem{briguglio98}
Briguglio S, Zonca F and Vlad G 1998 {\em Phys. Plasmas\/} {\bf 5} 3287

\bibitem{zhang12}
Zhang H~S, Lin Z and Holod I 2012 {\em Phys. Rev. Lett.\/} {\bf 109} 025001

\bibitem{chen95}
Chen L and Zonca F 1995 {\em Phys. Scr.\/} {\bf T60} 81

\bibitem{zonca96a}
Zonca F and Chen L 1996 {\em Phys. Plasmas\/} {\bf 3} 323

\bibitem{wang13}
Wang Z, Lin Z, Holod I, Heidbrink W~W, Tobias B, {\relax Van Zeeland} M and
  Austin M~E 2013 {\em Phys. Rev. Lett.\/} {\bf 111} 145003

\bibitem{chen13}
Chen L and Zonca F 2013 {\em Phys. Plasmas\/} {\bf 20} 055402

\bibitem{berk90a}
Berk H~L and Breizman B~N 1990 {\em Phys. Fluids B\/} {\bf 2} 2226

\bibitem{berk90b}
Berk H~L and Breizman B~N 1990 {\em Phys. Fluids B\/} {\bf 2} 2235

\bibitem{berk96}
Berk H~L, Breizman B~N and Pekker M 1996 {\em Phys. Rev. Lett.\/} {\bf 76} 1256

\bibitem{berk97}
Berk H~L, Breizman B~N and Petiashvili N~V 1997 {\em Phys. Lett. A\/} {\bf 234}
  213

\bibitem{breizman97}
Breizman B~N, Berk H~L, Pekker M, Porcelli F, Stupakov G~V and Wong K~L 1997
  {\em Phys. Plasmas\/} {\bf 4} 1559

\bibitem{berk99}
Berk H~L, Breizman B~N, Candy J, Pekker M and Petiashvili N~V 1999 {\em Phys.
  Plasmas\/} {\bf 6} 3102

\bibitem{wang12b}
{\relax Ge Wang} and Berk H~L 2012 {\em Nucl. Fusion\/} {\bf 52} 094003

\bibitem{wang13b}
{\relax Ge Wang} 2013 {\em {Ph. D. Thesis}\/} (Austin, TX: University of Texas)

\bibitem{breizman11a}
Breizman B~N 2011 {\em Fus. Sci. Technol.\/} {\bf 59} 549

\bibitem{mynick94}
Mynick H~E and Pomphrey N 1994 {\em Nucl. Fusion\/} {\bf 34} 1277

\bibitem{friedland06}
Friedland L, Khain P and Shagalov A~G 2006 {\em Phys. Rev. Lett.\/} {\bf 96}
  225001

\bibitem{deng10}
Deng W, Lin Z, Holod I, Wang X, Xiao Y and Zhang W 2010 {\em Phys. Plasmas\/}
  {\bf 17} 112504

\bibitem{deng12b}
Deng W, Lin Z, Holod I, Wang Z, Xiao Y and Zhang H 2012 {\em Nucl. Fusion\/}
  {\bf 52} 043006

\bibitem{tobias11}
Tobias B~J, Classen {\relax I G J}, Domier C~W, Heidbrink W~W, {\relax Luhmann
  Jr} N~C, Nazikian R, Park H~K, Spong D~A and {\relax Van Zeeland} M~A 2011
  {\em Phys. Rev. Lett.\/} {\bf 106} 075003

\bibitem{zonca00a}
Zonca F and Chen L 2000 {\em Phys. Plasmas\/} {\bf 7} 4600

\bibitem{briguglio02}
Briguglio S, Vlad G, Zonca F and Fogaccia G 2002 {\em Phys. Lett. A\/} {\bf
  302} 308

\bibitem{zonca02}
Zonca F, Briguglio S, Chen L, Dettrick S, Fogaccia G, Testa D and Vlad G 2002
  {\em Phys. Plasmas\/} {\bf 9} 4939

\bibitem{vlad04}
Vlad G, Briguglio S, Fogaccia G and Zonca F 2004 {\em Plasma Phys. Control.
  Fusion\/} {\bf 46} S81

\bibitem{briguglio07}
Briguglio S, Fogaccia G, Vlad G, Zonca F, Shinohara K, Ishikawa M and Takechi M
  2007 {\em Phys. Plasmas\/} {\bf 14} 055904

\bibitem{park92}
Park W, Parker S, Biglari H, Chance M, Chen L, Cheng C~Z, Hahm T~S, Lee W~W,
  Kulsrud R, Monticello D, Sugiyama L and White R~B 1992 {\em Phys. Fluids B\/}
  {\bf 4} 2033

\bibitem{white12}
White R~B 2012 {\em Commun. Nonlinear Sci. Numer. Simul.\/} {\bf 17} 2200

\bibitem{vlad99}
Vlad G, Zonca F and Briguglio S 1999 {\em Riv. Nuovo Cimento\/} {\bf 22} 1

\bibitem{vlad06}
Vlad G, Briguglio S, Fogaccia G, Zonca F and Schneider M 2006 {\em Nucl.
  Fusion\/} {\bf 46} 1

\bibitem{vlad09}
Vlad G, Briguglio S, Fogaccia G, Zonca F, {\relax Di Troia} C, Heidbrink W~W,
  {\relax Van Zeeland} M~A, Bierwage A and Wang X 2009 {\em Nucl. Fusion\/}
  {\bf 49} 075024

\bibitem{bierwage11}
Bierwage A, Todo Y, Aiba N, Shinohara K, Ishikawa M and Yagi M 2011 {\em Plasma
  Fus. Res.\/} {\bf 6} 2403109

\bibitem{bierwage12}
Bierwage A, Aiba N, Todo Y, Deng W, Ishikawa M, Matsunaga G, Shinohara K and
  Yagi M 2012 {\em Plasma Fus. Res.\/} {\bf 7} 2403081

\bibitem{qiu14d}
Qiu Z, Chen L and Zonca F 2014 {\em Proceedings of the 41.st EPS Conference on
  Plasma Physics, Berlin, Germany, 23 - 27 June, (2014), ECA\/} vol 38F (EPS) 
  Paper No. P4.004
  
\bibitem{chen88b}
Chen L, Vaclavik J and Hammett G~W 1988 {\em Nucl. Fusion\/} {\bf 28} 389

\bibitem{bergkvist04}
Bergkvist T and Hellsten T 2004 {\em Theory of Fusion Plasmas\/} ed Connor J~W,
  Sauter O and Sindoni E (Bologna, Italy: Editrice Compositori, Societ\`{a}
  Italiana di Fisica) p 123

\bibitem{bergkvist05}
Bergkvist T, Hellsten T, Johnson T and {\relax Lax{\aa}back} M 2005 {\em Nucl.
  Fusion\/} {\bf 45} 485

\bibitem{fu10}
Fu G~Y, Lang J, Chen Y, Berk H~L, Fredrickson E, Gorelenkov N and {\relax
  Podest\`a} M 2010 {\em Proceedings of the 23rd International Conference on
  Fusion Energy 2010\/} (Vienna: International Atomic Energy Agency) CD--ROM
  file THW/2--2Rb

\bibitem{lang11}
Lang J and Fu G~Y 2011 {\em Phys. Plasmas\/} {\bf 18} 055902

\bibitem{lu12}
Lu Z~X, Zonca F and Cardinali A 2012 {\em Phys. Plasmas\/} {\bf 19} 042104

\bibitem{dewar81}
Dewar R~L, Manickam J, Grimm R~C and Chance M~S 1981 {\em Nucl. Fusion\/} {\bf
  21} 493

\bibitem{dewar82}
Dewar R~L, Manickam J, Grimm R~C and Chance M~S 1982 {\em Nucl. Fusion\/} {\bf
  22} 307

\bibitem{guo09}
Guo Z, Chen L and Zonca F 2009 {\em Phys. Rev. Lett.\/} {\bf 103} 055002

\bibitem{chen12}
Chen L and Zonca F 2012 {\em Phys. Rev. Lett.\/} {\bf 109} 145002

\bibitem{kosuga12}
Kosuga Y and Diamond P~H 2012 {\em Phys. Plasmas\/} {\bf 19} 072307

\bibitem{zonca00b}
Zonca F, Briguglio S, Chen L, Fogaccia G and Vlad G 2000 {Theoretical Aspects
  of Collective Mode Excitations by Energetic Ions in Tokamaks} {\em Theory of
  Fusion Plasmas\/} ed Connor J~W, Sauter O and Sindoni E (Bologna: SIF) p~17

\bibitem{zonca07b}
Zonca F and Chen L 2007 {\em Proceedings of the 34.th EPS Conference on Plasma
  Physics, Warsaw, Poland, 2 - 6 July, (2007), ECA\/} vol 31F (EPS) CD--ROM
  file P4.071

\bibitem{mcguire83}
McGuire K, Goldston R and {Bell et al} M 1983 {\em Phys. Rev. Lett.\/} {\bf 50}
  891

\bibitem{kaku93}
Kaku M 1993 {\em Quantum Field Theory: A Modern Introduction\/} (New York:
  Oxford University Press, Inc.)

\bibitem{drummond62}
Drummond W~E and Pines D 1962 {\em Nucl. Fusion Suppl. Pt.\/} {\bf 3} 1049

\bibitem{vedenov61c}
Vedenov A~A, Velikhov E~P and Sagdeev R~Z 1961 {\em Nucl. Fusion\/} {\bf 1} 82

\bibitem{oneil65}
{O'Neil} T~M 1965 {\em Phys. Fluids\/} {\bf 8} 2255

\bibitem{oneil71}
{O'Neil} T~M, Winfrey J~H and Malmberg J~H 1971 {\em Phys. Fluids\/} {\bf 14}
  1204

\bibitem{berk97b}
Berk H~L, Breizman B~N and Pekker M~S 1997 {\em Plasma Phys. Rep.\/} {\bf 23}
  778

\bibitem{vann03}
Vann {\relax R G L}, Dendy R~O, Rowlands G, Arber T~D and {\relax d'Ambrumenil}
  N 2003 {\em Phys. Plasmas\/} {\bf 10} 623

\bibitem{vann05}
Vann {\relax R G L}, Dendy R~O and Gryaznevich M~P 2005 {\em Phys. Plasmas\/}
  {\bf 12} 032501

\bibitem{lesur09}
Lesur M, Idomura Y and Garbet X 2009 {\em Phys. Plasmas\/} {\bf 16} 092305

\bibitem{lesur12}
Lesur M and Idomura Y 2012 {\em Nucl. Fusion\/} {\bf 52} 094004

\bibitem{sagdeev69}
Sagdeev R~Z and Galeev A~A 1969 {\em {Nonlinear Plasma Theory}\/} (W. A.
  Benjamin Inc.)

\bibitem{connor78}
Connor J~W, Hastie R~J and Taylor J~B 1978 {\em Phys. Rev. Lett.\/} {\bf 40}
  396

\bibitem{zonca93a}
Zonca F and Chen L 1993 {\em Phys. Fluids B\/} {\bf 5} 3668

\bibitem{stix72}
Stix T~H 1972 {\em Plasma Phys.\/} {\bf 14} 367

\bibitem{zonca13b}
Zonca F and Chen L 2013 {\em Bull. Am. Phys. Soc.\/} {\bf 58}(16) 75

\bibitem{gross61}
Gross E~P 1961 {\em Nuovo Cimento\/} {\bf 20} 454

\bibitem{pitaevsky61}
Pitaevsky L~P 1961 {\em Sov. Phys. JETP\/} {\bf 13} 451

\bibitem{zakharov68}
Zakharov V~E 1968 {\em J. Appl. Mech. Tech. Phys.\/} {\bf 9} 190

\bibitem{cary09}
Cary J~R and Brizard A~J 2009 {\em Rev. Mod. Phys.\/} {\bf 81} 693

\bibitem{boozer81}
Boozer A~H 1981 {\em Phys. Fluids\/} {\bf 24} 1999

\bibitem{boozer82}
Boozer A~H 1982 {\em Phys. Fluids\/} {\bf 25} 520

\bibitem{white89b}
White R~B 1989 {\em Theory of Tokamak Plasmas\/} (Amsterdam: North Holland)

\bibitem{lichtenberg10}
Lichtenberg A~J and Lieberman M~A 2010 {\em Regular and Chaotic Dynamics\/} 2nd
  ed (Springer - Verlag)

\bibitem{levin72a}
Levin M~B, Lyubarskii M~G, Onishchenko I~N, Shapiro V~D and Shevchenko V~I 1972
  {\em Zh. Eksp. Teor. Fiz.\/} {\bf 62} 1725

\bibitem{levin72b}
Levin M~B, Lyubarskii M~G, Onishchenko I~N, Shapiro V~D and Shevchenko V~I 1972
  {\em Sov. Phys. JETP\/} {\bf 35} 898

\bibitem{wu95}
Wu Y, White R~B, Chen Y and Rosenbluth M~N 1995 {\em Phys. Plasmas\/} {\bf 2}
  4555

\bibitem{xiao11}
Xiao Y and Lin Z 2011 {\em Phys. Plasmas\/} {\bf 18} 110703

\bibitem{gustafson12a}
Gustafson K, Ricci P, Furno I and Fasoli A 2012 {\em Phys. Rev. Lett.\/} {\bf
  108} 035006

\bibitem{gustafson12b}
Gustafson K, Ricci P, Bovet A, Furno I and Fasoli A 2012 {\em Phys. Plasmas\/}
  {\bf 19} 062306

\bibitem{feng13}
Feng Z, Qiu Z and Sheng Z 2013 {\em Phys. Plasmas\/} {\bf 20} 122309

\bibitem{fredrickson09}
Fredrickson E~D, Crocker N~A, Bell R~E, Darrow D~S, Gorelenkov N~N, Kramer G~J,
  Kubota S, Levinton F~M, Liu S, Medley S~S, Podest\`{a} M, Tritz K, White R~B
  and Yuh H 2009 {\em Phys. Plasmas\/} {\bf 16} 122505

\bibitem{podesta11}
Podest\`{a} M, Bell R~E, Crocker N~A, Fredrickson E~D, Gorelenkov N~N,
  Heidbrink W~W, Kubota S, {\relax LeBlanc} B~P and Yuh H 2011 {\em Nucl.
  Fusion\/} {\bf 51} 063035

\bibitem{podesta12}
Podest\`{a} M, Bell R~E, Bortolon A, Crocker N~A, Darrow D~S, Diallo A,
  Fredrickson E~D, Fu {\relax G-Y}, Gorelenkov N~N, Heidbrink W~W, Kramer G~J,
  Kubota S, {\relax LeBlanc} B~P, Medley S~S and Yuh H 2012 {\em Nucl.
  Fusion\/} {\bf 52} 094001

\bibitem{zonca00c}
Zonca F and Chen L 2000 {Destabilization of Energetic Particle Modes by ICRF
  induced fast minority ion tails on TFTR} {\em Proceedings of the 6.th IAEA
  TCM on Energetic Particles in Magnetic Confinement Systems\/} (Naka, Japan:
  JAERI-Conf 2000-004) p~52

\bibitem{fu06b}
Fu G~Y, Park W, Strauss H~R, Breslau J, Chen J, Jardin S and Sugiyama L~E 2006
  {\em Phys. Plasmas\/} {\bf 13} 052517

\bibitem{vlad12}
Vlad G, Briguglio S, Fogaccia G, Zonca F, {\relax Di Troia} C, Fusco V and Wang
  X 2012 {Electron Fishbone Simulations in FTU-Like Equilibria Using XHMGC}
  {\em Proceedings of the 24th International Conference on Fusion Energy\/}
  (Vienna: International Atomic Energy Agency) CD--ROM
  file TH/P6-03 and 
  http://www--naweb.iaea.org/napc/physics/FEC/FEC2012/papers/77\_THP603.pdf

\end{thebibliography}

\providecommand{\newblock}{}

\end{document}